\documentclass[11pt]{article}
\usepackage[affil-it]{authblk}
\usepackage[utf8]{inputenc}
\usepackage{amsmath}
\usepackage{amssymb}
\usepackage[a4paper]{geometry}
\usepackage{amsthm}
\usepackage{subcaption}
\usepackage{graphicx}
\usepackage{tikz}
\usepackage{tikz-cd}
\usepackage{mathrsfs}
\usepackage{bm}
\usepackage{bbold}
\usepackage{slashed}
\usepackage[hidelinks]{hyperref}
\usepackage{listings}

\hypersetup{
    colorlinks = true,
    linkcolor = {black},
    citecolor = {blue},
    anchorcolor = {green},
    citecolor = {red},
    filecolor = {cyan},
    menucolor = {red},
    runcolor = {cyan},
    urlcolor = {magenta}
}
\newcommand{\twidth}{6in}
\textwidth=\twidth
\renewcommand{\AA}{{\mathcal{A}}}
\newcommand{\BB}{{\mathcal{B}}}
\newcommand{\CC}{{\mathcal{C}}}

\newcommand{\FF}{{\mathcal{F}}}
\newcommand{\GG}{{\mathcal{G}}}

\newcommand{\II}{{\mathcal{I}}}

\newcommand{\LL}{{\mathcal{L}}}

\newcommand{\OO}{{\mathcal{O}}}
\newcommand{\PP}{{\mathcal{P}}}

\newcommand{\RR}{{\mathcal{R}}}
\renewcommand{\SS}{{\mathcal{S}}}
\newcommand{\TT}{{\mathcal{T}}}

\newcommand{\f}{{\mathfrak{f}}}

\newcommand{\su}{{\mathfrak{su}}}

\renewcommand{\O}{{\mathrm{O}}}
\newcommand{\SU}{{\mathrm{SU}}}
\newcommand{\SO}{{\mathrm{SO}}}

\newcommand{\A}{{\mathbb{A}}}
\newcommand{\E}{{\mathbb{E}}}

\newcommand{\R}{{\mathbb{R}}}

\newcommand{\Z}{{\mathbb{Z}}}

\newcommand{\C}{{\mathbb{C}}}
\renewcommand{\H}{{\mathbb{H}}}

\newcommand{\ind}{{\mathbb{1}}}
\newcommand{\CP}{{\mathbb{C}}{\mathbb{P}}}

\renewcommand{\j}{{\bm{j}}}
\renewcommand{\k}{{\bm{k}}}
\newcommand{\1}{{\bm{1}}}
\newcommand{\beq}{\begin{equation}}
\newcommand{\eeq}{\end{equation}}
\newcommand{\bea}{\begin{eqnarray}}
\newcommand{\eea}{\end{eqnarray}}
\newcommand{\bal}{\begin{align}}
\newcommand{\eal}{\end{align}}
\newcommand{\bml}{\begin{multline}}
\newcommand{\eml}{\end{multline}}

\newcommand{\bdy}{\partial}

\newcommand{\wt}{\widetilde}
\newcommand{\wh}{\widehat}
\newcommand{\lto}{\longrightarrow}

\def \d{\mathrm{d}}
\newcommand{\ol}{\overline}
\newcommand{\tr}{{\rm tr}\, }

\newcommand{\im}{{\rm im}\, }

\newcommand{\id}{{\rm Id}}

\newcommand{\diag}{{\rm diag}}

\def\sfrac#1#2{{\textstyle\frac{#1}{#2}}}

\makeatletter
\newcommand\xleftrightarrow[2][]{%
  \ext@arrow 9999{\longleftrightarrowfill@}{#1}{#2}}
\newcommand\longleftrightarrowfill@{%
  \arrowfill@\leftarrow\relbar\rightarrow}
\makeatother

\newcommand{\bigslant}[2]{{\raisebox{.2em}{$#1$}\left/\raisebox{-.2em}{$#2$}\right.}}

\newtheorem{theorem}{Theorem}

\newtheorem{corollary}[theorem]{Corollary}

\newtheorem{lemma}[theorem]{Lemma}

\newtheorem{remark}[theorem]{Remark}

\makeatletter
\def\@maketitle{%
  \newpage
  \null
  \vskip 2em%
  \begin{center}%
  \let \footnote \thanks
    {\huge \@title \par}%
    \vskip 1.5em%
    {\normalsize
      \lineskip .5em%
      \begin{tabular}[t]{c}%
        \@author
      \end{tabular}\par}%
    \vskip 1em%
    {\normalsize \@date}%
  \end{center}%
  \par
  \vskip 1.5em}
\makeatother

\title{ADHM skyrmions}
\author{{\Large Josh Cork}%
  \thanks{Email address: \texttt{joshua.cork@itp.uni-hannover.de} and \texttt{jc893@leicester.ac.uk}}}
\affil{\textit{Institut f\"ur Theoretische Physik} and \textit{Riemann Center for Geometry and Physics}\\
Leibniz Universit\"at Hannover, Appelstraße 2, 30167 Hannover, Germany\vskip0.05cm
and\vskip0.05cm
\textit{School of Computing and Mathematical Sciences, The University of Leicester\\
University Road, Leicester, LE1 7RH, United Kingdom}}

\author{{\Large Chris Halcrow}%
  \thanks{Email address: \texttt{c.j.halcrow@leeds.ac.uk}}}
\affil{School of Mathematics, University of Leeds\\ Woodhouse Lane, Leeds, LS2 9JT, United Kingdom}
\date{\large \today}
\numberwithin{equation}{section}
\begin{document}

\maketitle
\begin{abstract}
\noindent We propose, via the Atiyah--Manton approximation, a framework for studying skyrmions on $\R^3$ using ADHM data for Yang--Mills instantons on $\R^4$. We provide a dictionary between important concepts in the Skyrme model and analogous ideas for ADHM data, and describe an efficient process for obtaining approximate Skyrme fields directly from ADHM data. We show that the approximation successfully describes all known skyrmions with charge $\BB \leq 8$, with energies reproduced within 2\% of the true minimisers. We also develop factorisation methods for studying clusters of instantons and skyrmions, generalising early work by Christ--Stanton--Weinberg, and describe some relatively large families of explicit ADHM data. These tools provide a unified framework for describing coalesced highly-symmetric configurations as well as skyrmion clusters, both of which are needed to study nuclear systems in the Skyrme model.
\end{abstract}
\hypersetup{
    linkcolor = {blue}
}
\newpage
\section{Introduction}

Instantons are topologically nontrivial solutions of pure Yang--Mills gauge theory on euclidean $\mathbb{R}^4$. They are classified by a conserved integral charge which we denote by $\BB$, and for each $\BB$ there is an $8\BB$-dimensional moduli space of instantons, each with the same energy. All $8\BB$ moduli may be found via the Atiyah--Drinfeld--Hitchin--Manin (ADHM) construction \cite{ADHM1978construction}, which reduces the self-dual Yang--Mills equations into a system of nonlinear algebraic equations. As such, instantons are one of the best understood and well-studied theories of topological solitons. Some time ago, Atiyah and Manton \cite{AtiyahManton1989} showed that one could use instantons to approximate configurations in a less understood system: the Skyrme model.

The Skyrme model \cite{Skyrme1962nucl} is a nonlinear theory of pions which also allows for topological solitons. These are also characterised by a topological charge, which we also call $\BB$, and an energy-minimising configuration in the $\BB$-sector is called a $\BB$-skyrmion. Physically, skyrmions are identified with nuclei and the topological charge with the nucleon (baryon) number. Separated 1-skyrmions have forces between them and can be oriented to attract one another; generally the 1-skyrmions coalesce to form the $\BB$-skyrmion, a compact object often with high symmetry. Unlike instantons, there is usually a unique energy minimiser (up to isometries) and it costs energy to break skyrmions into smaller clusters, just as it costs energy to break large nuclei into smaller ones.

Witten showed \cite{witten1979baryons} that the Skyrme model is an effective theory of QCD in the large $N_C$ limit. Soon after, Adkins--Nappi--Witten used the model to calculate properties of the nucleon, with reasonable agreement to experimental data \cite{AdkinsNappiWitten1983static}. They used a moduli space quantisation where only the zero modes of the skyrmion are allowed. Physically, the skyrmion can rotate and translate but not deform. Since deformations cost energy, this approximation is valid at low energies. Following this work, moduli space quantisation was implemented for larger nuclei \cite{Braaten:1985np,Irwin:1998bs} but few authors have gone beyond this approximation. Moreover, the papers which try to take account of deformations do so in a variety of ways: using the Rational Map approximation (RMA) \cite{Lin:2008uf}, modelling skyrmions as point particles \cite{Halcrow:2016spb}, and restricting skyrmions to important one-dimensional paths from a larger space \cite{Halcrow:2015rvz, Rawlinson:2017rcq}. This patchwork of methods has arisen because there is still no agreed unified approach for describing skyrmions as they deform and break into clusters. Techniques such as the RMA can describe the coalesced symmetric skyrmions while the Product Approximation can describe configurations which look like widely separated clusters. The only known approach which can describe both types of configurations, and how they are related, is the instanton approximation.

Atiyah and Manton suggested that instantons could be used to approximate skyrmions simply by noting that a Skyrme field could be constructed easily from the gauge field of an instanton \cite{AtiyahManton1989}. This is done by defining a Skyrme field at each point in $\R^3$ by the holonomy taken along the complementary fourth dimension in $\R^4$. The construction can describe well-separated clusters, compact symmetric skyrmions, and what happens in between. Atiyah--Manton used this freedom to model the classical 2-skyrmion configuration space \cite{AtiyahManton1993}, while Leese--Manton--Schroers built a restricted 2-skyrmion space where they could model the deuteron as a quantum state on this space \cite{Leese:1994hb}. Others constructed known symmetric skyrmions from instantons for $\BB = 3,4$ \cite{LeeseManton1994stable,houghton1999-3Skyrme}, $7$ \cite{SingerSutcliffe1999}, and $17$ \cite{sutcliffe2004Buckyball}. The approximation works well, reproducing skyrmion energies to an accuracy of 2\%. Recently Sutcliffe demonstrated why the approximation works so well \cite{sutcliffe2010skyrmions}. Inspired by the holographic model of Sakai--Sugimoto \cite{SakaiSugimoto2005}, Sutcliffe showed that the standard Skyrme model is the first term in an infinite series of Skyrme models coupled to towers of vector mesons. Recently, it was conjectured that the full moduli space of instantons can be understood as low energy modes of a skyrmion, provided at least one vector meson is included in the Skyrme theory \cite{HalcrowWinyard2021consistent}.

Given that the instanton approximation works well, is more powerful than other common approximations, and has a solid theoretical basis, one may ask: why is it not used more widely? One reason is that there is no simple dictionary between instanton and skyrmion configurations. We hope to make this map more transparent using ADHM data. Another problem is the numerical difficulty in generating skyrmions from instantons. Na\"ively one must solve an ODE, which depends on the gauge field of the instanton, at every point in space. Using ADHM data, calculating the gauge field is non-trivial and it can contain gauge singularities. We bypass both of these issues by generating the Skyrme field using a finite-difference approximation to parallel transport, as recently developed in \cite{CorkHarlandWinyard2021gaugedSkyrmelowbinding}. This reduces the problem of generating skyrmions from instantons to calculating the kernel of a linear operator, which one can do efficiently. We take advantage of this new technique to explore instanton-generated-skyrmions on larger spaces; this allows us to find, for example, the energy-minimising $\BB = 5,6,$ and $8$ skyrmions generated from instantons for the first time.

One obstacle to using instantons to model real nuclei, not addressed in this paper, is that the skyrmions decay polynomially, which occurs in theories with massless pions. In reality, pions have mass and the skyrmions should decay exponentially. There has been a suggestion to include the mass by considering the holonomy of instantons along circles (instead of lines) to approximate skyrmions in hyperbolic space, which in turn may approximate skyrmions in euclidean space with massive pions \cite{AtiyahSutcliffe2005skyrmions}. However, this approach has not been well-tested, and furthermore, recent work relating calorons to skyrmions \cite{CorkHarlandWinyard2021gaugedSkyrmelowbinding,cork2018skyrmions} suggests that to do this consistently requires the inclusion of gauge fields -- i.e. \textit{gauged Skyrme models}. Nevertheless, if this problem of including the pion mass in the instanton approximation can be overcome, we hope our results show that the instanton approximation can become the standard method for studying skyrmions in the future.

This paper is organised as follows. In Section \ref{sec:Sky-and-inst} we review the instanton approximation of skyrmions and the moduli space of ADHM data, discuss how to efficiently approximate skyrmions from ADHM data, and how to study symmetric solutions. In Section \ref{sec:decompositions} we lay out the theoretical framework and phenomenology for describing clusters of skyrmions via ADHM data. Sections
\ref{sec:database} and \ref{sec:ADHM-zoo} are dedicated to studying explicit examples of ADHM data. The former is a database of ADHM data corresponding to the minimal energy skyrmions of charge $1\leq\BB\leq8$, along with the energies of the energy-minimising instanton-generated-skyrmions. In the latter we construct a variety of ADHM data, including all charge $\BB$ tori, and large families which interpolate between well-separated clusters and highly-symmetric configurations. We conclude Section \ref{sec:ADHM-zoo} with a numerical calculation of a potential function on the $\BB=4$ moduli space, showcasing the power of this tool for applications in nuclear physics. Some important results regarding symmetric ADHM data are stated in Section \ref{sec:symmetric-data}, which we use throughout, however their proofs are rather technical. These, along with other prerequisite properties of ADHM data, are therefore given in Appendix \ref{appendix:properties-ADHM}.
\renewcommand{\i}{{\bm{i}}} 
\section{Skyrmions and instantons}\label{sec:Sky-and-inst}
The Skyrme model consists of a single $\SU(2)$-valued field $U:\R^3\lto\SU(2)$ called the Skyrme field. Introducing the left-invariant $\su(2)$-current $\LL=U^{-1}\d U\in\Omega^1(\R^3,\su(2))$, we may express the static energy of a Skyrme field as
\begin{align}
    E[U]&=-\int_{\R^3}\tr\left(c_2\LL\wedge\star_3\LL+c_4(\LL\wedge\LL)\wedge\star_3(\LL\wedge\LL)\right)\notag\\
    &=-\int_{\R^3}\tr\left(c_2\LL_i\LL_i+\frac{c_4}{2}[\LL_i,\LL_j][\LL_i,\LL_j]\right)\;\d^3x,\label{Skyrme-energy}
\end{align}
where $c_2,c_4>0$ are arbitrary constants which represent a choice of length and energy units. In this paper we use Skyrme units, and fix $c_2=\sfrac{1}{24\pi^2}$ and $c_4=\sfrac{1}{96\pi^2}$.

One imposes the boundary condition $U\to\id$ as $|\vec{x}|\to\infty$ on the Skyrme field, which means that $U$ descends to a map $\wt{U}:S^3\to\SU(2)$. Such maps are classified by the homotopy group $\pi_3(\SU(2))=\Z$, with each classifying integer given by the degree of $\wt{U}$. The degree of a Skyrme field is physically identified as the nucleon/baryon number, which we denote by $\BB$. The baryon number may be calculated via an integral, namely
\begin{align}
    \BB=\frac{1}{24\pi^2}\int_{\R^3}\tr(\LL\wedge\LL\wedge\LL).\label{baryon-charge-int}
\end{align}
We are also interested in instantons on $\R^4$. Instantons are the minima of the pure Yang--Mills energy on euclidean $\R^4$. Explicitly, let $\AA$ be an $\SU(2)$ gauge field on $\R^4$, with field strength (curvature) $\FF=\d\AA+\AA\wedge\AA\in\Omega^2(\R^4,\su(2))$. $\AA$ is called an instanton if its curvature is anti-self-dual, that is
\begin{align}
    \star_4\FF=-\FF,\label{ASD}
\end{align}
and has finite energy. The Yang--Mills energy is given by
\begin{align}
    E_{\rm YM}[\AA]=-\int_{\R^4}\tr(\FF\wedge\star_4\FF)=-\frac{1}{2}\int_{\R^4}\tr(\FF_{\mu\nu}\FF_{\mu\nu})\;\d^4x.\label{YM-energy}
\end{align}
By completing the square, one obtains
\begin{equation}
    E_{\rm YM}[\AA]=-\frac{1}{2}\int_{\R^4}\tr((\FF+\star_4\FF)\wedge\star_4(\FF+\star_4\FF))+\int_{\R^4}\tr(\FF\wedge\FF)\notag
\end{equation}
\begin{equation}
    \Rightarrow E_{\rm YM}\geq 8\pi^2|Q|,
\end{equation}
where
\begin{align}
    Q=\frac{1}{8\pi^2}\int\tr(\FF\wedge\FF),\label{topological-charge}
\end{align}
with equality if and only if $\FF$ is anti-self-dual \eqref{ASD}. For finite-energy configurations, the quantity \eqref{topological-charge} is a topological invariant. In fact, \eqref{topological-charge} is an integer $-\BB$, where $\BB\in\Z$ is called the instanton number in this context. This integer may be understood in a variety of ways, but the most invariant way is as follows. By conformal invariance of \eqref{ASD}, one may impose a boundary condition where the gauge field extends smoothly to the conformal compactification $\R^4\cup\{\infty\}=S^4$; in fact finite-energy is equivalent to this condition \cite{Uhlenbeck1982}. One then identifies \eqref{topological-charge} as the second Chern number $c_2(S^4)$ of the associated bundle over $S^4$. The instanton number may then be realised as the degree of a corresponding transition function between patches of the four-sphere, for instance $g_{NS}:S^3=U_N\cap U_S\lto\SU(2)$ between the northern and southern hemispheres; this is often referred to as the `gauge transformation at infinity' in the physics literature.
\subsection{The Atiyah--Manton construction}
Many years ago, Atiyah and Manton proposed a relatively simple ansatz for a Skyrme field by using instantons on $\R^4$ \cite{AtiyahManton1989}. In short, a Skyrme field is determined as the holonomy of an instanton along all lines in $\R^4$ parallel to a particular direction. Without loss of generality we can take this direction to be $x_4$, and to identify this clearly, we shall label this coordinate by $h$. The holonomy $U:\R^3\lto\SU(2)$ is calculated explicitly by solving the parallel transport equation
\begin{align}
    \bdy_hH+\AA_hH=0,\quad \lim_{h\to-\infty}H=\id,\label{pt-eqn}
\end{align}
where $H:\R^3\times\R\lto\SU(2)$, and then setting $U(\vec{x})=\lim_{h\to\infty}H(\vec{x},h)$. The unique solution to the initial-value-problem \eqref{pt-eqn} is given formally by the path-ordered exponential
\begin{align}
    H(\vec{x},h)=\PP\exp\left(-\int_{-\infty}^{h}\AA_h(\vec{x},z)\;\d z\right).
\end{align}
The holonomy $U$ is a reasonable candidate for a Skyrme field; it respects the correct boundary conditions, namely $U\to\id$ as $|\vec{x}|\to\infty$, and furthermore, this construction is topological, with the baryon number $\BB$ of the Skyrme field exactly equal to the instanton number $\BB$, hence the identification in notation. We shall refer to a Skyrme field $U$ generated from the holonomy of an instanton as an \textbf{i-skyrmion} (instanton-generated-skyrmion). This approach has proven to provide a remarkably good approximation to skyrmions, with the energies of the i-skyrmions within $2\%$ of the energies of skyrmions obtained through direct numerical minimisation of \eqref{Skyrme-energy}.

\subsection{Instanton moduli spaces and ADHM data}
The group $\GG$ of gauge transformations $g:\R^4\lto\SU(2)$ acts on instantons via
\begin{align}
    \AA\mapsto g\AA g^{-1}-\d gg^{-1}.
\end{align}
The boundary conditions for instantons require fixing an isomorphism $\f:P_\infty\lto\SU(2)$ which identifies the fiber of a principal bundle $P\to\R^4\cup\{\infty\}\cong S^4$ with a fixed trivialisation at infinity; this identification is known as a framing (see e.g. \cite{donaldson1992boundary}). The moduli space $\II_\BB$ of framed $\BB$-instantons \cite{donaldson1984instantons} consists of equivalence classes of pairs $(\f,\AA)$, namely framings and instantons with instanton number $\BB$. This may be viewed equivalently as the quotient $\II_\BB=\bigslant{\CC_\BB}{\GG^0}$, where $\CC_\BB$ is the space of all (possibly gauge-equivalent) $\BB$-instantons, and
\begin{align}
    \GG^0=\{g:\R^4\lto\SU(2)\::\:g\to\id\text{ as }|x|\to\infty\}\label{framed-gauge-group}
\end{align}
is the gauge group of framed gauge transformations. Physically, those gauge transformations which are not identity at infinity (and therefore act non-trivially on the framing) account for a change in the global gauge orientation of the instanton. It is important to include these additional modes in order to have a full description of the moduli spaces. The moduli spaces $\II_\BB$ are $8\BB$-dimensional manifolds with various nice structures, for example they always admit a hyperk\"ahler metric.
\subsubsection{ADHM data and transform}
The moduli spaces $\II_\BB$ are parameterised fully by a moduli space of maps on hermitian vector bundles \cite{ADHM1978construction,atiyah1978geometry}. However, the most convenient description packages these into a moduli space of matrices called ADHM data, as described in \cite{christWeinbergStanton1978general,CorriganFairlieTempletonGoddard1978green}. This consists of a pair $(L,M)$ where $L$ is a length $\BB$ row vector of quaternions\footnote{We define the quaternions $\H$ as the vector space $\R^4$ with basis $\{\i,\j,\k,\1\}$ endowed with the multiplication $\i^2=\j^2=\k^2=\i\j\k=-\1$, with multiplicative unit $\1$. We denote by $\Re(\H)=\mathrm{sp}_\R\{\1\}$ and $\Im(\H)=\mathrm{sp}_\R\{\i,\j,\k\}$ the real and imaginary quaternions respectively.}, and $M$ is a symmetric $\BB\times\BB$ matrix of quaternions. The ADHM data is required to satisfy two main consistency conditions:
\begin{enumerate}
    \item The \textbf{reality condition}:
    \begin{align}
        \Im(L^\dagger L+M^\dagger M)=0,\label{reality-cond}
    \end{align}
    where $\Im$ denotes the quaternionic imaginary part, and ${}^\dagger$ denotes matrix transposition composed with the quaternionic conjugation $x\mapsto\ol{x}$, namely $\Im(x)\mapsto-\Im(x)$.
    \item The \textbf{irreducibility condition}: for all $x=x_1\i+x_2\j+x_3\k+x_4\1\in\H$,
    \begin{align}
        \det(\Delta_x^\dagger\Delta_x)\neq0,\label{non-singularity-cond}
    \end{align}
    where
    \begin{align}
        \Delta_x=\begin{pmatrix}L\\M\end{pmatrix}-x\begin{pmatrix}0\\\ind_\BB\end{pmatrix}.\label{ADHM-operator}
    \end{align}
\end{enumerate}
The first condition is the most important for guaranteeing anti-self-duality \eqref{ASD}. The second condition is less important, playing the role of removing singularities, along with ensuring the ADHM data is not inefficiently describing an instanton of lower charge.

The orthogonal group $\O(\BB)$ acts naturally on the set of all ADHM data via
\begin{align}
    O\cdot(L,M)=(LO^{-1},OMO^{-1}),\quad O\in\O(\BB).\label{orthog-gt}
\end{align}
The moduli space $\mathbb{A}_\BB$ of rank $\BB$ ADHM data is the space of all solutions $(L,M)$ to \eqref{reality-cond}-\eqref{non-singularity-cond} modulo the action \eqref{orthog-gt}.

The \textit{ADHM transform}, namely the process to obtain an instanton from given ADHM data $(L,M)$, is a simple construction. For each $x\in\H$ one chooses a length $\BB+1$ column vector $\Psi_x$ of quaternions satisfying
\begin{align} 
    \Delta_x^\dagger\Psi_x=0,\quad\text{and}\quad\Psi_x^\dagger\Psi_x=1,\label{ADHM-eqn}
\end{align}
where $\Delta_x$ is the matrix operator defined in \eqref{ADHM-operator}. Using these $\Psi_x$, one then defines a gauge field $\AA=\AA_\mu\;\d x^\mu$ on $\R^4$ pointwise as
\begin{align}
    \AA_\mu(x)=\Psi_x^\dagger\bdy_\mu\Psi_x.\label{ADHM-inst}
\end{align}
By identifying $\i=\tau^1,\;\j=\tau^2,\;\k=\tau^3,$ and $\1=\ind_2$, where $\tau^i=-{\rm i}\sigma^i$ are the $\su(2)$ spin matrices, \eqref{ADHM-inst} is found to be an $\su(2)$-valued $1$-form at $x=(x_1,x_2,x_3,x_4)\in\R^4$. Due to the constraints \eqref{reality-cond} and \eqref{non-singularity-cond}, the resulting $1$-form is a gauge field for an instanton with instanton number $\BB$, and every element of the moduli space $\II_\BB$ arises uniquely in this way \cite{ADHM1978construction,CorriganGoddard1984}.

One subtle point is as follows. To ensure that the resulting instanton is framed, we need to carefully consider the behaviour as $|x|\to\infty$. From \eqref{ADHM-operator}, we see that for $|x|$ large the ADHM equation \eqref{ADHM-eqn} is well-approximated by the limiting equation
\begin{align}
    \begin{pmatrix}
    0&-\ol{x}\ind_\BB
    \end{pmatrix}\Psi_x=0,\quad\Psi_x^\dagger\Psi_x=\1.\label{framing-eqn}
\end{align}
A framing is defined by a fixed choice of solution to \eqref{framing-eqn}, and in this paper we make the canonical choice
\begin{align}
    \Psi_\infty=\begin{pmatrix}
    \1\\0\\\vdots\\0
    \end{pmatrix}.\label{framing-Psi}
\end{align}
Finally, it is a straightforward exercise in parameter counting to verify that the moduli space $\A_\BB$ of ADHM data has real dimension $8\BB$.
\subsubsection{Approximating skyrmions with ADHM data}\label{sec:approximation-method}
In order to make meaningful comparisons with the Skyrme model, we should calculate i-skyrmions and their energies. In principle, given the ADHM data for an instanton, one can perform the ADHM transform to obtain an explicit expression for the gauge field, from which one may solve the parallel transport equation \eqref{pt-eqn} and obtain the associated Skyrme field. However, many of these steps, especially the latter which requires solving a differential equation, are analytically impractical, and so it is reasonable to settle for numerical approximations for the Skyrme field, its energy, etc.

Constructing the instanton explicitly and solving the equation \eqref{pt-eqn} is computationally expensive, and can also be tricky to implement due to the possible appearance of gauge-singularities in the gauge field. We can bypass this completely by using a finite-difference approach analogous to one used recently in the context of constructing gauged skyrmions from caloron Nahm data \cite{CorkHarlandWinyard2021gaugedSkyrmelowbinding}. The approach rests on the fact that the linear map determined by the $2\times 2$ complex matrix
\begin{align}
    \Omega_{x,\delta x}=\Psi^\dagger_{x+\delta x}\Psi_{x},\label{orthog-proj}
\end{align}
with $\Psi_x$ defined in \eqref{ADHM-eqn}, approximates parallel transport generated by \eqref{ADHM-inst} at $x$ along the straight line between $x$ and  $x+\delta x$. Here we have identified $\H$ with ${\rm sp}_\R(\tau^1,\tau^2,\tau^3,\ind_2)$ in the usual way so that now $\Psi_x$ is a $(2\BB+2)\times 2$ complex matrix.

To understand this approximation, it is useful to gain some geometric intuition for the ADHM transform. At each point $x\in\R^4$, the columns of $\Psi_x$ form an orthonormal basis for a $2$-dimensional complex inner product space. These in turn form a smooth orthonormal frame for a rank $2$ sub-bundle $E$ of the trivial rank $2\BB+2$ vector bundle $V=\R^4\times\C^{2\BB+2}$ over $\R^4$. Viewed in this way, the gauge field \eqref{ADHM-inst} is an expression in this basis for the connection on $E$ induced by the trivial connection on $V$, and the matrix \eqref{orthog-proj} describes orthogonal projection between fibers $E_x\lto E_{x+\delta x}$.

More explicitly, let $v:\R^4\to\C^2$ be such that $v(x+\delta x)=\Omega_{x,\delta x}v(x)$ for all $x,\delta x\in\R^4$. Then
\begin{align}
   \delta x_{\mu}\bdy_\mu v(x)\approx v(x+\delta x)-v(x)=\left(\Omega_{x,\delta x}-\id\right)v(x)=\left(\Psi_{x+\delta x}-\Psi_x\right)^\dagger \Psi_{x}v(x).
\end{align}
We also have
\begin{align}
    \left(\Psi_{x+\delta x}-\Psi_x\right)^\dagger \Psi_{x}v(x)\approx \delta x_\mu\bdy_\mu \Psi_x^\dagger \Psi_xv(x)=-\delta x_\mu \AA_\mu(x) v(x).
\end{align}
In each case, the final equality follows from the orthonormality condition $\Psi_x^\dagger\Psi_x=1$.

To compute the approximate holonomy from these operators practically, first we map the real line $\R$ bijectively to the finite interval $(0,\pi)$ via the reparameterisation $h\mapsto\tau$, $h=\tan(\tau-\sfrac{\pi}{2})$ \cite{LeeseManton1994stable}. Then discretising the interval as $\delta\tau/2=\tau_0<\tau_1<\cdots<\tau_n=\pi-\delta\tau/2$, with $\tau_j=\tau_0+j\delta\tau$, we approximate the holonomy via
\begin{align}
    U(\vec{x})\approx\Omega_{\vec{x}}(\tau_n,\tau_{n-1})\Omega_{\vec{x}}(\tau_{n-1},\tau_{n-2})\cdots\Omega_{\vec{x}}(\tau_1,\tau_0),\label{approximate-holonomy}
\end{align}
where $\Omega_{\vec{x}}(s_2,s_1):=\Omega_{(\vec{x},s_1),(\vec{x},s_2-s_1)}$, with the approximation improving as $\delta\tau\to0$. Note that the expression \eqref{approximate-holonomy} is more explicit when expanded out in terms of the $\Psi_x$ as
\begin{align}
\begin{aligned}
    U(\vec{x})&\approx\Psi_{(\vec{x},\tau_n)}^\dagger\Psi_{(\vec{x},\tau_{n-1})}\Psi_{(\vec{x},\tau_{n-1})}^\dagger\Psi_{(\vec{x},\tau_{n-2})}\Psi_{(\vec{x},\tau_{n-2})}^\dagger\cdots\Psi_{(\vec{x},\tau_{1})}\Psi^\dagger_{(\vec{x},\tau_1)}\Psi_{(\vec{x},\tau_0)}\\
    &=\Psi_{(\vec{x},\tau_n)}^\dagger P_{\tau_{n-1}}P_{\tau_{n-2}}\cdots P_{\tau_1}\Psi_{(\vec{x},\tau_0)},
    \end{aligned}\label{approximate-holonomy-Psis}
\end{align}
where $P_{\tau_k}:=\Psi_{(\vec{x},\tau_k)}\Psi_{(\vec{x},\tau_k)}^\dagger$ is the projector. In particular, this shows that $U$ is insensitive to choices of gauge for $\tau_0<\tau<\tau_n$ as the projectors are invariant under the unitary transformations $\Psi_x\mapsto\Psi_xg(x)^{-1}$. The values at $\tau_0$ and $\tau_n$ are dictated by the behaviour at $h=\pm\infty$, namely the framing, and this is fixed by the choice \eqref{framing-Psi}. Thus $U$ is invariant under the action of the gauge group $\GG^0$.

A caveat to this approach is that the operator on the right-hand-side of \eqref{approximate-holonomy} is only approximately unitary. This may be remedied without affecting its properties by making the replacement $U\mapsto U(U^\dagger U)^{-\sfrac{1}{2}}$.

In practice, we also calculate $\Psi_x$ numerically. To do this we follow \cite{LeeseManton1994stable} by using the ansatz for a non-normalised vector
\begin{equation}
\wt{\Psi}_{x} = \Psi_\infty - \Delta_x \phi(x),
\end{equation}
with $\Psi_\infty$ defined in \eqref{framing-Psi}. Substituting this into \eqref{ADHM-eqn} gives the linear equation
\begin{align}
    \Delta_x^\dagger\Delta_x\phi(x)=L^\dagger,
\end{align}
which can be solved uniquely for the vector $\phi(x)$ since, by \eqref{reality-cond}-\eqref{non-singularity-cond}, $\Delta_x^\dagger\Delta_x$ is real and invertible. A normalised vector is then found by dividing $\wt{\Psi}_x$ by the square root of $\wt{\Psi}^\dagger_x\wt{\Psi}_x$. This method avoids having to deal directly with quaternionic algebra. By finding $\Psi_{(\vec{x},\tau_i)}$ at each $\tau_i$ we can generate an approximation for the Skyrme field $U(\vec{x})$ using \eqref{approximate-holonomy-Psis}. The i-skyrmions have polynomially decaying tails, which must be considered in the numerical scheme. We are able to capture their contribution accurately by mapping $\mathbb{R}^3$ bijectively to $[-1,1]^3$ via $x_i \mapsto y_i$ where $x_i = \beta y_i/(1-y_i^2)$ with constant $\beta \in [1,2]$, and find $U(\vec{y})$ on a $100\times100\times100$ lattice spaced evenly on $[-1,1]^3$. We then calculate the energy \eqref{Skyrme-energy}. All derivatives were calculated using central fourth-order finite-difference operators.

To summarise, our numerical scheme reduces the Atiyah--Manton construction to the repeated calculation of a kernel: a simple numerical problem which can be performed with extreme efficiency.
\subsection{Symmetries}
The Skyrme energy \eqref{Skyrme-energy} is invariant under rotations and translations in $\R^3$, namely
\begin{align}
    U(\vec{x})\mapsto U(R\vec{x}),\quad\text{and}\quad U(\vec{x})\mapsto U(\vec{x}+\vec{a}),
\end{align}
for $R\in\SO(3)$ and $\vec{a}\in\R^3$. It also has the $\SO(4)$ chiral symmetry of $S^3\cong\SU(2)$, given by $U\mapsto q_1Uq_2^{-1}$, where $(q_1,q_2)\in(\SU(2)\times\SU(2))/\{\pm\id\}$. However, the boundary condition $U\to\id$ is only preserved by the diagonal subgroup, corresponding to the $\SO(3)$ isospin symmetry, namely
\begin{align}
    U\mapsto pUp^{-1},\quad p\in\bigslant{\SU(2)}{\{\pm\id\}}\cong\SO(3).
\end{align}
Finally, the energy, and boundary conditions are invariant under the parity reversing transformation
\begin{align}
    U(\vec{x})\mapsto U(-\vec{x})^{-1},\label{parity-reversal}
\end{align}
resulting in a full symmetry group $\E(3)\times\SO(3)$ consisting of translations, rotations, reflections, and isorotations.

In contrast, Yang--Mills theory on $\R^4$ has a lot more symmetries. The anti-self-duality equations \eqref{ASD} (and hence the energy \eqref{YM-energy}) are invariant under all orientation-preserving conformal transformations $\phi:\R^4\lto\R^4$, via the pullback $\AA\mapsto\phi^\ast\AA$. In particular, the energy \eqref{YM-energy} is scale-invariant, a property which is not shared by the Skyrme energy \eqref{Skyrme-energy}. Finally, \eqref{ASD}-\eqref{YM-energy} are $\GG$-invariant, and therefore there is an action of the residual symmetry group
\begin{align}
    \bigslant{\GG}{\GG^0}\cong\SU(2)\label{residual}
\end{align}
on the moduli space $\II_\BB$, corresponding to a change in the choice of framing $\f$.

The $\E(3)\times\SO(3)$ symmetry group of the Skyrme model is recovered in the Atiyah--Manton construction as a subgroup of the symmetry group acting on instantons. The rotations and translations are those which fix the holographic direction (in our case, the $h:=x_4$-axis), the parity reversing transformation arises from the element in $\SO(4)$ acting on $\R^4$ via $x\mapsto-x$, and finally the isorotations correspond to the global change in gauge orientation, whose universal cover is the residual group \eqref{residual}.
\subsubsection{Actions on ADHM data}
It is possible to describe the action of the full orientation-preserving conformal symmetry group of instantons on ADHM data. However, since we are ultimately only interested in the interpretation of ADHM data in terms of the Skyrme model, we shall mostly only consider the $\E(3)\times\SO(3)$ symmetry group. It will be important to keep track of the instanton scale and position, so we shall also describe dilations and translations in $\R^4$. To remain consistent with the language used in the Skyrme model, we shall refer to changes in gauge orientation of instantons as `isorotations'. The actions of interest are summarised in Table \ref{tab:symmetries}. Understanding how these actions on ADHM data are equivalent to the actions on instantons amounts to straightforward manipulation of the ADHM transform, details of which may be found in \cite{SingerSutcliffe1999,FurutaHashimoto1990,AllenSutcliffe2013}.

\begin{table}[ht]
    \centering
    \begin{tabular}{l|l}
        Symmetry & Action on ADHM data \\\hline
        Translation by $a\in\R^4$ & $(L,M)\mapsto(L,M+a\ind_\BB)$\\
        Dilation by $\lambda>0$ & $(L,M)\mapsto(\lambda L,\lambda M)$\\
        Rotation $R(\vec{n},\theta)\in\SO(3)$&$(L,M)\mapsto(Lq(\vec{n},\theta)^{-1},q(\vec{n},\theta)Mq(\vec{n},\theta)^{-1})$\\
        Parity reversal $\vec{x}\mapsto-\vec{x}$&$(L,M)\mapsto(-L,-M)$\\
        Isorotation $p\in\SU(2)$&$(L,M)\mapsto(pL,M)$
    \end{tabular}
    \caption{A summary of symmetry group actions on ADHM data.}
    \label{tab:symmetries}
\end{table}

The notation in Table \ref{tab:symmetries} may be understood as follows. Firstly, we have identified $a\in\R^4$ by $a=a_1\i+a_2\j+a_3\k+a_4\1\in\H$. Secondly a rotation $R(\vec{n},\theta)$ of angle $\theta$ around a fixed unit axis $\vec{n}\in S^2\subset\R^3$ is represented by a unit quaternion
\begin{align}
    q(\vec{n},\theta)=\1\cos\sfrac{\theta}{2}+(n_1\i+n_2\j+n_3\k)\sin\sfrac{\theta}{2},\label{rotation-unit-qua}
\end{align}
corresponding to a choice of preimage of $R(q)$ in the double cover $\SU(2)\lto\SO(3)$. Finally, we think of the isorotations, analogously to the rotations, as unit quaternions $p=p(\vec{n},\theta)$, defined by a fixed unit axis $\vec{n}$ and angle of rotation $\theta$. It is clear here that the rotations and isorotations only correspond to an action of $\bigslant{\SU(2)}{\pm\1}\cong\SO(3)$ since in each case the action of $q,p=-\1$ is the same as the gauge transformation \eqref{orthog-gt} given by $O=-\ind_\BB$.
\subsubsection{Symmetric ADHM data and skyrmions}\label{sec:symmetric-data}
For a given subgroup $K$ of the group of isometries of $\II_\BB$, an instanton is said to be $K$-symmetric if its gauge-equivalence class is invariant under the action of $K$, that is, for all $h\in K$, there exists $g\in\GG^0$ such that $h\cdot\AA=g\AA g^{-1}-\d gg^{-1}$. We can cast a similar definition for ADHM data \cite{SingerSutcliffe1999,FurutaHashimoto1990,AllenSutcliffe2013}, although here we are only interested in symmetries which could correspond to symmetries of skyrmions.

We say that ADHM data $(L,M)\in\A_\BB$ is invariant under a rotation-isorotation pair $(q(\vec{n}_1,\theta_1),p(\vec{n}_2,\theta_2))\in\SU(2)\times\SU(2)$ if there exists a \textbf{compensating gauge transformation} (c.g.t.) $O\in\O(\BB)$ such that
\begin{align}
    \begin{aligned}
        L&=p(\vec{n}_2,\theta_2)Lq(\vec{n}_1,\theta_1)^{-1}O^{-1},&M&=Oq(\vec{n}_1,\theta_1)Mq(\vec{n}_1,\theta_1)^{-1}O^{-1}.
    \end{aligned}\label{ADHM-symm}
\end{align}
We may also include inversion symmetries by writing a minus-sign on the the right-hand-sides of \eqref{ADHM-symm}. For any group $\RR\subset\O(3)\times\SU(2)$ of rotations, reflections, and isorotations, we say $(L,M)$ is $\RR$-invariant (or $\RR$-symmetric) if there is a compensating gauge transformation for every element $(R,p)$ of $\RR$.

Any Skyrme field $U$ generated from ADHM data satisfying \eqref{ADHM-symm} will satisfy the invariance condition
\begin{align}
    U(\vec{x})=pU(R\vec{x})p^{-1},
\end{align}
where $R\in\SO(3)$ is the rotation corresponding to $q$. The same holds for ADHM data and skyrmions with inversion symmetries. In this way, there is a direct correspondence between symmetries of skyrmions and ADHM data.

Before we move on, we shall state and discuss some important results regarding invariant ADHM data. We shall use these throughout, and the proofs may be found in Appendix \ref{appendix:properties-ADHM}.
\begin{lemma}\label{lem:free-action}
ADHM data is irreducible with respect to isorotations and gauge transformations, i.e. for all $(L,M)\in\A_\BB$,
\begin{align}
    (L,M)=(\omega L\Omega^{-1},\Omega M\Omega^{-1})\quad\iff\quad(\omega,\Omega)=\pm(\1,\ind_\BB).
\end{align}
\end{lemma}
\begin{lemma}\label{lem:reps}
Let $G\subset\SU(2)$ be the binary group of some subgroup $\wt{G}\subset\SO(3)$, $p:G\to\SU(2)$ be a representation of $G$ with sign\footnote{Every such representation $p:G\to\SU(2)$ satisfies $p(-\1)=\pm\1=:\varepsilon$, and this is known as the sign.} $\varepsilon$, and let $\RR=\{(q,p(q))\::\:q\in G\}$. Let $(L,M)\in\A_{\BB}$ be $\RR$-invariant. Then the assignment $q\mapsto O_q\in\O(\BB)$ for the compensating gauge transformations is a $\BB$-dimensional real representation of $G$ with sign $-\varepsilon$.
\end{lemma}

Note that the correspondence between ADHM and skyrmion symmetries does not care about the compensating gauge transformations, and this is because these depend on a choice of gauge. Indeed, if $L'=L\Omega^{-1}$ and $M'=\Omega M\Omega^{-1}$ for some $\Omega\in\O(\BB)$, and $(L,M)$ satisfies \eqref{ADHM-symm}, then it is straightforward to see that $(L',M')$ satisfies \eqref{ADHM-symm} with compensating gauge transformation $O'=\Omega O\Omega^{-1}$. In this way, if the set of compensating gauge transformations form a representation of some subgroup $G\subset\SU(2)$, then equivalence of representations implies gauge-equivalence of ADHM data.

The understanding in terms of representations as in Lemma \ref{lem:reps} is useful for classifying which symmetries of ADHM data are possible, and for explicitly calculating the invariant data \cite{SingerSutcliffe1999,AllenSutcliffe2013}. To see why, note that the condition \eqref{ADHM-symm} for all $q\in G$ may be interpreted as saying that $L\in{\rm Hom}_G(V\otimes H,W)$ and $M\in{\rm Hom}_G(V\otimes H,H\otimes V)$, where $V$ and $W$ are the representations of $G$ corresponding to $O$ and $p$ respectively, and $H$ is the defining quaternionic representation of $G\subset\SU(2)$. Since every subgroup $G\subset\SU(2)$ is compact, every real representation $V$ may be decomposed as a direct sum $V=V_1\oplus\cdots\oplus V_n$ of irreducible representations $V_i$ of $G$. Therefore, by splitting $L$ and $M$ into blocks $L_i$ and $M_{ij}$ corresponding to the dimensions of the $V_i$, the invariance condition decomposes to:
\begin{itemize}
    \item $L_i$ is in a trivial subrep of $V_i\otimes W\otimes H$;
    \item $M_{ij}$ is in a trivial subrep of $V_i\otimes V_j\otimes H\otimes H$.
\end{itemize}
These tensor products may in turn be decomposed as direct sums of irreps. If there is no trivial subrep, then that block is identically zero. Otherwise, the invariant data may be calculated by hand in the chosen gauge. Note that there is also the further requirement that $M$ is a symmetric matrix, but this can also be imposed by hand. One only needs to consider all pairs $(V_i,V_j)$ for $i\leq j$ since $M_{ji}=M_{ij}^t$ due to the c.g.ts being orthogonal matrices. We shall use these ideas regularly without comment, however an example of how it works in practice is detailed later in Section \ref{sec:tori}, where we calculate the ADHM data for charge $\BB$ toroidally-symmetric solutions.

Everything outlined above may be applied similarly when considering inversion symmetries, but with some minor adjustments. Since every subgroup $K$ of $\O(3)$ which doesn't contain $-\ind_3$ is isomorphic to a subgroup $K'\subset\SO(3)$, one considers instead the binary group $G=2K'$ in the above. The invariance conditions are replaced by asking for $L_i$ in a trivial subrep of $V_i\otimes W\otimes H'$ and $M_{ij}$ in a trivial subrep of $V_i\otimes V_j\otimes H'\otimes H$, where $H'=A\otimes H$, with $A$ corresponding to the alternating representation of $G$, namely the assignment $\kappa\mapsto\det \kappa$, with $\kappa\in K\subset\O(3)$.

\section{Decomposing ADHM data}\label{sec:decompositions}
In the Skyrme model, it is easy to approximate a set of well-separated skyrmions $U_1,\dots,U_n$ by a product ansatz
\begin{align}
    U=U_1\cdots U_n.
\end{align}
Each skyrmion $U_i$ has freedom to be rotated, isorotated, and translated, namely by writing $U_i=p_iU_i(R_i\vec{x}-\vec{a}_i)p_i^{-1}$, and this approximation is good at describing $n$ such widely separated skyrmions. It is therefore natural to ask how to describe these configurations using ADHM data, and in this section we shall discuss such a framework. As we shall see in Section \ref{sec:ADHM-zoo}, an advantage of this framework, in contrast to the product ansatz, is that one may describe both well-separated clusters and \textit{central configurations}, where the separation is small, within one unified picture.
\subsection{Well-separated clusters}\label{sec:CSW}
In \cite{christWeinbergStanton1978general}, Christ, Stanton, and Weinberg give a formalism which describes $\BB$ well-separated $1$-instantons. Specifically, one looks for a gauge in which ADHM data $(L,M)\in\A_\BB$ takes the form
\begin{align}
\begin{aligned}
    L&=\begin{pmatrix}
    \lambda_1\omega_1&\lambda_2\omega_2&\cdots&\lambda_\BB\omega_\BB\end{pmatrix}\\
    M&=\begin{pmatrix}
    0&\sigma_{12}&\cdots&\sigma_{1\BB}\\
    \sigma_{12}&0&\ddots&\vdots\\
    \vdots&\ddots&\ddots&\sigma_{(\BB-1)\BB}\\
    \sigma_{1\BB}&\cdots&\sigma_{(\BB-1)\BB}&0
    \end{pmatrix}+\diag\{r_1,r_2,\dots,r_\BB\},
\end{aligned}\label{CSW-diag}
\end{align}
where $\lambda_i>0$, $\omega_i\in\SU(2)$, and $r_i\in\H$ are interpreted as scales, orientations, and positions respectively, and $\sigma_{ij}$ are fixed by the reality condition. For this picture to describe well-separated instantons, one requires the scales $\lambda_i$ to be small compared to the `separations' $|r_i-r_j|$. Formally, one lets the diagonal elements $r_i$ be fixed, distinct quaternions for all $i$, chooses $0<\epsilon\ll1$, and writes $\lambda_i=\epsilon\kappa_i$ for all $i$. Considering the limit $\epsilon\to0$, one may show \cite{christWeinbergStanton1978general} that the reality condition may be solved approximately (up to order $\epsilon^2$) by
\begin{align}
    \sigma_{ij}\sim\dfrac{\lambda_i\lambda_j}{2}\dfrac{r_j-r_i}{|r_i-r_j|^2}(\ol{\omega}_i\omega_j-\ol{\omega}_j\omega_i)+O(\epsilon^4).
\end{align}
In particular, this formula allows for an iterative solution to the reality condition for \eqref{CSW-diag}, where $\sigma_{ij}$ are determined as a power series in even powers of $\epsilon$ which converges for $\epsilon$ sufficiently small.

Inspired by the Christ--Stanton--Weinberg formalism, we now propose a description of well-separated clusters. To do this, we first need to define the \textit{diagonal ADHM moduli spaces} $\A_{\BB_1}\oplus\cdots\oplus\A_{\BB_n}$. These consist of matrices $(\wh{L},\wh{M})$, where $\wh{L}$ is a $\BB=\BB_1+\cdots+\BB_n$ row vector of quaternions, and $\wh{M}$ is a symmetric $\BB\times\BB$ matrix of quaternions, each decomposed as
\begin{align}
    \begin{aligned}
        \wh{L}&=\begin{pmatrix}L_1&\cdots&L_n\end{pmatrix},&
        \wh{M}&=\diag\{M_1,\dots,M_n\},
    \end{aligned}\label{diagonal-ADHM-data}
\end{align}
with $(L_i,M_i)\in\A_{\BB_i}$ rank $\BB_i$ ADHM data. The moduli space is given by the quotient of such matrices with respect to the action of $\bigoplus_{i=1}^n\O(\BB_i)$ on each block.

In general, an element $(\wh{L},\wh{M})\in\A_{\BB_1}\oplus\cdots\oplus\A_{\BB_n}$ will not be ADHM data. However, in analogy with the picture above, we propose that any true ADHM data which is `close' to such diagonal data will give a good description of well-separated clusters of the instantons described by the ADHM data $(L_i,M_i)$, and likewise for the corresponding i-skyrmions. Due to how translations and scalings act on ADHM data, as seen in Table \ref{tab:symmetries}, we define the \textbf{location} $r_i$ of the cluster $(L_i,M_i)$ as the quaternion
    \begin{align}
      r_i=\frac{1}{\BB_i}\tr(M_i),\label{positions}
    \end{align}
and the \textbf{scale} $\lambda_i>0$ of the cluster $(L_i,M_i)$ is defined by
\begin{align}
    \lambda_i=\sqrt{\frac{1}{\BB_i}\tr\left(L_i^\dagger L_i\right)}.\label{scales}
\end{align}
These quantities are $\bigoplus_{i=1}^n\O(\BB_i)$-invariant, but only in the case $n=1$ do they make sense as gauge-invariant quantities for ADHM data. For each cluster $(L_i,M_i)$, it is useful to extract the data $(l_i,m_i)\in\A_{\BB_i}$ via
\begin{align}
    L_i=\lambda_il_i,\quad M_i=\lambda_im_i+r_i\ind_{\BB_i}.
\end{align}
Using this, we consider an ansatz for ADHM data $(L,M)\in\A_{\BB_1+\cdots+\BB_n}$ by writing
\begin{align}
\begin{aligned}
    L&=\begin{pmatrix}\lambda_1l_1&\cdots&\lambda_nl_n\end{pmatrix},\\
    M&=\begin{pmatrix}\mu_1m_1&\Sigma_{12}&\cdots&\Sigma_{1n}\\
        \Sigma_{12}^t&\ddots&\ddots&\vdots\\
        \vdots&\ddots&\ddots&\Sigma_{(n-1)n}\\
        \Sigma_{1n}^t&\cdots&\Sigma_{(n-1)n}^t&\mu_nm_n\end{pmatrix}+\diag\{r_1\ind_{\BB_1},\dots,r_n\ind_{\BB_n}\}.
\end{aligned}\label{cluster-decomposition}
\end{align}
The parameters $\mu_i\in\R$ have been introduced here since the reality condition is a nonlinear constraint, and in general cannot be resolved with $\mu_i=\lambda_i$. These, and the off-diagonal components $\Sigma_{ij}\in\mathrm{Mat}_{\BB_i\times\BB_j}(\H)$, should be fixed by the reality condition. We think of any ADHM data which is gauge-equivalent to the form \eqref{cluster-decomposition} as describing well-separated clusters corresponding to the data $(l_i,m_i)$ if the scales $\lambda_i$ are small compared to the separations $|r_i-r_j|$.

To realistically extract this interpretation of \eqref{cluster-decomposition}, we ideally need to resolve the reality condition \eqref{reality-cond}. Since $(l_i,m_i)\in\A_{\BB_i}$, this is equivalent to the conditions
\begin{align}
    \Im\left((\lambda_i^2-\mu_i^2)l_i^\dagger l_i+\sum_{j=1}^{i-1}\Sigma_{ji}^\dagger\Sigma_{ji}+\sum_{j=i+1}^n\ol{\Sigma}_{ij}\Sigma_{ij}^t\right)=0,\label{RC-decomposition1}
\end{align}
for all $i=1,\dots,n$, and
\begin{multline}
    \left(\ol{(r_i-r_j)}+\mu_im_i^\dagger\right)\Sigma_{ij}-\mu_j(m_j^\dagger\Sigma_{ij}^t)^t+\frac{\lambda_i\lambda_j}{2}\left(l_i^\dagger l_j-({l_j}^\dagger l_i)^t\right)\\
    +\frac{1}{2}\sum_{k=1}^{i-1}(\Sigma_{ki}^\dagger\Sigma_{kj}-(\Sigma_{kj}^\dagger\Sigma_{ki})^t)+\frac{1}{2}\sum_{k=i+1}^{j-1}(\ol{\Sigma}_{ik}\Sigma_{kj}-(\Sigma_{kj}^\dagger\Sigma_{ik}^t)^t)\\
    +\frac{1}{2}\sum_{k=j+1}^n(\ol{\Sigma}_{ik}\Sigma_{jk}^t-(\ol{\Sigma}_{jk}\Sigma_{ik}^t)^t)=\mathfrak{R}_{ij},\label{RC-decomposition2}
\end{multline}
where $\mathfrak{R}_{ij}\in\mathrm{Mat}_{\BB_i\times\BB_j}(\R)$, for each $1\leq i<j\leq n$. We cannot resolve these in general, however it is clear from \eqref{RC-decomposition2} that as $|r_i-r_j|\to\infty$, these are only consistent if $\Sigma_{ij}\to0$, and furthermore, it then holds from \eqref{RC-decomposition1} that generically $\mu_i^2\to\lambda_i^2$. In this way, the approximation in terms of diagonal data informally holds in a limit of large separation. This informal understanding will be important later when we demonstrate various examples of ADHM data which exhibit such cluster decompositions, and we shall consider some explicit examples formally.

It would be nice to obtain an iterative formula for approximating $\Sigma_{ij}$, analogous to the $\BB_i=1$ case above, however it is unclear how to deal with the term $\mu_im_i^\dagger\Sigma_{ij}-\mu_j(m_j^\dagger\Sigma_{ij}^t)^t$ in \eqref{RC-decomposition2} which was not relevant in the $\BB_i=1$ case. Another possibility is to develop a numerical method for constructing ADHM data which is close to the diagonal form \eqref{diagonal-ADHM-data} which works well for describing arbitrary clusters at large enough separation. We will report on such a method in future work.
\subsection{Symmetric decomposition}\label{sec:sym-decom}
One of the problems with resolving the reality condition \eqref{RC-decomposition1}-\eqref{RC-decomposition2} is that there are generally no further constraints which may be used to fix the off-diagonal components $\Sigma_{ij}$, and in particular there is rarely a unique solution. A useful approach to combating this is to arrange the clusters in such a way that the overall system has some shared symmetry. This has previously been considered where the constituents are arranged in the shape of regular polyhedra \cite{LeeseManton1994stable,SingerSutcliffe1999,AllenSutcliffe2013}.

We will consider situations where the constituents lie on a shared axis of symmetry. This is less restrictive than polyhedral symmetry as it allows for the description of larger moduli spaces of solutions. Suppose we know ADHM data $(L_1,M_1)\in\A_{\BB_1}$ and $(L_2,M_2)\in\A_{\BB_2}$, which both have a symmetry around shared axes of rotation and isorotation $\vec{n}_1,\vec{n}_2$. As a simple corollary to Lemma \ref{lem:free-action}, the isorotations can never be of greater order than the rotations, so we may always arrange it so that the shared symmetry is of the form
\begin{align}
    \begin{aligned}
         L_i&=p(\vec{n}_2,k_i\theta_i)L_iq(\vec{n}_1,\theta_i)^{-1}O_i^{-1},\\
         M_i&=O_iq(\vec{n}_1,\theta_i)M_iq(\vec{n}_1,\theta_i)^{-1}O_i^{-1},
    \end{aligned}\label{aligned-symm}
\end{align}
for some angles $\theta_i$, integers $k_i\in\Z$, and compensating gauge transformations $O_i\in\O(\BB_i)$, for $i=1,2$. If one symmetry is a subgroup of the other, one can aim to build rank $\BB=\BB_1+\BB_2$ data by imposing the shared symmetry with the block-diagonal compensating transformation $O=\diag\{O_1,O_2\}$. More generally, one may find that whilst each symmetry is broken, there is some shared unbroken symmetry. A sufficient condition for this is when there exists $n\in\Z$ such that
\begin{align}
    \theta_2=n\theta_1\equiv\theta,\quad\text{and}\quad k_1\theta=k_2\theta\;(\text{mod }2\pi).\label{shared-iso-rot}
\end{align}
The case $k_1\theta+4k\pi=k_2\theta$ is equivalent to the situation described above, but in the cases where $k_1\theta+2(2k+1)\pi=k_2\theta$, we can obtain data invariant under a $\theta$ rotation and $k_1\theta$ isorotation, by considering the alternating block-diagonal matrix
\begin{align}
    O=\diag\{O_1,-O_2\}.\label{block-cgt}
\end{align}
This works since all isorotations satisfy $p(\vec{n},\phi+2l\pi)=(-1)^lp(\vec{n},\phi)$. It should be noted that in the case of continuous axial symmetries, equation \eqref{shared-iso-rot} will only be solved by a specific choice of $\theta$. Furthermore, the compensating gauge transformations will likely depend on $\theta$, and should thus be evaluated at the solution to \eqref{shared-iso-rot}; one may choose a gauge such that the compensating transformation along one axis of symmetry decomposes (up to a sign) as a direct sum of irreps of $\SO(2)$, which corresponds either to the trivial representation, or one of the $2$-dimensional reps\footnote{Note that the action of $\SO(2)$ is generated by a rotation of angle $\theta/2$ by definition of $q(\vec{n},\theta)$ in \eqref{rotation-unit-qua}, hence the factor of $\sfrac{1}{2}$ in the representation.}
\begin{align}
    \theta\mapsto Q_k(\theta)=\begin{pmatrix}\cos\sfrac{k\theta}{2}&-\sin\sfrac{k\theta}{2}\\
    \sin\sfrac{k\theta}{2}&\cos\sfrac{k\theta}{2}
    \end{pmatrix},\quad k\in\Z^+.\label{rep-SO(2)}
\end{align}

A subtly different scenario involves a way of manipulating the shared symmetry \eqref{aligned-symm} in order to enable \eqref{shared-iso-rot}. It is possible that whilst \eqref{shared-iso-rot} doesn't hold, the modification
\begin{align}
    \theta_2=n\theta_1\equiv\theta,\quad\text{and}\quad k_1\theta=-k_2\theta\;(\text{mod }2\pi),\label{inverted-orientation-shared-iso-rot}
\end{align}
does. In which case, let $\omega\in\SU(2)$ be such that
\begin{align}
    \omega(\vec{n}_2\cdot\vec{\bm e})\omega^{-1}=-\vec{n}_2\cdot\vec{\bm e},\label{axis-inversion}
\end{align}
where $\vec{\bm e}=(\i,\j,\k)$, that is, $\omega$ is a unit quaternion which inverts the axis of isorotation. Then, noting that $p(-\vec{n},\phi)=p(\vec{n},-\phi)$, we see that the inverted ADHM data $(\omega L_2,M_2)$ will satisfy \eqref{aligned-symm} with $k_2$ replaced by $-k_2$, and thus the same procedure may be applied as explained above, by instead considering solutions to \eqref{inverted-orientation-shared-iso-rot}.

Sometimes we will want to impose more symmetry than that which may be derived from a shared axial symmetry of the individual constituents. To do this we appeal to the block decomposition described in Section \ref{sec:CSW}: for configurations $(l_i,m_i)$ with positions $r_i$, orientations $(p_i,q_i)$, and scales $\lambda_i$, we may impose the expected symmetry on the diagonal data
\begin{align}
    \begin{aligned}
        \wh{L}&=\begin{pmatrix}\lambda_1p_1l_1q_1^{-1}&\cdots&\lambda_np_nl_nq_n^{-1}\end{pmatrix},\\
        \wh{M}&=\diag\{q_1m_1q_1^{-1}+r_1\ind_{\BB_1},\dots,q_nm_nq_n^{-1}+r_n\ind_{\BB_n}\},
    \end{aligned}\label{general-cluster}
\end{align}
to determine the compensating gauge transformations for each generating symmetry.
\subsection{Colour matching: the attractive channel}\label{sec:colouring}
Some of the technical discussion in the previous sections can be understood pictorially using basic Skyrme phenomenology. This is particularly useful for identifying the rotations and isorotations compatible with a given symmetry, as required for fixing the ansatz \eqref{general-cluster}.

It is helpful to define a colouring scheme for plotting skyrmions. By writing the Skyrme field as a unit quaternion, the coefficients may be identified with the pion field
\begin{align}
    U = \pi_0 \1 + \pi_1 \i + \pi_2 \j + \pi_3 \k.
\end{align}
One then assigns a colour at each point in space based on the pion field directions, with white/black corresponding to $\pi_3 = \pm 1$, and red, green, or blue given when $\pi_1 + {\rm i} \pi_2 = 1$, $\exp(2\pi{\rm i}/3)$, or $\exp(4\pi {\rm i}/3)$ respectively. We use a mapping from $(\pi_1,\pi_2,\pi_3)$ to the Runge colour sphere, first defined in \cite{FeistLauManton2013skyrmions}. For example, the $1$-skyrmion with ADHM data $(L,M) = (\1,0)$ has pion field $\pi_i$ proportional to $x_i$. Hence the colour sphere is mapped bijectively on to each spherical shell in $\mathbb{R}^3$. We may reorient the skyrmion by taking  ADHM data with $L = p \in \mathbb{H}$, giving a pion field proportional to $R_{ij}(p)\pi_j$, and a corresponding skyrmion with a new colouring.

The energy of two well-separated $1$-skyrmions with orientations $L_1 = p_1, L_2 = p_2$ depends on the relative orientation $p_1^{-1}p_2$. The energy is minimal, and hence the attraction is maximal, when $p_1^{-1}p_2$ is orthogonal to the axis joining the skyrmions. This circle of configurations is known as the attractive channel. In terms of the colouring scheme, the attractive channel occurs when the colours of closest contact match.

\begin{figure}[htbp]
	\begin{center}
		\includegraphics[scale=0.5]{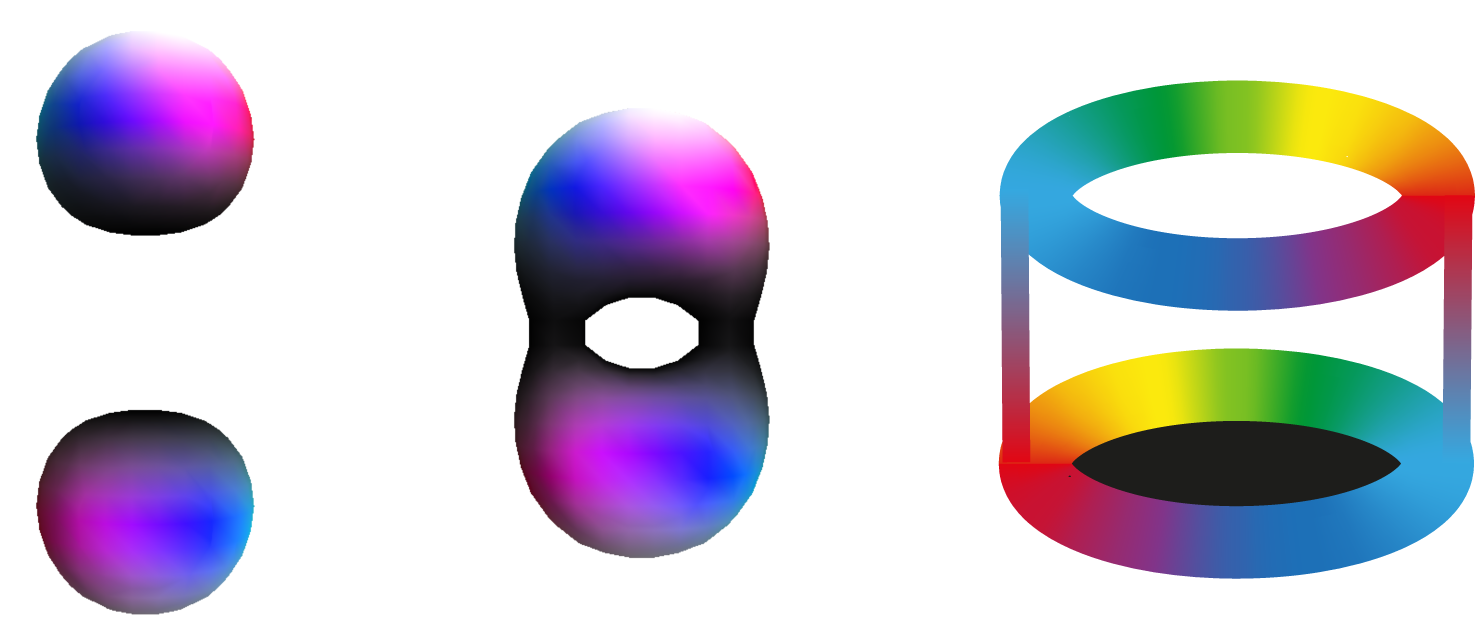}
		\caption{Left, center: Two skyrmions in the attractive channel with separation $3$ and separation $2$. Right: A schematic view of the skyrmions as discs of colour, which help understand where energy density is created: at the points where the colours on the skyrmion equators are opposite (red and teal, in this case). }
		\label{fig:2model}
	\end{center}
\end{figure}

In Figure \ref{fig:2model} we consider two skyrmions in the attractive channel, moving towards one another. Without loss of generality, we have taken the colour of closest contact to be black. This means that the colour wheel winds once around the equator of each $1$-skyrmion, in opposite directions. Note that each colour has an opposite: the colour on the antipodal point of the sphere. Red, green and blue are contrasted by teal, magenta and yellow. The colours on the equators match at two points and opposite colours meet at two points (red and teal). As the skyrmions are brought together, the energy density is concentrated in the region where the equator colours are opposite. Conversely, no energy is created where the equator colours match. Using this simple phenomenology, we can quickly determine the orientation of the central torus which is created when two skyrmions are brought together in the attractive channel. We can also easily determine the orientations required to make, for example, the ${\rm D}_{2d}$-symmetric $\BB=5$ skyrmion which we consider later in Section \ref{B=5}.

\vspace{10mm}

\begin{figure}[htbp]
	\begin{center}
		\includegraphics[scale=0.35]{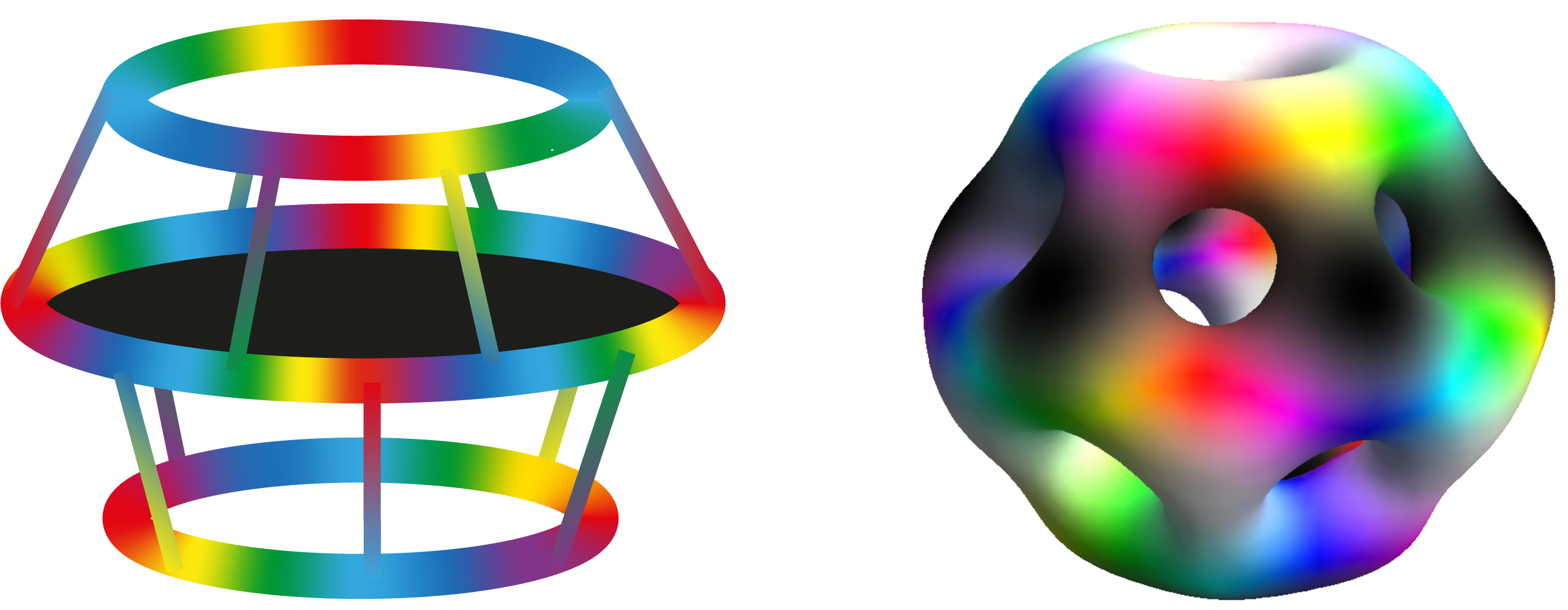}
		\caption{Left: A schematic model of the $\BB=8$ skyrmion as stacked $2$-, $4$- and $2$- tori. Energy density is created between the tori where the colours on the  neighbouring tori are opposite. Right: the ${\rm D}_{6d}$-symmetric $\BB=8$ i-skyrmion.}
		\label{fig:6model}
	\end{center}
\end{figure}

These ideas extend naturally to larger skyrmions. In Section \ref{sec:tori} we derive ADHM data for charge $\BB$ tori. These are oriented so that the colour wheel winds $\BB$ times around the equator and the top and bottom of the tori are coloured white or black. In this orientation, tori attract if they lie on parallel planes with closest colours matching. For their colours to match, one torus must be rotated by $\pi$ around an axis perpendicular to the symmetry axis. Hence pairs of tori can be thought of as two discs, one whose colour wheel winds $\BB_1$ times around the equator and the other $-\BB_2$ times. Similar to the $\BB=1$ case, additional energy density is formed at $\BB_1+\BB_2$ points, wherever the colours on the torus equators are opposite. This is helpful, for example, when trying to construct the ${\rm D}_{6d}$-symmetric $\BB=8$ skyrmion. This is known to look like three stacked tori with charges $2$, $4$, and $2$. Between a $2$- and $4$-torus, six additional lumps of energy density are formed. To obtain ${\rm D}_{6d}$-symmetry, the lumps must be symmetrically placed; the bottom torus must be isorotated by $\pi$ with respect to the top torus, so that every colour is sent to its opposite. A schematic plot of this is shown in Figure \ref{fig:6model}, next to the i-skyrmion with minimal energy which we generate in the next section. We note that the i-skyrmion is very similar to the true energy-minimising skyrmion.
\section{Energy-minimising instanton-generated-skyrmions}\label{sec:database}
Here we provide a list, sorted by topological charge $1\leq\BB\leq8$, of ADHM data matching the symmetries corresponding to the accepted minimal energy skyrmions of the massless Skyrme model \eqref{Skyrme-energy}. These highly-symmetric data are the quintessential examples of \textit{central configurations}, which are important for understanding the framework for cluster decompositions as outlined in Section \ref{sec:CSW}. 

The data in the cases of charge $\BB=1,2,3,4,$ and $7$ were known previously \cite{LeeseManton1994stable,houghton1999-3Skyrme,SingerSutcliffe1999,christWeinbergStanton1978general}, but the data for charges $\BB=5,6,$ and $8$ are derived for the first time here. The difficulty in these new cases is that there is a multidimensional space of instantons with the same symmetry. Hence we must find the energy minimiser within a large family. In these cases we perform a Newton--Raphson algorithm to find a point in the moduli space which minimises the Skyrme energy. In more detail, let $\boldsymbol{a}\in\R^n$ parameterise the $n$-dimensional moduli space and $E(\boldsymbol{a})$ denote the energy of the i-skyrmion. We start with a well motivated initial point $\boldsymbol{a}_0$, typically where the separations and scales of any constituent clusters are of a comparable size, and perform the iteration
\begin{equation}
\boldsymbol{a}_1 = \boldsymbol{a}_0 - H_E^{-1} \nabla E \rvert_{\boldsymbol{a}_0},\label{Newton-Raphson}
\end{equation}
where $H_E$ is the Hessian of $E(\boldsymbol{a})$. This iterative process is repeated until $|\nabla E| < \epsilon$ for some suitably low tolerance $\epsilon>0$.

All i-skyrmions depend on an overall scale factor $\lambda>0$. We normalise the data so that the scale is defined by the gauge-invariant quantity
\begin{align}
    \lambda=\sqrt{\frac{1}{\BB}\tr(L^\dagger L)},\label{instanton-scale}
\end{align}
matching the convention of \eqref{scales}. For each i-skyrmion, we determine the value of $\lambda$ which minimises the Skyrme energy \eqref{Skyrme-energy}.

\begin{figure}[htbp]
	\captionsetup[subfigure]{labelformat=empty}
	\begin{center}
		\mbox{\subfloat[1: $\O(3)$]{\includegraphics[scale=0.65]{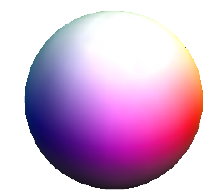}} \quad
			\subfloat[2: ${\rm D}_{\infty h}$]{\includegraphics[scale=0.65]{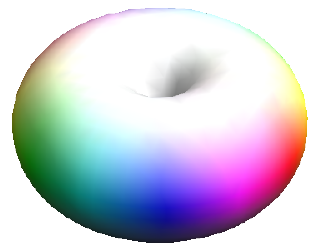}} \quad
			\subfloat[3: ${\rm T}_d$]{\includegraphics[scale=0.65]{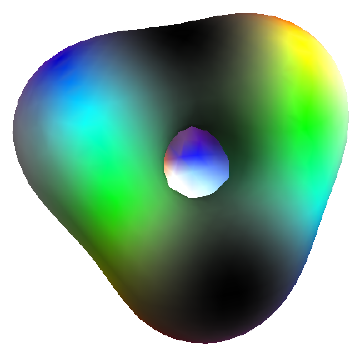}} \quad
			\subfloat[4: $\O_h$]{\includegraphics[scale=0.65]{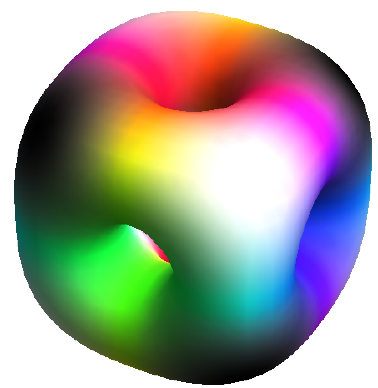}} \quad
			\subfloat[5: ${\rm D}_{2d}$]{\includegraphics[scale=0.65]{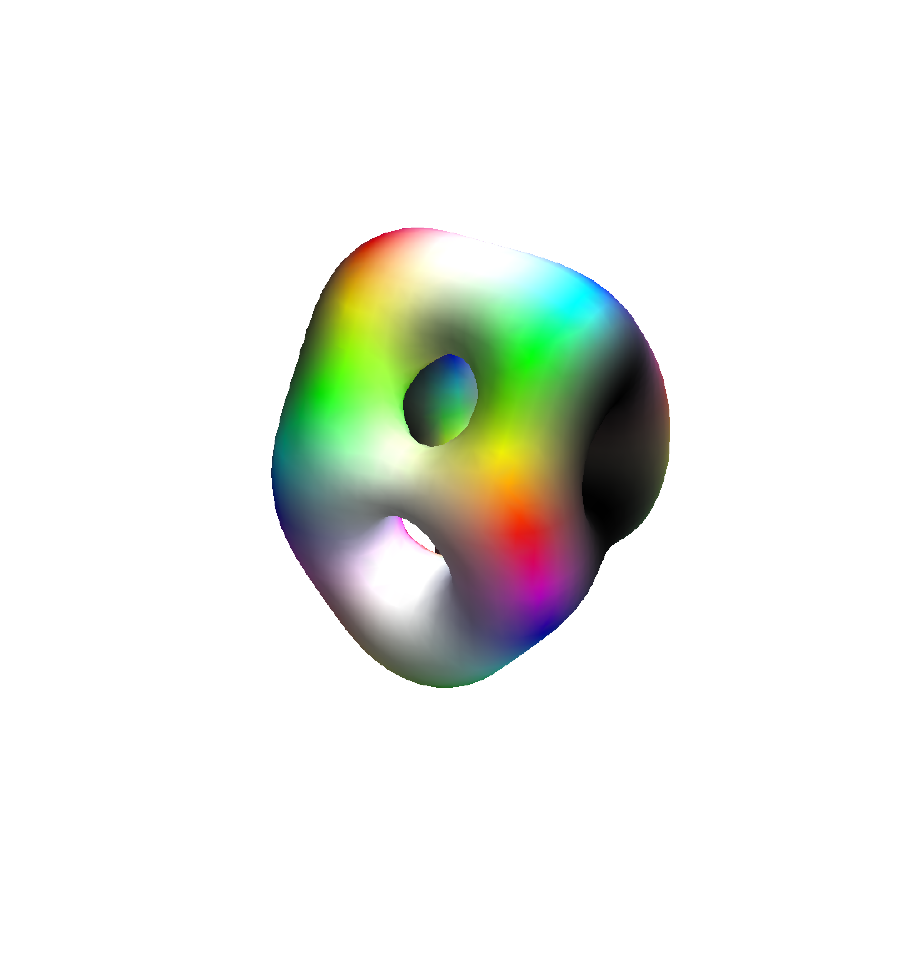}}}\quad \mbox{
			\subfloat[6: ${\rm D}_{4d}$]{\includegraphics[scale=0.65]{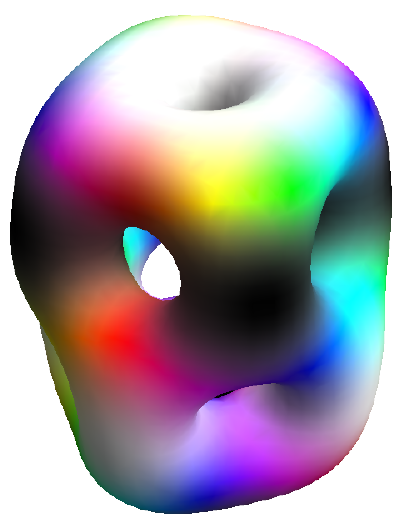}} \quad
				\subfloat[7: ${\rm Y}_h$]{\includegraphics[scale=0.65]{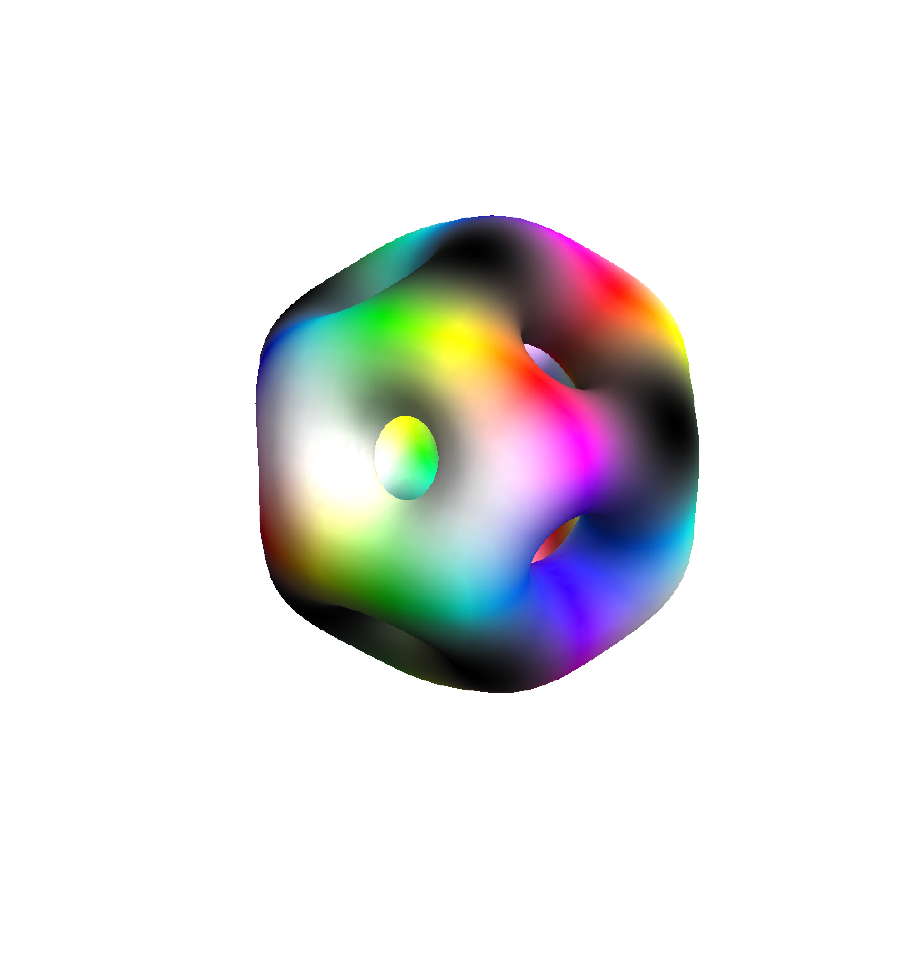}} \quad
			\subfloat[8: ${\rm D}_{6d}$]{\includegraphics[scale=0.65]{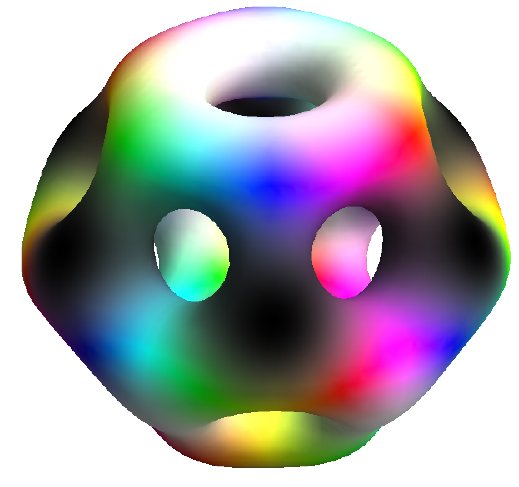}}} 
		\caption{The small-$\BB$ i-skyrmions and their symmetry groups, generated from the ADHM data throughout this section. All i-skyrmions are plotted on the same grid, and we plot an isosurface of constant energy density. All surfaces represent the constant energy density  $\mathcal{E} = 0.05$.}
		\label{fig:Skyrmions}
	\end{center}
\end{figure}

Energy isosurface plots of the corresponding energy-minimising i-skyrmions are given in Figure \ref{fig:Skyrmions}. The colouring scheme matches \cite{FeistLauManton2013skyrmions}, as described in Section \ref{sec:colouring}.
\subsection{\texorpdfstring{$\BB=1$}{B=1}}
The $\BB=1$ central configuration is a hedgehog, which is spherically-symmetric. This is described by the ADHM data
\begin{align}
    \begin{pmatrix}
    L\\
    M
    \end{pmatrix}=\begin{pmatrix}
    \lambda\1\\0
    \end{pmatrix},\quad\lambda>0.\label{B=1}
\end{align}
This has $\SO(4)$-symmetry, but from the perspective of the i-skyrmion, there is $\O(3)$-symmetry manifested by
\begin{align}
    \begin{aligned}
    L&=\pm p(\vec{n},\theta)Lq(\vec{n},\theta)^{-1}O_\pm,&M&=\pm O_\pm q(\vec{n},\theta)Mq(\vec{n},\theta)^{-1}O_\pm^{-1},
    \end{aligned}\label{hedgehog-symmetry}
\end{align}
for any $\vec{n}\in S^2$, and where $O_\pm=\pm\ind$.

The energy-minimising i-skyrmion is given when $\lambda=1.45$ with energy $E=1.243$.
\subsection{\texorpdfstring{$\BB=2$}{B=2}}
The central $\BB=2$ configuration is a torus, which may be represented by the ADHM data
\begin{align}
    \begin{aligned}
    L&=\lambda\begin{pmatrix}
    \1&\k\end{pmatrix},&
    M&=\frac{\lambda}{\sqrt{2}}\begin{pmatrix}
    \i&\j\\
    \j&-\i\end{pmatrix},
    \end{aligned}\quad \lambda>0.\label{torus}
\end{align}
The ${\rm D}_{\infty h}$-symmetry is given by
\begin{align}
    \begin{aligned}
    L&=p(\vec{e}_3,2\theta)Lq(\vec{e}_3,\theta)^{-1}O_\theta^{-1},&M&=O_\theta q(\vec{e}_3,\theta)Mq(\vec{e}_3,\theta)^{-1}O_\theta^{-1},\\
    L&=p(\vec{e}_1,\pi)Lq(\vec{e}_1,\pi)^{-1}O_2^{-1},&M&=O_2q(\vec{e}_1,\pi)Mq(\vec{e}_1,\pi)^{-1}O_2^{-1},\\
    L&=-p(\vec{e}_3,\pi)LO_-^{-1},&M&=-O_-MO_-^{-1}.
    \end{aligned}\label{torus-symmetry-B=2}
\end{align}
with c.g.ts
\begin{align}
    O_\theta=Q_1(\theta)=\begin{pmatrix}\cos\sfrac{\theta}{2}&-\sin\sfrac{\theta}{2}\\
    \sin\sfrac{\theta}{2}&\cos\sfrac{\theta}{2}\end{pmatrix},\quad O_2=\begin{pmatrix}1&0\\
    0&-1\end{pmatrix},\quad O_-=\begin{pmatrix}0&1\\
    -1&0\end{pmatrix}.\label{torus-gts}
\end{align}
This is the first example in a sequence of charge $\BB$ tori, which we detail later in Section \ref{sec:tori}.

The energy-minimising i-skyrmion is given when $\lambda=1.31$ with energy $E=2.384$.
\subsection{\texorpdfstring{$\BB=3$}{B=3}}
The minimal energy $\BB=3$ skyrmion has tetrahedral symmetry, and may be approximated by a ${\rm T}_d$-symmetric instanton \cite{LeeseManton1994stable}. This is described by the ADHM data \cite{houghton1999-3Skyrme}
\begin{align}
    \begin{aligned}
    L&=\lambda\begin{pmatrix}
    \i&\j&\k
    \end{pmatrix},&
    M&=\lambda\begin{pmatrix}
    0&\k&\j\\
    \k&0&\i\\
    \j&\i&0
    \end{pmatrix},
    \end{aligned}\quad\lambda>0,\label{tetrahedron}
\end{align}
and the ${\rm T}_d$-symmetry is manifested in this gauge and orientation by
\begin{align}
    \begin{aligned}
    L&=-p(\vec{e}_3,\sfrac{\pi}{2})Lq(\vec{e}_3,\sfrac{\pi}{2})^{-1}O_{2d}^{-1},&M&=-O_{2d}q(\vec{e}_3,\sfrac{\pi}{2})Mq(\vec{e}_3,\sfrac{\pi}{2})^{-1}O_{2d}^{-1},\\
    L&=p(\vec{r},\sfrac{2\pi}{3})Lq(\vec{r},\sfrac{2\pi}{3})^{-1}O_3^{-1},&M&=O_3q(\vec{r},\sfrac{2\pi}{3})Mq(\vec{r},\sfrac{2\pi}{3})^{-1}O_3^{-1},
    \end{aligned}\label{tetrahedral-symmetry}
\end{align}
where $\vec{r}=\sfrac{\sqrt{3}}{3}(\vec{e}_1+\vec{e}_2+\vec{e}_3)$, and the c.g.ts are
\begin{align}
    O_{2d}=\begin{pmatrix}
    0&1&0\\
    -1&0&0\\
    0&0&-1
    \end{pmatrix},\quad O_3=\begin{pmatrix}
    0&0&1\\
    1&0&0\\
    0&1&0
    \end{pmatrix}.
\end{align}
The energy-minimising i-skyrmion is given when $\lambda=1.11$ with energy $E=3.488$. We remark that the dual tetrahedron may be obtained by replacing $M\mapsto-M$, which is equivalent to reversing parity \eqref{parity-reversal}, and this leaves the Skyrme energy \eqref{Skyrme-energy} invariant.
\subsection{\texorpdfstring{$\BB=4$}{B=4}}
The minimal energy $\BB=4$ skyrmion has cubic symmetry, and this is well-approximated by a corresponding symmetric instanton \cite{LeeseManton1994stable}. The ADHM data may be written as
\begin{align}
    \begin{aligned}
    L&=\lambda\wp\begin{pmatrix}
    \1&\i&\j&\k
    \end{pmatrix},\\
    M&=\frac{\lambda}{\sqrt{2}}\begin{pmatrix}
    0&-(\j+\k)&-(\k+\i)&-(\i+\j)\\
    -(\j+\k)&0&\j-\i&\i-\k\\
    -(\k+\i)&\j-\i&0&\k-\j\\
    -(\i+\j)&\i-\k&\k-\j&0
    \end{pmatrix},
    \end{aligned}\quad\lambda>0,\label{cube}
\end{align}
with the unit quaternion $\wp$ defined by
\begin{align}
\begin{aligned}
    \wp
    &=\sqrt{\frac{(2-\sqrt{2})(3+\sqrt{3})}{24}}\left(\sfrac{\sqrt{3}-1}{\sqrt{2}}+\i+(1+\sqrt{2})\j-\sfrac{1}{2}(\sqrt{3}-1)(\sqrt{2}+2)\k\right),
\end{aligned}\label{cube-orientation}
\end{align}
to match the orientation in \cite{mankomantonwood2007}; this orientation is nice as each pair of opposite faces is coloured either red, green, or blue. In this orientation the $\O_h$-symmetry is realised by invariance under
\begin{align}
    \begin{aligned}
    L&=p(\vec{e}_3,-\sfrac{2\pi}{3})Lq(\vec{r},\sfrac{2\pi}{3})^{-1}O_3^{-1},&M&=O_3q(\vec{r},\sfrac{2\pi}{3})Mq(\vec{r},\sfrac{2\pi}{3})^{-1}O_3^{-1},\\
    L&=p(\vec{e}_1,-\pi)Lq(\vec{e}_3,\sfrac{\pi}{2})^{-1}O_4^{-1},&M&=O_4q(\vec{e}_3,\sfrac{\pi}{2})Mq(\vec{e}_3,\sfrac{\pi}{2})^{-1}O_4^{-1},\\
    L&=-p(\vec{e}_3,\pi)LO_-^{-1},&M&=-O_-MO_-^{-1}.
    \end{aligned}\label{cubic-symmetry}
\end{align}
Here we have denoted as before $\vec{r}=\sfrac{1}{\sqrt{3}}(\vec{e}_1+\vec{e}_2+\vec{e}_3)$, and the c.g.ts are
\begin{align}
    O_3=\begin{pmatrix}
    1&0&0&0\\
    0&0&0&1\\
    0&1&0&0\\
    0&0&1&0
    \end{pmatrix},\; O_4=\begin{pmatrix}
    0&0&1&0\\
    1&0&0&0\\
    0&0&0&-1\\
    0&1&0&0
    \end{pmatrix},\; O_-=\frac{\sqrt{3}}{3}\begin{pmatrix}
    0&-1&-1&-1\\
    1&0&-1&1\\
    1&1&0&-1\\
    1&-1&1&0
    \end{pmatrix}.\label{cube-gts}
\end{align}
The energy-minimising i-skyrmion is given when $\lambda=1.04$ with energy $E=4.532$.
\subsection{\texorpdfstring{$\BB=5$}{B=5}}\label{B=5}
Unlike the previous examples, ADHM data for the minimal energy $\BB=5$ skyrmion has not been considered before. The main reason for this is that the expected minimal energy $\BB=5$ skyrmion has ${\rm D}_{2d}$-symmetry \cite{battye2001solitonic}, which is significantly less symmetry than the previously well-studied examples.

In order to construct an i-skyrmion with ${\rm D}_{2d}$-symmetry, we require ADHM data $(L,M)\in\A_5$ such that
\begin{align}
    \begin{aligned}
        L&=-p(\vec{e}_3,\sfrac{\pi}{2})Lq(\vec{e}_3,\sfrac{\pi}{2})^{-1}O_\rho^{-1}&M&=-O_\rho q(\vec{e}_3,\sfrac{\pi}{2})Mq(\vec{e}_3,\sfrac{\pi}{2})^{-1}O_{\rho}^{-1},\\
        L&=-p(\vec{e}_1,\pi)Lq(\vec{e}_1,\pi)^{-1}O_\sigma^{-1},&M&=-O_\sigma q(\vec{e}_1,\pi)Mq(\vec{e}_1,\pi)^{-1}O_\sigma^{-1},
    \end{aligned}\label{B=5-D2d-symm}
\end{align}
with $O_\rho$ and $O_\sigma$ to be determined. An immediate consequence of these symmetries follows from Lemma \ref{lem:reps}: since the isorotations form a negative representation of the dicyclic group ${\rm Dic}_4$, we must have that $(O_\rho,O_\sigma)$ form a positive representation, i.e. 
\begin{align}
    O_\rho^4=O_\sigma^2=(O_\sigma O_\rho)^2=\ind_5,
\end{align}
meaning they really represent the dihedral group ${\rm D}_4$. There are several possible cases to consider, and a systematic search through these is not realistic. So in order to fix the compensating transformations, we shall form an ansatz for what we expect the $\BB=5$ data to look like by using the colouring phenomenology outlined in Section \ref{sec:colouring}. Asymptotically, one may think of the $\BB=5$ skyrmion as five $1$-skyrmions, all aligned along the $\vec{e}_3$ axis, with positions $(R_1,R_2,0,-R_2,-R_1)$, sizes $(\lambda_1,\lambda_2,\lambda_3,\lambda_2,\lambda_1)$, and orientations $(\k,\i,\1,\j,\k)$. The minimal energy version should occur when the scales and separations are of a similar order of magnitude. This configuration may be represented by the diagonal ADHM data
\begin{align}
    \wh{L}=\begin{pmatrix}
    \lambda_1\k&\lambda_2\i&\lambda_3\1&\lambda_2\j&\lambda_1\k
    \end{pmatrix},\quad \wh{M}=\diag\{R_1\k,R_2\k,0,-R_2\k,-R_1\k\}.\label{diagonal-B=5}
\end{align}
Imposing the symmetry \eqref{B=5-D2d-symm} on \eqref{diagonal-B=5} forces
\begin{align}
    O_\rho=\begin{pmatrix}
    0&0&0&0&-1\\
    0&0&0&1&0\\
    0&0&-1&0&0\\
    0&-1&0&0&0\\
    -1&0&0&0&0
    \end{pmatrix},\quad\text{and}\quad O_\sigma=\diag\{1,-1,-1,1,1\}.\label{D2d-gts}
\end{align}
This allows us to form an ansatz for ADHM data with this symmetry given by
\begin{align}
    \begin{aligned}
        L&=\begin{pmatrix}
    \lambda_1\k&\lambda_2\i&\lambda_3\1&\lambda_2\j&\lambda_1\k
    \end{pmatrix},\\
    M&=\begin{pmatrix}
    R_1\k&a_1\i&\eta_1\1&a_2\j&0\\
    a_1\i&R_2\k&a_3\j&\eta_2\1&-a_2\i\\
    \eta_1\1&a_3\j&0&a_3\i&-\eta_1\1\\
    a_2\j&\eta_2\1&a_3\i&-R_2\k&-a_1\j\\
    0&-a_2\i&-\eta_1\1&-a_1\j&-R_1\k
    \end{pmatrix},
    \end{aligned}\label{D2d-5-gen-orientation}
\end{align}
and imposing the reality condition \eqref{reality-cond} yields the equations
\begin{align}
    \begin{aligned}
(a_1+a_2)\eta_1+a_3(\eta_2+R_2) + \lambda_2\lambda_3&=0,\\
a_1(R_2-R_1-\eta_2)+a_2(R_2+R_1-\eta_2)- 2a_3\eta_1 &=0,\\
a_1(\eta_2+R_2-R_1) - a_2(\eta_2+R_1+R_2) - 2\lambda_1\lambda_2&=0,\\
(a_1-a_2)a_3 - R_1\eta_1 - \lambda_1\lambda_3&=0,\\
a_3^2-2a_1a_2 - \lambda_2^2 - 2\eta_2R_2&=0.
    \end{aligned}
\end{align}
There are several solutions to this system satisfying \eqref{non-singularity-cond}. We shall only consider the very general solution with open constraints, where the first three equations are an invertible system in $(a_1,a_2,a_3)$. This occurs when
\begin{align}
    \Delta:=(R_2+\eta_2)(R_1^2-2\eta_1^2-R_2^2+\eta_2^2)\neq0,
\end{align}
with the solution conveniently written as
\begin{align}
    \begin{pmatrix}
    a_1-a_2\\a_1+a_2\\a_3
    \end{pmatrix}=-\dfrac{\lambda_2}{\Delta}\begin{pmatrix}
    2((2\eta_1^2+R_2^2-\eta_2^2)\lambda_1 - R_1\eta_1\lambda_3)\\
    2(R_2+\eta_2)(R_1\lambda_1-\eta_1\lambda_3)\\
    (R_1^2-R_2^2+\eta_2^2)\lambda_3 - 2R_1\eta_1\lambda_1
    \end{pmatrix}.
\end{align}
The penultimate equation then defines $\lambda_2$, namely
\begin{align}
    \lambda_2^2=\dfrac{\Delta}{2}\dfrac{R_1\eta_1+\lambda_1\lambda_3}{\left((2\eta_1^2+R_2^2-\eta_2^2)\lambda_1 - R_1\eta_1\lambda_3\right)\left((R_1^2-R_2^2+\eta_2^2)\lambda_3 - 2R_1\eta_1\lambda_1\right)}.
\end{align}
The final equation is extremely complicated, so we omit it from here, but one may verify that it is reduced to a homogeneous, degree eight polynomial in the remaining variables $(R_1,R_2,\eta_1,\eta_2,\lambda_1,\lambda_3)$, which means we cannot guarantee a closed-form solution in all variables. However the variables $\lambda_1,\lambda_3$ only appear up to order three, so one may resolve this condition explicitly for one of these. The number of real solutions to this cubic varies across the moduli space; when resolving the iteration \eqref{Newton-Raphson}, we always choose the root which gives the lowest energy i-skyrmion. The most general solution to the reality condition is a five-parameter family. Note that this is the number of expected physical parameters (two positions $R_1, R_2$ and three scales $\lambda_1, \lambda_2, \lambda_3$).

We find the minimal energy configuration arises when
 \begin{equation}
(R_1, R_2, \lambda_1, \lambda_2, \lambda_3) = (2.05, 0.99, 0.95, 1.12, 1.07)
 \end{equation}
which corresponds to $(\eta_1,\eta_2,a_1, a_2, a_3) = (-0.015, -0.079,-0.87,-0.26,1.05)$. This configuration has energy $E=5.667$. We can then calculate the scale of the solution using \eqref{instanton-scale}, which gives $\lambda=1.05$.
\subsection{\texorpdfstring{$\BB=6$}{B=6}}
The symmetry of the $\BB=6$ skyrmion is ${\rm D}_{4d}$, which may be represented by ADHM data satisfying
\begin{align}
    \begin{aligned}
    L&=-p(\vec{e}_3,\sfrac{\pi}{2})Lq(\vec{e}_3,\sfrac{5\pi}{4})^{-1}O_\rho^{-1},&M&=-O_\rho q(\vec{e}_3,\sfrac{5\pi}{4})Mq(\vec{e}_3,\sfrac{5\pi}{4})^{-1}O_\rho^{-1},\\
    L&=-p(\vec{e}_2,\pi)Lq(\vec{e}_1,\pi)^{-1}O_\sigma^{-1},&M&=-O_\sigma q(\vec{e}_1,\pi)Mq(\vec{e}_1,\pi)^{-1}O_\sigma^{-1},
    \end{aligned}\label{B=6-symmetry}
\end{align}
for some c.g.ts $O_\rho,O_\sigma\in\O(6)$. We have simplified the expressions here, but to see the symmetries more clearly, one may verify that, in terms of rotations, reflections, isorotations, and isoreflections, the conditions \eqref{B=6-symmetry} translate as
\begin{itemize}
    \item A $\sfrac{\pi}{4}$ rotation and $-\sfrac{\pi}{2}$ isorotation around $\vec{e}_3$, coupled with the reflection $\vec{e}_3\mapsto-\vec{e}_3$ and isoreflection ${\pi}_3\mapsto-{\pi}_3$;
    \item A $\pi$ rotation and isorotation around $\vec{e}_1$ and $\vec{e}_2$ respectively, coupled with the reflection $\vec{x}\mapsto-\vec{x}$ and isoreflection $\vec{\pi}\mapsto-\vec{\pi}$.
\end{itemize}
Like the ${\rm D}_{2d}$-symmetric $\BB=5$ solution above, there is not enough symmetry to easily constrain the general form of the ADHM data (for example, by using representation theory). Instead, we again make an ansatz to describe the configuration as in \eqref{general-cluster}. The $\BB=6$ solution looks like three stacked $\BB=2$ tori with positions $(R, 0, -R)$ and sizes $(\lambda_1, \lambda_2, \lambda_1)$. The relative orientations are fixed by the ${\rm D}_{4d}$-symmetry as outlined in Section \ref{sec:colouring}. For $R\gg 1$, the three tori may be described by the diagonal data
\begin{align}
\begin{aligned}
    \wh{L}&=\begin{pmatrix}
    \lambda_1\k&-\lambda_1\1&\lambda_2\j&\lambda_2\i&\lambda_1\1&\lambda_1\k
    \end{pmatrix},\\\wh{M}&=\diag\{\mu_1m_{T^2}+R\k\,\ind_2,\mu_2m_{T^2},\mu_1m_{T^2}-R\k\,\ind_2\},
\end{aligned}
\end{align}
where $m_{T^2}=\begin{pmatrix}
\i&\j\\
\j&-\i
\end{pmatrix}$ is the $M$ matrix for the $\BB=2$ torus \eqref{torus}. Imposing \eqref{B=6-symmetry} on this diagonal data yields the compensating gauge transformations
\begin{align}
\begin{aligned}
    O_\rho&=\dfrac{\sqrt{2+\sqrt{2}}}{2}\begin{pmatrix}
    0&0&0&0&1&1-\sqrt{2}\\
    0&0&0&0&\sqrt{2}-1&1\\
    0&0&1&1-\sqrt{2}&0&0\\
    0&0&\sqrt{2}-1&1&0&0\\
    -1&\sqrt{2}-1&0&0&0&0\\
    1-\sqrt{2}&-1&0&0&0&0
    \end{pmatrix},\\
    O_\sigma&=\diag\{\sigma^1,-\sigma^1,-\sigma^1\},\quad\sigma^1=\begin{pmatrix}
    0&1\\
    1&0
    \end{pmatrix}.
\end{aligned}
\end{align}
The most general solution to \eqref{B=6-symmetry} with these c.g.ts takes the form
\begin{align}
    \begin{aligned}
    L&=\begin{pmatrix}
    \lambda_1\k&-\lambda_1\1&\lambda_2\j&\lambda_2\i&\lambda_1\1&\lambda_1\k
    \end{pmatrix},\\
    M&=\begin{pmatrix}
    \mu_1\i+R\k&\mu_1\j&\nu\j&\nu\i&\eta\1&0\\
    \mu_1\j&-\mu_1\i+R\k&\nu\i&-\nu\j&0&\eta\1\\
    \nu\j&\nu\i&\mu_2\i&\mu_2\j&\nu\i&-\nu\j\\
    \nu\i&-\nu\j&\mu_2\j&-\mu_2\i&-\nu\j&-\nu\i\\
    \eta\1&0&\nu\i&-\nu\j&\mu_1\i-R\k&\mu_1\j\\
    0&\eta\1&-\nu\j&-\nu\i&\mu_1\j&-\mu_1\i-R\k
    \end{pmatrix},
    \end{aligned}
\end{align}
and imposing the reality condition \eqref{reality-cond} yields the equations
\begin{align}
    \begin{aligned}
    \mu_1^2&=\frac{\lambda_1^2}{2}+\nu^2,&\mu_2^2&=\frac{\lambda_2^2}{2}+2\nu^2,\\
    \lambda_1^2&=2(\nu^2-R\eta),&\lambda_1\lambda_2&=-\nu(R+\eta).
    \end{aligned}\label{B=6-reality}
\end{align}
The top two equations may be used to determine $\mu_1,\mu_2$, with both roots equivalent up to gauge choice and isometries. In order to guarantee irreducibility \eqref{non-singularity-cond}, we must have $\nu\neq0$, so we may also determine
\begin{align}
    \eta=-R-\frac{\lambda_1\lambda_2}{\nu}.
\end{align}
Finally, we may determine $\lambda_2$ in terms of the non-zero parameters $R,\lambda_1$, and $\nu$ from the remaining equation. Hence, we have a three-parameter family of ${\rm D}_{4d}$-symmetric $\BB=6$ ADHM data. Again, the number of parameters match the expected physical parameters: two scales $\lambda_1, \lambda_2$ and a separation $R$.

We find that the minimum energy skyrmion has energy $E=6.736$ when the parameters are
\begin{equation}
(\lambda_1, \lambda_2, R ) = (0.98,  1.03, 2.0), \quad (\mu_1, \mu_2, \nu, \eta) = (0.88,1.05,0.54,-0.10) .
\end{equation}
Using the normalisation \eqref{instanton-scale}, the optimal scale is thus $\lambda=1.00$.
\subsection{\texorpdfstring{$\BB=7$}{B=7}}
The accepted minimal energy $7$-skyrmion has dodecahedral symmetry. An instanton approximation is given in \cite{SingerSutcliffe1999} via the ADHM data
\begin{align}
    \begin{aligned}
    L&=\lambda\frac{\sqrt{7}}{2}\begin{pmatrix}
    \1&\i&\j&\k&0&0&0
    \end{pmatrix},\\
    M&=\lambda\frac{\sqrt{7}}{2}\left(\begin{array}{c|c}
    \bm{0}&\II\\\hline
    \II^t&\bm{0}\end{array}\right),\; \lambda>0,
    \end{aligned}\quad\II=\begin{pmatrix}
    \i&\j&\k\\
    0&\tau\k&\tau^{-1}\j\\
    \tau^{-1}\k&0&\tau\i\\
    \tau\j&\tau^{-1}\i&0
    \end{pmatrix},\label{dodecahedron}
\end{align}
where $\tau=\sfrac{1}{2}(1+\sqrt{5})$ is the golden ratio. The full ${\rm Y}_h$-symmetry is realised here via\footnote{These rotations and isorotations may be written more transparently as
\begin{align*}
    \begin{aligned}
    q(\vec{n}_5^+,\sfrac{2\pi}{5})&=\sfrac{1}{2}(\tau\1+\tau^{-1}\i+\k),&p(\vec{n}_5^-,\sfrac{4\pi}{5})&=\sfrac{1}{2}(\tau^{-1}\1+\tau\i-\k),\\
    q(\vec{n}_3^+,\sfrac{2\pi}{3})&=\sfrac{1}{2}(\1+\tau^{-1}\j+\tau\k),&p(\vec{n}_3^-,-\sfrac{2\pi}{3})&=\sfrac{1}{2}(\1-\tau\j-\tau^{-1}\k).
    \end{aligned}
\end{align*}
}
\begin{align}
    \begin{aligned}
    L&=p(\vec{n}_5^-,\sfrac{4\pi}{5})Lq(\vec{n}_5^+,\sfrac{2\pi}{5})^{-1}O_5^{-1},&M&=O_5q(\vec{n}^+_5,\sfrac{2\pi}{5})Mq(\vec{n}_5^+,\sfrac{2\pi}{5})^{-1}O_5^{-1},\\
    L&=p(\vec{n}_3^-,-\sfrac{2\pi}{3})Lq(\vec{n}_3^+,\sfrac{2\pi}{3})^{-1}O_3^{-1},&M&=O_3q(\vec{n}_3^+,\sfrac{2\pi}{3})Mq(\vec{n}_3^+,\sfrac{2\pi}{3})^{-1}O_3^{-1},\\
    L&=-LO_-^{-1},&M&=-O_-MO_-^{-1},
    \end{aligned}\label{dodecahedron-symmetry}
\end{align}
where $\vec{n}_5^\pm=\sqrt{\sfrac{5\pm\sqrt{5}}{10}}(\tau^{\mp1}\vec{e}_1\pm\vec{e}_3)$, $\vec{n}_3^\pm=\sfrac{1}{\sqrt{3}}(\tau^{\mp1}\vec{e}_2+\tau^{\pm1}\vec{e}_3)$, and
\begin{align}
\begin{aligned}
    O_5&=\dfrac{1}{2}\begin{pmatrix}
    \sfrac{1}{2}&-\sfrac{\sqrt{5}}{2}&\sfrac{\sqrt{5}}{2}&\sfrac{\sqrt{5}}{2}&0&0&0\\
    \sfrac{\sqrt{5}}{2}&\sfrac{3}{2}&\sfrac{1}{2}&\sfrac{1}{2}&0&0&0\\
    \sfrac{\sqrt{5}}{2}&-\sfrac{1}{2}&\sfrac{1}{2}&-\sfrac{3}{2}&0&0&0\\
    -\sfrac{\sqrt{5}}{2}&\sfrac{1}{2}&\sfrac{3}{2}&-\sfrac{1}{2}&0&0&0\\
    0&0&0&0&-1&-\tau^{-1}&\tau\\
    0&0&0&0&\tau^{-1}&\tau&1\\
    0&0&0&0&\tau&-1&\tau^{-1}
    \end{pmatrix},\\
    O_3&=\dfrac{1}{2}\begin{pmatrix}
    -\sfrac{1}{2}&\sfrac{\sqrt{5}}{2}&\sfrac{\sqrt{5}}{2}&\sfrac{\sqrt{5}}{2}&0&0&0\\
    \sfrac{\sqrt{5}}{2}&\sfrac{3}{2}&-\sfrac{1}{2}&-\sfrac{1}{2}&0&0&0\\
    -\sfrac{\sqrt{5}}{2}&\sfrac{1}{2}&\sfrac{1}{2}&-\sfrac{3}{2}&0&0&0\\
    -\sfrac{\sqrt{5}}{2}&\sfrac{1}{2}&-\sfrac{3}{2}&\sfrac{1}{2}&0&0&0\\
    0&0&0&0&-1&\tau^{-1}&-\tau\\
    0&0&0&0&-\tau^{-1}&\tau&1\\
    0&0&0&0&\tau&1&-\tau^{-1}
    \end{pmatrix},\\
    O_-&=\diag\{-1,-1,-1,-1,1,1,1\}.\label{dodecahdedron-gts}
\end{aligned}
\end{align}
The energy-minimising i-skyrmion is given when $\lambda=0.98$ with energy $E=7.766$.
\subsection{\texorpdfstring{$\BB=8$}{B=8}}
In the massless Skyrme model \eqref{Skyrme-energy}, the minimal energy $\BB=8$ configuration has ${\rm D}_{6d}$-symmetry. The ${\rm D}_{6d}$-symmetry may be imposed via the conditions
\begin{align}
    \begin{aligned}
    L&=-p(\vec{e}_3,\sfrac{\pi}{3})Lq(\vec{e}_3,\sfrac{7\pi}{6})^{-1}O_\rho^{-1},&M&=-O_\rho q(\vec{e}_3,\sfrac{7\pi}{6})Mq(\vec{e}_3,\sfrac{7\pi}{6})^{-1}O_\rho^{-1},\\
    L&=-p(\vec{e}_2,\pi)Lq(\vec{e}_1,\pi)^{-1}O_\sigma^{-1},&M&=-O_\sigma q(\vec{e}_1,\pi)Mq(\vec{e}_1,\pi)^{-1}O_\sigma^{-1}.
    \end{aligned}\label{D_6d}
\end{align}
This may be understood in terms of rotations, isorotations, reflections, and isoreflections analogously to the $\BB=6$ case, but with the angles of rotation and isorotation around $\vec{e}_3$ here replaced by $\sfrac{\pi}{6}$ and $-\sfrac{2\pi}{3}$ respectively.

As with the previous cases, it is convenient to build the data by considering its deformation into smaller clusters. This $\BB=8$ skyrmion is like the $\BB=6$, but with the middle torus replaced by a $\BB=4$ torus. This may be described asymptotically by ADHM data which is approximately of the form
\begin{align}
\begin{aligned}
    \wh{L}&=\begin{pmatrix}
    \lambda_1\k&-\lambda_1\1&\lambda_2\j&\lambda_2\i&0&0&\lambda_1\1&\lambda_1\k
    \end{pmatrix},\\\wh{M}&=\diag\{\mu_1m_{T^2,2}+R\k\,\ind_2,\mu_2m_{T^2,4},\mu_1m_{T^2,2}-R\k\,\ind_2\},
\end{aligned}
\end{align}
where $m_{T^2,n}$ denotes the $\BB=n$ torus \eqref{B-even-torus} for $n=2,4$, which we go on to explain in greater detail in Section \ref{sec:tori}. This forces c.g.ts given by
\begin{align}
    \begin{aligned}
    O_\rho&=\dfrac{\sqrt{2}}{4}\begin{pmatrix}
    0&0&0&0&0&0&\sqrt{3}+1&1-\sqrt{3}\\
    0&0&0&0&0&0&\sqrt{3}-1&\sqrt{3}+1\\
    0&0&2&-2&0&0&0&0\\
    0&0&2&2&0&0&0&0\\
    0&0&0&0&\sqrt{3}+1&1-\sqrt{3}&0&0\\
    0&0&0&0&\sqrt{3}-1&\sqrt{3}+1&0&0\\
    -(\sqrt{3}+1)&\sqrt{3}-1&0&0&0&0&0&0\\
    1-\sqrt{3}&-(\sqrt{3}+1)&0&0&0&0&0&0
    \end{pmatrix},\\
    O_\sigma&=\diag\{\sigma^1,-\sigma^1,\sigma^1,-\sigma^1\},\quad\sigma^1=\begin{pmatrix}
    0&1\\
    1&0
    \end{pmatrix}.
    \end{aligned}
\end{align}
Imposing \eqref{D_6d} with these c.g.ts yields the ansatz
\begin{align*}
    \begin{aligned}
    L&=\begin{pmatrix}
    \lambda_1\k&-\lambda_1\1&\lambda_2\j&\lambda_2\i&0&0&\lambda_1\1&\lambda_1\k
    \end{pmatrix},\\
    M&=\begin{pmatrix}
    \mu_1\i+R\k&\mu_1\j&\nu\j&\nu\i&0&0&\eta\1&0\\
    \mu_1\j&-\mu_1\i+R\k&\nu\i&-\nu\j&0&0&0&\eta\1\\
    \nu\j&\nu\i&0&0&\mu_2\i&-\mu_2\j&\nu\i&-\nu\j\\
    \nu\i&-\nu\j&0&0&\mu_2\j&\mu_2\i&-\nu\j&-\nu\i\\
    0&0&\mu_2\i&\mu_2\j&\chi\i&\chi\j&0&0\\
    0&0&-\mu_2\j&\mu_2\i&\chi\j&-\chi\i&0&0\\
    \eta\1&0&\nu\i&-\nu\j&0&0&\mu_1\i-R\k&\mu_1\j\\
    0&\eta\1&-\nu\j&-\nu\i&0&0&\mu_1\j&-\mu_1\i-R\k
    \end{pmatrix}.
    \end{aligned}
\end{align*}
The reality condition \eqref{reality-cond} forces $\chi=\mu_2$, and, unsurprisingly, the same equations \eqref{B=6-reality} as in the $\BB=6$ case. Thus in the same way as there, we have a three-parameter family, determined by the sizes and separation $\lambda_1,\lambda_2,R$. The energy-minimising i-skyrmion within this family is found to have energy $E=8.933$ when
\begin{align}
    (\lambda_1,\lambda_2,R)=(1.07,1.35,2.03),\quad (\mu_1,\mu_2,\nu,\eta)=(-1.05,1.40,-0.72,-0.026).
\end{align}
According to the normalisation \eqref{instanton-scale}, the scale of this solution is $\lambda=1.01$.
\newpage 
\subsection{Comparison with numerically-generated-skyrmions}
The minimal energy skyrmions with $1\leq\BB\leq8$ and their energies are well known. In Table \ref{tab:energy-comparison}, we compare the i-skyrmions to the numerically-generated-skyrmions from \cite{battye2001solitonic}. We see that the error in the energy calculation is never more than $2\%$. This confirms that the instanton approximation works well for a wide range of skyrmions. In the table we also record the optimal normalised scale \eqref{instanton-scale} for each minimiser. It is worth remarking that $\lambda$ appears to be approximately $1$ as $\BB$ grows. This observation is useful for calculating energy-minimising i-skyrmions for larger $\BB$ as it gives a good idea of what scale to set for the initial configuration.

\begin{table}[h!]
	\centering
	\begin{tabular}{ | c | c | c | c | c | c | c | c | c | } \hline
	 	$\BB$ &1 &  2 & 3 & 4 & 5 & 6 & 7 & 8 \\ \hline
	 	Symmetry & $\O(3)$ & ${\rm D}_{\infty h}$ & ${\rm T}_d$ & $\O_h$ & ${\rm D}_{2d}$ & ${\rm D}_{4d}$ & ${\rm Y}_h$ & ${\rm D}_{6d}$ \\
	 	$E_\text{inst.}$ & 1.243 & 2.384 & 3.488 & 4.532 & 5.667 & 6.736 & 7.766  & 8.933 \\
	 	$E_\text{num.}$ & 1.232 & 2.358 & 3.438 & 4.480 & 5.586  & 6.647 & 7.663 & 8.768 \\
	 	Error & 0.89\% & 1.10\% & 1.45\% & 1.16\% & 1.45\% & 1.34\% & 1.34\% & 1.88\% \\  
	 	$\lambda$ &  1.45 & 1.31 & 1.11 & 1.04 & 1.05 & 1.00 &   0.98 & 1.01  \\ \hline
	\end{tabular}
	\caption{A comparison between the energies of i-skyrmions ($E_\text{inst.}$) and numerically minimised skyrmions from \cite{battye2001solitonic} ($E_\text{num.}$). We also tabulate the size $\lambda$ defined by \eqref{instanton-scale}, and the symmetries of the energy-minimising skyrmions.}
	\label{tab:energy-comparison}
\end{table}
\section{The ADHM zoo: exploring the moduli space}\label{sec:ADHM-zoo}
In the preceding sections we have established a framework for viewing rank $\BB$ ADHM data in terms of lower charge clusters, and have presented ADHM data for the standard charge $1\leq\BB\leq8$ skyrmions. There is a whole $8\BB$-dimensional moduli space to look at, and in this section we highlight some important examples. In particular we describe some relatively large analytic families which interpolate between well-separated cluster configurations and highly-symmetric central configurations.
\subsection{Tori}\label{sec:tori}
An interesting sequence of instantons are tori. As we shall show, for all $\BB\geq2$, there exist ADHM data with the ${\rm D}_{\infty h}$-symmetry of the torus. These are important, for example for constructing minimal energy i-skyrmions, as we have already seen with the ${\rm D}_{6d}$-symmetric $\BB=8$. Furthermore, the larger tori play a role in some generalised Skyrme models, where they are found as minimal energy solutions \cite{GudnasonNitta2015baryonic}.

Approaches to obtain charge $\BB$ tori have already been considered elsewhere, for example using JNR data \cite{Jackiw1977} with equally weighted poles the vertices of a $\BB$-gon, or by thinking about axially-symmetric hyperbolic monopoles \cite{BraamAustin1990boundary,cockburn2014symmetric}. Here we present a more direct construction which does not rely on indirect methods, and allows for a more systematic analysis.

The tori are described by ADHM data $(L,M)\in\A_\BB$ satisfying the ${\rm D}_{\infty h}$-invariance conditions
\begin{align}
    \begin{aligned}
    L&=p(\vec{e}_3,\BB\theta)Lq(\vec{e}_3,\theta)^{-1}O_\theta^{-1},&M&=O_\theta q(\vec{e}_3,\theta)Mq(\vec{e}_3,\theta)^{-1}O_\theta^{-1},\\
    L&=p(\vec{e}_1,\pi)Lq(\vec{e}_1,\pi)^{-1}O_2^{-1},&M&=O_2q(\vec{e}_1,\pi)Mq(\vec{e}_1,\pi)^{-1}O_2^{-1},\\
    L&=-p_-LO_-^{-1},&M&=-O_-MO_-^{-1}.
    \end{aligned}\label{torus-symmetry}
\end{align}
The choice of $p_-$ and the compensating gauge transformations depend on the parity of $\BB$. For $\BB$ even, we consider \eqref{torus-symmetry} with $p_-=p(\vec{e}_3,\pi)=\k$, and compensating transformations
\begin{align}
    \begin{aligned}
    O_\theta&=\bigoplus_{k=0}^{\sfrac{\BB}{2}-1} Q_{\BB-2k-1},&
    O_2&=Q_\sigma\oplus\cdots\oplus Q_\sigma,&
    O_-&=\bigoplus_{k=1}^{\sfrac{\BB}{2}}(-1)^kQ_-,
    \end{aligned}\label{torus-gts-B-even}
\end{align}
and for $\BB$ odd, we consider \eqref{torus-symmetry} with $p_-=\1$, and compensating transformations
\begin{align}
    \begin{array}{c}
    O_\theta=\left(\bigoplus_{k=0}^{\sfrac{\BB-1}{2}-1} Q_{\BB-2k-1}\right)\oplus(1),\quad
    O_2=Q_\sigma\oplus\cdots\oplus Q_\sigma\oplus(1),\\
    O_-=\left(\bigoplus_{k=1}^{\sfrac{\BB-1}{2}}(-1)^k\ind_2\right)\oplus(-1)^{\BB-1}.
    \end{array}\label{torus-gts-B-odd}
\end{align}
Here $Q_k$ denotes the irreducible representation \eqref{rep-SO(2)} of $\SO(2)$, and we have denoted
\begin{align}
    Q_\sigma=\begin{pmatrix}1&0\\
    0&-1\end{pmatrix}\quad\text{and}\quad Q_-=\begin{pmatrix}0&-1\\
    1&0\end{pmatrix}.
\end{align}
The invariant data may be determined by appealing to the formalism discussed at the end of Section \ref{sec:symmetric-data}. Here we lay out this process in detail. In each case, the c.g.t $O_\theta$ for the axial symmetry in \eqref{torus-symmetry} is decomposed as a direct sum of the form $V=Q_{k_1}\oplus\cdots\oplus Q_{k_{\lfloor{\BB/2}\rfloor}}$, consisting of the representation defined in \eqref{rep-SO(2)}, with the addition of a trivial component in the case $\BB$ odd. Since $Q_m\otimes Q_n\cong Q_{m+n}\oplus Q_{m-n}$, it follows that each $1\times 2$ block $L_i$ of $L$, and $2\times 2$ block $M_{ij}$ of $M$, must lie in a trivial subrep of
\begin{align}
    Q_{k_i-\BB+1}\oplus Q_{k_i+\BB+1}\oplus Q_{k_i-\BB-1}\oplus Q_{k_i+\BB-1},\label{invariant-L}
\end{align}
and
\begin{align}
    2(Q_{k_i-k_j}\oplus Q_{k_i+k_j})\oplus Q_{k_i+k_j-2}\oplus Q_{k_i+k_j+2}\oplus Q_{k_i-k_j-2}\oplus Q_{k_i-k_j+2},\label{invariant-M}
\end{align}
respectively. Since we are considering $\SO(2)$ here, this occurs if and only if at least one of the indices here is zero.\footnote{This may be generalised for finite cyclic groups of order $n$, by replacing this condition with the requirement of indices being zero modulo $n$.} Given the choice in \eqref{torus-gts-B-even}-\eqref{torus-gts-B-odd}, up to a sign, for each block there is a unique choice for the corresponding component of $O_2$ and $O_-$, and we may use these to further fix the invariant blocks.

From \eqref{invariant-L}, and the choice \eqref{torus-gts-B-even}-\eqref{torus-gts-B-odd}, only one block in $L$ will be non-zero, corresponding to the component $Q_{\BB-1}$. Combining this with the first $2\times 2$ block for $O_2$ and $O_-$ respectively, and imposing \eqref{torus-symmetry}, yields the invariant block
\begin{align}
    L_1=\alpha\begin{pmatrix}
    \1&\k
    \end{pmatrix},
\end{align}
and we may always choose a gauge where $\alpha>0$. For the matrix $M$, from \eqref{invariant-M} we see that there are only two possible cases for invariant $2\times 2$ blocks. These occur when $k_i=k_j$ and when $k_i=k_j\pm2$; the cases $\pm$ lead to data which are the transpose of the other. Due to the choice \eqref{torus-gts-B-even}-\eqref{torus-gts-B-odd}, the case $k_i=k_j$ only affects the diagonal $2\times 2$ blocks, which are necessarily symmetric matrices. It is a straightforward exercise to check that, when combined with the action of $O_2$ and $O_-$, the only solution to \eqref{torus-symmetry} which is also a symmetric matrix occurs when $k_i=1$, and takes the form
\begin{align}
    M_{ii}=\beta_i m_{T^2},\quad m_{T^2}=\begin{pmatrix}
    \i&\j\\
    \j&-\i
    \end{pmatrix},\label{block_ii}
\end{align}
which is precisely the form of the data for the $\BB=2$ torus \eqref{torus}. Similarly, the case $k_i=k_j\pm2$ only affects the immediate off-diagonal blocks, and the invariant data, after imposing the full symmetry \eqref{torus-symmetry} takes the form
\begin{align}
    M_{i(i+1)}=\mu_i\Sigma,\quad\Sigma=\begin{pmatrix}
    \i&-\j\\
    \j&\i
    \end{pmatrix}.\label{block_ii+1}
\end{align}
Finally, in the case $\BB$ odd, the top left $(\BB-1)\times(\BB-1)$ block is fixed by the above analysis, and there remains a $1\times\BB$ block, its transpose, and the bottom $1\times 1$. The latter is easily seen to be $0$ after imposition of the full symmetry \eqref{torus-symmetry}. To classify the $1\times\BB$ block, it suffices to determine $k_i$ such that there is a trivial component in \eqref{invariant-M} for $k_j=0$. The only block in the choice \eqref{torus-gts-B-odd} is when $k_i=2$, and imposing the full symmetry leads to the block $\gamma\begin{pmatrix}
\i&\j
\end{pmatrix}$. For uniqueness up to gauge equivalence and discrete changes in orientation, it suffices to choose $\beta_i,\mu_i,\gamma>0$.

It remains to resolve the reality condition \eqref{reality-cond} in each case. For $\BB$ even, the solution takes the form
\begin{align}
    \begin{aligned}
        L&=\lambda\sqrt{\frac{\BB}{2}}\begin{pmatrix}
        \1&\k&0&\cdots&0
        \end{pmatrix},&
        M&=\lambda\frac{\sqrt{\BB}}{2}\left(\begin{array}{c|c|c|c|c}
            \bm{0} & \Sigma & \bm{0} & \cdots & \bm{0}\\\hline
            \Sigma^t & \bm{0} & \Sigma & \ddots & \vdots\\\hline
            \bm{0} & \Sigma^t&\ddots&\ddots&\bm{0}\\\hline
            \vdots&\ddots&\ddots&\bm{0}&\Sigma\\\hline
            \bm{0}&\cdots&\bm{0}&\Sigma^t&m_{T^2}
        \end{array}\right),
    \end{aligned}\label{B-even-torus}
\end{align}
and for $\BB>1$ odd, the solution is
\begin{align}
    \begin{aligned}
        L&=\lambda\sqrt{\frac{\BB}{2}}\begin{pmatrix}
        \1&\k&0&\cdots&0
        \end{pmatrix},&
        M&=\lambda\frac{\sqrt{\BB}}{2}\left(\begin{array}{c|c}\begin{array}{c|c|c|c|c}
            \bm{0} & \Sigma & \bm{0} & \cdots & \bm{0}\\\hline
            \Sigma^t & \bm{0} & \Sigma & \ddots & \vdots\\\hline
            \bm{0} & \Sigma^t&\ddots&\ddots&\bm{0}\\\hline
            \vdots&\ddots&\ddots&\bm{0}&\Sigma\\\hline
            \bm{0}&\cdots&\bm{0}&\Sigma^t&\bm{0}\end{array}&\vec{\nu}^t\\\hline
            \vec{\nu}&0
        \end{array}\right),\label{B-odd-torus}
    \end{aligned}
\end{align}
where $\vec{\nu}=\sqrt{2}\begin{pmatrix}
0&\cdots&0&\i&\j
\end{pmatrix}$, and we have normalised according to \eqref{instanton-scale}. It is straightforward to check that both of these cases satisfy \eqref{non-singularity-cond}.

It is reasonable to conjecture that, up to gauge-equivalence and isometries, these are the only possible irreducible data satisfying \eqref{torus-symmetry}. This is supported by the relationship to hyperbolic monopoles \cite{BraamAustin1990boundary,cockburn2014symmetric}, but also the representation theory. Indeed, the c.g.t $O_\theta$ may still be decomposed as a direct sum of $Q_m$s and so it remains to determine all combinations $(k_i,k_j)$ which lead to trivial subreps of \eqref{invariant-L}-\eqref{invariant-M}. The choice made above is only one of the possible solutions to this problem, however the possibilities for $(k_i,k_j)$ are still limited in a similar way as above. Similarly, the c.g.ts $O_2$ and $O_-$ must decompose as a direct sum of the matrices $\ind_2$, $Q_\sigma$ or $Q_-$, further restricting the possibilities. It is very likely that the choices made above are the only cases which allow for both \eqref{reality-cond}-\eqref{non-singularity-cond} to be satisfied.

\begin{figure}[htbp]
	\centering
	\includegraphics[scale=0.7]{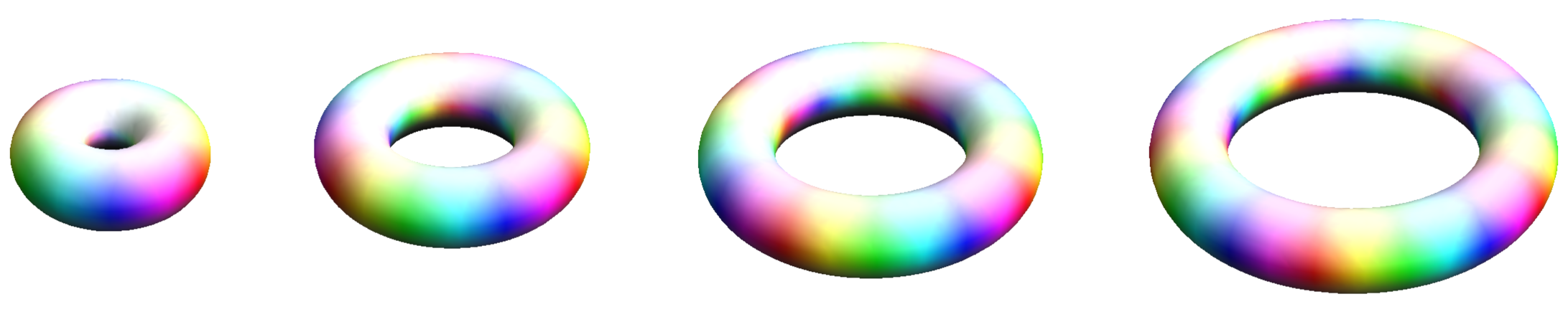}
	\caption{The toroidal i-skyrmions for $\BB=2,3,4,5$. As $\BB$ increases, the tori become larger. We plot an isosurface of constant energy density, with $\mathcal{E} = 0.1$. Note that for the $\BB$-torus, the colour wheel winds $
	\BB$ times around its equator.}
	\label{fig:tori}
\end{figure}

To illustrate these solutions, we have found the energy-minimising toroidal i-skyrmions numerically for $\BB = 2,3,4$, and $5$. We compute their optimal scales to be  $1.28, 1.20, 1.23$, and $1.28$, which give energies of $2.238$, $3.619$, $4.904$, and $6.226$ respectively. The energies are well-approximated by the linear function $E_\BB =  -0.198 + 1.2805 \BB$. We plot energy isosurfaces of the tori in Figure \ref{fig:tori}.

\subsection{Spinning tops}
In this section we consider charge $\BB>3$ configurations which look asymptotically like two $\BB=1$ hedgehogs around an internal $(\BB-2)$-torus: a spinning top! The three constituents are aligned along a common axis of symmetry, with the outer two having the same orientation along that axis, and the middle one oppositely oriented. We shall fix this shared axis of symmetry as $\vec{e}_3$. By comparing \eqref{torus-symmetry} with \eqref{hedgehog-symmetry}, we see that this breaks the symmetry as discussed in Section \ref{sec:sym-decom}; we may orient the system so that the outer $1$-skyrmions are invariant under a $\theta$ rotation and $\theta$ isorotation, whereas the inverted central $(\BB-2)$-skyrmion is invariant under a $\theta$ rotation and $-(\BB-2)\theta$ isorotation. The unbroken symmetry, namely the solution to \eqref{inverted-orientation-shared-iso-rot}, is the $(\BB-1)$-fold cyclic symmetry $\mathrm{C}_{\BB-1}$. To describe this configuration in the most generality, we require invariance under
\begin{align}
    \begin{aligned}
    L&=p(\vec{e}_3,\sfrac{2\pi}{\BB-1})Lq(\vec{e}_3,\sfrac{2\pi}{\BB-1})^{-1}O_{1,\BB-2,1}^{-1},\\
    M&=O_{1,\BB-2,1}q(\vec{e}_3,\sfrac{2\pi}{\BB-1})Mq(\vec{e}_3,\sfrac{2\pi}{\BB-1})^{-1}O_{1,\BB-2,1}^{-1},
    \end{aligned}\label{C_B-1}
\end{align}
where $O_{1,\BB-2,1}=\diag\{1,-O_{\BB-2},1\}$ is a direct sum of the c.g.ts for the constituents, with $O_{\BB-2}$ determined by the formulae \eqref{torus-gts-B-even}-\eqref{torus-gts-B-odd} for the $\BB-2$ torus, evaluated at $\theta=\sfrac{2\pi}{\BB-1}$.
\subsubsection{\texorpdfstring{$\BB=3$}{B=3} prototype}
To get to grips with the spinning tops described above, it is useful to discuss a simpler example when $\BB=3$. In this case, the only difference is that the middle cluster is not a torus, but a hedgehog, and so one considers \eqref{C_B-1} with c.g.t $O_{1,1,1}=\diag\{1,-1,1\}$. Up to gauge, orientation, and translation,\footnote{The orientation is fixed around the central skyrmion, we have fixed the gauge so that $\Re(M)$ is diagonal, and the position so that the central skyrmion is at the origin.} this yields the $\mathrm{C}_2$-symmetric $\BB=3$ ansatz
\begin{align}
    \begin{aligned}
    L&=\begin{pmatrix}
    \lambda_1\omega(\phi)&\kappa\i&\lambda_2\omega(\psi)
    \end{pmatrix},\\
    M&=\begin{pmatrix}
    R_1\k+\eta_1\1&c_1\i+c_2\j&                         \chi\k\\
    c_1\i+c_2\j&0&d_1\i+d_2\j\\
    \chi\k&d_1\i+d_2\j&- R_2\k-\eta_2\1
    \end{pmatrix}.
    \end{aligned}\label{B=3-spinning-top}
\end{align}
where $\omega(\vartheta)=\1\cos\vartheta+\k\sin\vartheta$. Imposing the reality condition leads to the equations
\begin{align}
\begin{aligned}
    \eta_1c_1+R_1c_2 + \chi d_2 + \kappa\lambda_1\cos\phi&=0,\\
    R_1c_1 - \eta_1c_2 + \chi d_1 + \kappa\lambda_1\sin\phi&=0,\\
    \chi c_2 - \eta_2d_1 - R_2d_2   + \kappa\lambda_2\cos\psi&=0,\\
    \chi c_1-R_2d_1 + \eta_2d_2 + \kappa\lambda_2\sin\psi&=0,\\
 c_2d_1 - c_1d_2 + \chi(\eta_1 + \eta_2) + \lambda_1\lambda_2\sin(\psi-\phi)&=0.
\end{aligned}\label{rc-B=3-top}
\end{align}
It is convenient to think of the first four as a linear system in $(c_1,c_2,d_1,d_2)$, namely
\begin{align}
    \begin{pmatrix}
    -\eta_1&-R_1&0&-\chi\\
    -R_1&\eta_1&-\chi&0\\
    0&-\chi&\eta_2&R_2\\
    -\chi&0&R_2&-\eta_2
    \end{pmatrix}\begin{pmatrix}
    c_1\\c_2\\d_1\\d_2
    \end{pmatrix}=\kappa\begin{pmatrix}
    \lambda_1\cos\phi\\
    \lambda_1\sin\phi\\
    \lambda_2\cos\psi\\
    \lambda_2\sin\psi
    \end{pmatrix}.\label{linear-system-spinningtop}
\end{align}
The operator of this system has determinant
\begin{align}
    \Delta=(R_1R_2+\chi^2-\eta_1\eta_2)^2+(R_1\eta_2+R_2\eta_1)^2\geq0,
\end{align}
which is thus invertible if and only if
\begin{align}
    \chi^2\neq\eta_1\eta_2-R_1R_2,\quad\text{or}\quad R_1\eta_2+R_2\eta_1\neq0.
\end{align}
There are solutions besides these cases, but we shall not consider those here. So assuming $\Delta\neq0$, the solution is
\begin{align}
\begin{aligned}
    \begin{pmatrix}
    c_1\\c_2
    \end{pmatrix}&=-\dfrac{\kappa}{\Delta}\begin{pmatrix}
    \lambda_1\TT(\phi)&\lambda_2\TT(\psi)
    \end{pmatrix}\begin{pmatrix}
    \eta_1(R_2^2+\eta_2^2)-\eta_2\chi^2\\R_1(R_2^2+\eta_2^2)+R_2\chi^2\\\chi(R_1\eta_2 + R_2\eta_1)\\\chi(\chi^2+R_1R_2-\eta_1\eta_2)
    \end{pmatrix},\\
    \begin{pmatrix}
    d_1\\d_2
    \end{pmatrix}&=-\dfrac{\kappa}{\Delta}\begin{pmatrix}
    -\lambda_2\TT(\psi)&\lambda_1\TT(\phi)
    \end{pmatrix}\begin{pmatrix}
    \eta_2(R_1^2+\eta_1^2)-\eta_1\chi^2\\R_2(R_1^2+\eta_1^2)+R_1\chi^2\\\chi(R_1\eta_2 + R_2\eta_1)\\\chi(\chi^2+R_1R_2-\eta_1\eta_2)
    \end{pmatrix},
\end{aligned}\label{spinning-top-lin-system-result}
\end{align}
where we have introduced the shorthand, $\TT(x)=\begin{pmatrix}
    \cos x&\sin x\\
    -\sin x&\cos x
\end{pmatrix}$. There then remains one equation to solve, which takes the form
\begin{multline}
    \kappa^2(\lambda_1\lambda_2(\cos(\psi - \phi)|\vec{r}_1\times\vec{r}_2|+ \sin(\psi - \phi)(\chi^2 + \vec{r}_1\cdot\vec{r}_2))-\chi(\eta_1\lambda_2^2 + \eta_2\lambda_1^2))\\
    -\Delta(\chi(\eta_1+\eta_2)+\lambda_1\lambda_2\sin(\psi-\phi))=0,\label{final-reality-condition-spin3}
\end{multline}
where we have denoted $\vec{r}_i=(R_i,\eta_i,0)^t$ for the relative position vectors of the outer two constituents in the $(x_3,x_4)$-plane. There are special cases where this is resolved identically (for example, whenever $\eta_1=\eta_2=0$ and $\psi=\phi\;(\text{mod }\pi)$), but in the most general case, this may be resolved as a condition for $\kappa$, namely
\begin{align}
    \kappa^2=\dfrac{\Delta(\chi(\eta_1+\eta_2)+\lambda_1\lambda_2\sin(\psi-\phi))}{\lambda_1\lambda_2(\cos(\psi - \phi)|\vec{r}_1\times\vec{r}_2|+ \sin(\psi - \phi)(\chi^2 + \vec{r}_1\cdot\vec{r}_2))-\chi(\eta_1\lambda_2^2 + \eta_2\lambda_1^2)},\label{kappa-fix-B=3}
\end{align}
in which case we have assumed that the denominator in \eqref{kappa-fix-B=3} is non-zero. This general case, up to overall position and orientation, is a nine-parameter family of $\mathrm{C}_2$-symmetric, rank $3$ ADHM data. This coincides with the number of expected physical parameters as follows. There are three scales $\lambda_1,\lambda_2$ and $\kappa$, two relative positions $r_i=R_i\k+\eta_i\1$, and two angles $\phi,\psi$ controlling the relative orientations around the axis of symmetry.
\subsubsection{Charge \texorpdfstring{$\BB$}{B} case}
The generalisation of the above to charge $\BB$ is similar, but has some subtle technical differences. One may again fix a gauge so that $\Re(M)$ is diagonal, and, up to a choice of global orientation, the general solution to \eqref{C_B-1} takes the form\footnote{This calculation is analogous to the example of the tori in Section \ref{sec:tori} since ${\rm gcd}(\BB-1,\BB-2)=1$.}
\begin{align}
    \begin{aligned}
        L&=\begin{pmatrix}
        \lambda_1\omega(\phi)&\kappa\i&-\kappa\j&0&\cdots&0&\lambda_2\omega(\psi)
        \end{pmatrix},\\
        M&=\begin{pmatrix}
        R_1\k+\eta_1\1&\sigma_1&\chi\k\\
        \sigma_1^t&m_{\BB-2}+\mathfrak{r}_{\BB-2}&\sigma_2^t\\
        \chi\k&\sigma_2&-R_2\k-\eta_2\1
        \end{pmatrix},
    \end{aligned}\label{general-top}
\end{align}
where $\omega(\vartheta)=\1\cos\vartheta+\k\sin\vartheta$,
\begin{align}
\begin{aligned}
    \sigma_1&=\begin{pmatrix}
    c_1\i+c_2\j&c_2\i-c_1\j&0&\cdots&0
    \end{pmatrix},\\
    \sigma_2&=\begin{pmatrix}
    d_1\i+d_2\j&d_2\i-d_1\j&0&\cdots&0
    \end{pmatrix},\\
    \mathfrak{r}_{\BB-2}&=\left\{\begin{array}{ll}\diag\{v_1\ind_2,\dots,v_{\sfrac{\BB}{2}-1}\ind_2\},&\text{if }\BB\text{ is even},\\
    \diag\{v_1\ind_2,\dots,v_{\sfrac{\BB-1}{2}-1}\ind_2,v_{\sfrac{\BB-1}{2}}\},&\text{if }\BB\text{ is odd},\end{array}\right.
\end{aligned}
\end{align}
with $v_i=a_i\k+b_i\1$, and we have chosen a gauge so that $m_{\BB-2}$ is like the general form for $M$ in the $(\BB-2)$-torus (as in \eqref{B-even-torus} and \eqref{B-odd-torus}), but where each off-diagonal block contains a different factor $\mu_i\in\R$. By translating in the $(x_3,x_4)$-plane we may fix $\sum_iv_i=0$. The reality condition \eqref{reality-cond} then fixes many of these parameters very straightforwardly, namely
\begin{align}
    \mu_i\equiv\mu,\quad v_i=0,\quad\text{for all }i.
\end{align}
This partial imposition of the reality condition shows that the ansatz \eqref{general-top} looks like the block data \eqref{cluster-decomposition} describing two charge $1$ hedgehogs positioned at $r_1=R_1\k+\eta_1\1$ and $-r_2=-R_2\k-\eta_2\1$, and an inverted $(\BB-2)$-torus at the origin.

It is straightforward to check that the remaining conditions are almost identical to \eqref{rc-B=3-top}, but with the final equation replaced by the two equations
\begin{align}
    \begin{aligned}
    c_1^2 + c_2^2 + d_1^2 + d_2^2 + \kappa^2 - 2\mu^2&=0,\\
    2c_2d_1 - 2c_1d_2 + \chi(\eta_1+\eta_2) + \lambda_1\lambda_2\sin(\psi-\phi)&=0.
    \end{aligned}\label{rc-general-top}
\end{align}
Since the first four equations of \eqref{rc-B=3-top} are required here, we may resolve for $(c_1,c_2,d_1,d_2)$ as in \eqref{spinning-top-lin-system-result}, assuming again that the determinant of the system is non-vanishing. The first equation of \eqref{rc-general-top} is the only equation here involving $\mu$, and thus defines it, with both roots gauge-equivalent. There then remains one equation to solve, which differs by a factor of $2$ to the equation \eqref{final-reality-condition-spin3} in the $\BB=3$ case, leading to the subtly different solution
\begin{align}
    \kappa^2=\dfrac{\sfrac{\Delta}{2}(\chi(\eta_1+\eta_2)+\lambda_1\lambda_2\sin(\psi-\phi))}{\lambda_1\lambda_2(\cos(\psi - \phi)|\vec{r}_1\times\vec{r}_2|+ \sin(\psi - \phi)(\chi^2 + \vec{r}_1\cdot\vec{r}_2))-\chi(\eta_1\lambda_2^2 + \eta_2\lambda_1^2)}.\label{kappa-fix-B-general}
\end{align}
Again, there are cases where the denominator is zero, some of which we discuss further below, but in general the ADHM data as in \eqref{general-top} yields again a nine-parameter family, with the same interpretation as in the $\BB=3$ case. Note that we can easily extend these to $12$-parameter families by including translations and rotations in the $(x_3,x_4)$-plane, and this describes the fixed-point-set under the invariance condition \eqref{C_B-1}, in the representation dictated by the given c.g.ts, away from the solutions where $\Delta$ and the denominator in \eqref{kappa-fix-B-general} are zero.

\begin{figure}[htbp]
	\centering
	\includegraphics[scale=0.6]{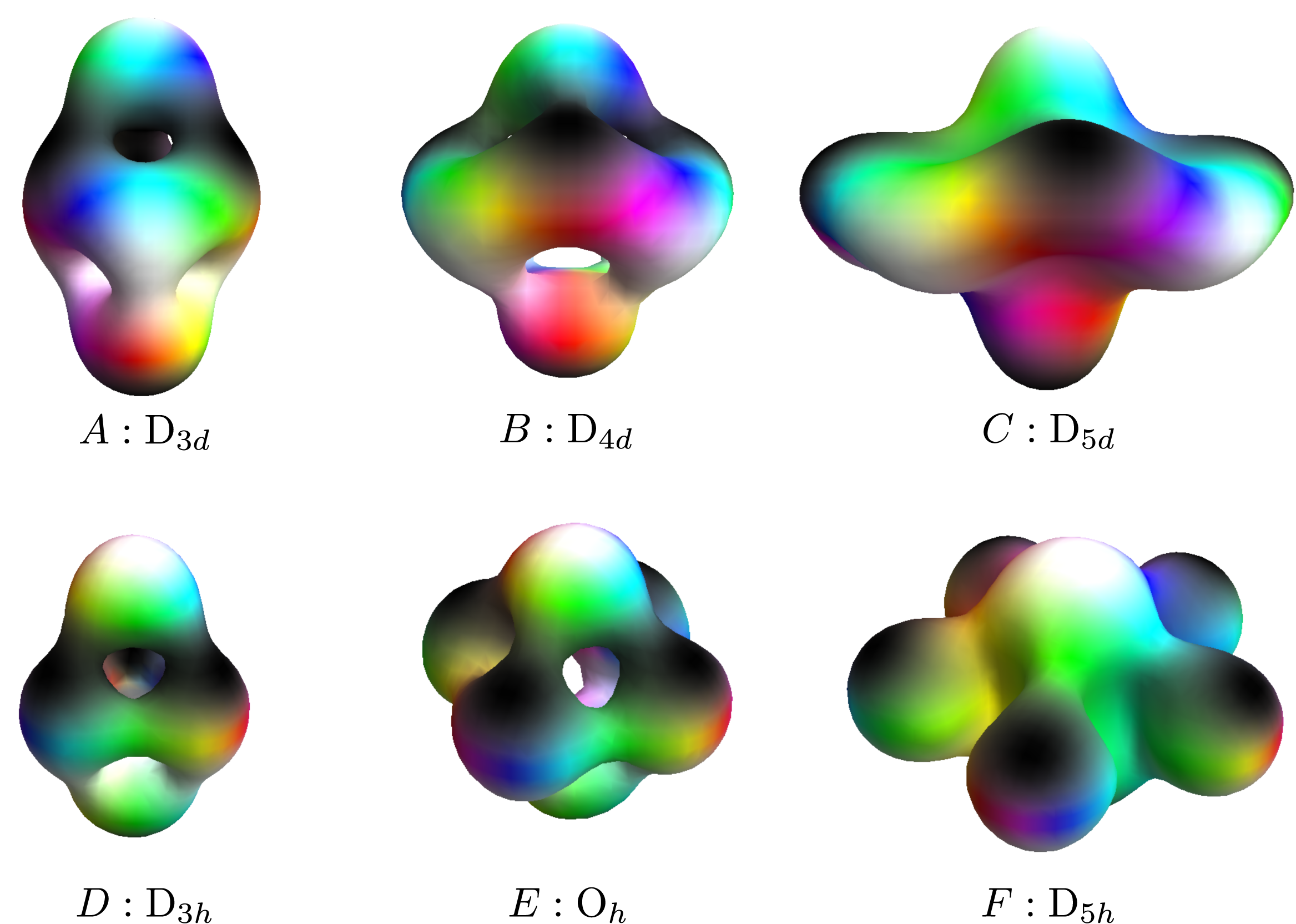}
	\caption{Examples of spinning top i-skyrmions. In the top row we view the skyrmions from the side. In the bottom row we view the i-skyrmions from a steeper angle.}
	\label{fig:spinningtopcs}
\end{figure}

We plot some examples of the spinning top configurations in Figure \ref{fig:spinningtopcs} for $\BB=4,5,$ and $6$. In each case we restrict to dihedral solutions, for which we set $\lambda_1=\lambda_2=\lambda$, $R_1=R_2=R$, and $\eta_1=\eta_2=\eta$. The top row have D$_{(\BB-1)d}$-symmetry, which arises by setting $\psi=\phi-\sfrac{\pi}{2}=0$. In terms of the remaining parameters $(\lambda,R,\eta,\chi)$, the configurations in Figure \ref{fig:spinningtopcs} $A$, $B$, and $C$ correspond to
\begin{align}
	A: \left(1,2,-\sfrac{1}{2},-\sfrac{1}{2}\right),\quad
	B:(1,2,0,0), \quad C:\left(2,\sfrac{5}{2},0,\sfrac{\sqrt{7}}{2}\right).
\end{align}
The bottom row have D$_{(\BB-1)h}$-symmetry, for which we set $\psi=\phi=\sfrac{\pi}{2}$. We also set $\eta=0$, and so these all correspond to the degenerate case where the numerator and denominator in \eqref{kappa-fix-B-general} are zero. So here there are three free parameters $(\lambda,R,\kappa)$, as in this case $\chi$ may be gauged to $0$. In terms of these, the configurations plotted in Figure \ref{fig:spinningtopcs} $D$, $E$, and $F$ correspond to
\begin{align}
	D: (1,2,1),\quad
	E: \left(1,2,\sqrt{2}\right), \quad F: \left(2,\sfrac{5}{2},2\right) .\label{D_d-tops}
\end{align}
We note that configuration $E$ has octahedral symmetry, which will be discussed in detail later. Each energy-minimising i-skyrmion in these families is a candidate for an approximate solution to the static field equations of \eqref{Skyrme-energy}, most likely a saddle point.

\subsubsection{Extracting central configurations}\label{sec:Central-data-in-tops}
Due to how the symmetries act, there will be points in the moduli spaces of $\mathrm{C}_{\BB-1}$-symmetric spinning tops which correspond to highly-symmetric ``central" configurations; we have already seen examples with dihedral symmetry, and in this section we extract some other important examples. In general, doing this is equivalent to to determining points $\bm{a}$ in the moduli space such that
\begin{align}
    (L_c,M_c)=(\omega L|_{\bm{a}}\rho^{-1}\Omega^{-1},\Omega\rho M|_{\bm{a}}\rho^{-1}\Omega^{-1)}),\label{central-config-gt}
\end{align}
where $(L_c,M_c)$ denotes the highly-symmetric configuration, $\Omega\in\O(\BB)$, $\omega\in\SU(2)$, and $\rho\in\SU(2)$ is chosen so as to reorient the corresponding axis of rotation, that is
\begin{align}
    \rho(\vec{e}_3\cdot\vec{\bm{e}})\rho^{-1}=\vec{n}_q
\end{align}
with $\vec{n}_q$ the axis of the $\mathrm{C}_{\BB-1}$ rotation for the configuration $(L_c,M_c)$. In most cases, explicit expressions for $(\Omega,\omega,\rho)$ are cumbersome, and we therefore omit them, however they may be derived using the following straightforward strategy. Let $O_{\BB-1}$ and $\vec{n}_p$ denote the c.g.t and axis of isorotation respectively for the $\mathrm{C}_{\BB-1}$-symmetry of the $(L_c,M_c)$. Then it is sufficient (and by Lemma \ref{lem:free-action} necessary) that $(\Omega,\omega)$ satisfy
\begin{align}
    \Omega O_{1,\BB-2,1}\Omega^{-1}=\pm O_{\BB-1},\quad \omega(\vec{e}_3\cdot\vec{\bm{e}})\omega^{-1}=\pm\vec{n}_p.\label{constraints-on-gts-and-rots}
\end{align}
\paragraph{$\BB=3$.}  It is easy to show that the data $(L,M)$ given by \eqref{B=3-spinning-top} with $\kappa>0$ and the parameters
\begin{align}
\begin{aligned}
    (R_i,\eta_i,\lambda_i,\chi,\psi,\phi)&=(\lambda\cos2\tau,0,\lambda,(-1)^m\lambda\sin2\tau,\tau-\sfrac{\pi}{4},\tau+\sfrac{\pi}{4}+m\pi),\label{tetra-fix-spin}
\end{aligned}
\end{align}
for any $\tau\in\R$, $m\in\Z$, are all gauge equivalent to a tetrahedron. Explicitly, letting $(L_{{\rm T}_d},M_{{\rm T}_d})$ denote the data \eqref{tetrahedron}, at these points we have
\begin{align}
    L_{{\rm T}_d}=\j L\Omega^{-1},\quad M_{{\rm T}_d}=\Omega M\Omega^{-1},\label{gt-spin-tetra}
\end{align}
where
\begin{align}
    \Omega=\begin{pmatrix}
    (-1)^{m}\cos(\tau-\sfrac{\pi}{4})&0&\sin(\tau-\sfrac{\pi}{4})\\
    (-1)^{m+1}\sin(\tau-\sfrac{\pi}{4})&0&\cos(\tau-\sfrac{\pi}{4})\\
    0&-1&0
    \end{pmatrix}.
\end{align}
It is straightforward to check that these are the only cases which give a tetrahedron, except for replacing $R_1,R_2,\chi$ by their negatives in \eqref{tetra-fix-spin}, from which we obtain the dual tetrahedron.

It is well-known that both the tetrahedron \eqref{tetrahedron} and its dual appear in the centre of the so-called `twisted-line scattering' of three skyrmions. This was first discussed for monopoles by Houghton--Sutcliffe in \cite{HoughtonSutcliffe1996monopole}, then soon after applied to skyrmions by Walet \cite{walet1996quantising}. It is also mentioned in the context of ADHM data by Houghton \cite{houghton1999-3Skyrme}. However all of these cases only considered the ${\rm D}_{2d}$-symmetric version where the two outer skyrmions have the same size and separation, and have fixed orientation. Our solution generalises this, and allows for all three constituents to vary independently in size, axial orientation, and separation.
\paragraph{$\BB=4$.}
The $\mathrm{C}_3$-symmetry for the $\BB=4$ case of \eqref{general-top} acts in a similar way to the $\mathrm{C}_3$-symmetry in the cube \eqref{cube}, except there is a discrepancy in the axis of symmetry. Choosing a gauge where $\lambda_i,\kappa,\mu>0$, we find that the data
\begin{align}
    (R_i,\eta_i,\lambda_i,\chi,\psi,\phi)=(\sqrt{2}\lambda\cos2\tau,0,\lambda,(-1)^{n+1}\lambda\sin2\tau,\sfrac{\pi}{2}+n\pi+\tau,\tau)
\end{align}
for $\tau\in\R$ and $n\in\Z$ are all gauge-equivalent to a rotated version of the cube \eqref{cube}. This may be shown explicitly by constructing $(\Omega,\omega,\rho)$ satisfying \eqref{central-config-gt}-\eqref{constraints-on-gts-and-rots}.
\paragraph{$\BB=5$.} By comparing the symmetry equations \eqref{B=5-D2d-symm} and \eqref{C_B-1}, and using Lemma \ref{lem:free-action}, it is straightforward to see that the ${\rm D}_{2d}$-symmetric $\BB=5$ data \eqref{D2d-5-gen-orientation} will not appear within the family \eqref{general-top}. However, there is an interesting central configuration which does appear, namely an octahedron, which is known to exist as a saddle-point solution in the Skyrme model \cite{mantonPiette2001SkyRat}. The corresponding i-skyrmion is already plotted in Figure \ref{fig:spinningtopcs} E, with parameters found in \eqref{D_d-tops}.

The full octahedral symmetry is manifested via
\begin{align}
    \begin{aligned}
    L&=p(\vec{r},\sfrac{2\pi}{3})Lq(\vec{r},\sfrac{2\pi}{3})^{-1}O_3^{-1},&M&=O_3q(\vec{r},\sfrac{2\pi}{3})Mq(\vec{r},\sfrac{2\pi}{3})^{-1}O_3^{-1},\\
    L&=p(\vec{e}_3,\sfrac{\pi}{2})Lq(\vec{e}_3,\sfrac{\pi}{2})^{-1}O_4^{-1},&M&=O_4q(\vec{e}_3,\sfrac{\pi}{2})Mq(\vec{e}_3,\sfrac{\pi}{2})^{-1}O_4^{-1},\\
    L&=-LO_-^{-1},&M&=-O_-MO_-^{-1},
    \end{aligned}\label{octahedral-symmetry}
\end{align}
where $O_4=O_{1,3,1}$, and
\begin{align}
\begin{aligned}
    O_3&=\dfrac{1}{4}\begin{pmatrix}
-1& 0& -2\sqrt{2}&-\sqrt{6}& 1\\
2\sqrt{2}& 0&0&  0& 2\sqrt{2}\\
0& 4&   0&     0&  0\\
\sqrt{6}& 0&  0&-2& -\sqrt{6}\\
1& 0&-2\sqrt{2}&\sqrt{6}&-1
    \end{pmatrix},&O_-=\begin{pmatrix}
    0&0&0&0&-1\\
    0&-1&0&0&0\\
    0&0&-1&0&0\\
    0&0&0&1&0\\
    -1&0&0&0&0
    \end{pmatrix}.
\end{aligned}
\end{align}
We remark that this data is gauge-equivalent to data which appears in the infinite-period limit of a recently studied octahedrally-symmetric caloron \cite{kato2021symmetric}. Similar limits also hold \cite{Ward2004} for the torus \eqref{torus}, tetrahedron \eqref{tetrahedron}, cube \eqref{cube}, and dodecahedron \eqref{dodecahedron}.

Note that the octahedral symmetry here is different to that of the cubic $4$-instanton \eqref{cubic-symmetry} due to the $\mathrm{C}_4$ rotation symmetry being coupled to different isorotations. There is also a ${\rm D}_{2d}$-symmetry here, found by the embedding of ${\rm D}_{2d}\subset \O_h$, however it is a different realisation of the D$_{2d}$-symmetry \eqref{B=5-D2d-symm} of the minimal data from the previous section, as the compensating gauge transformations \eqref{D2d-gts} belong to inequivalent representations of ${\rm Dic}_4$. Imposing \eqref{B=5-D2d-symm} with this representation leads to data which is very similar to \eqref{D2d-5-gen-orientation}, but the asymptotic form will be different as the constituent at the origin is forced to have orientation $\k$. On the other hand, it is possible to show that there are no octahedrally-symmetric versions of the data \eqref{D2d-5-gen-orientation} as there are no compensating gauge transformations which allow for ADHM data satisfying \eqref{octahedral-symmetry} which are compatible with \eqref{D2d-gts}.

Finally, to reinforce the idea that the ${\rm D}_{2d}$-symmetric data gives rise to something which best approximates the minimal energy solution, we have calculated the energy-minimising i-skyrmion for this octahedron. There is freedom to rescale the solution, and within the normalisation \eqref{instanton-scale}, the optimal scale is given when $\lambda=0.904$, with energy $E=5.80$. This is $2.35\%$ more than the ${\rm D}_{2d}$ i-skyrmion, and is thus is not a global minimum, as expected.
\paragraph{Higher charge central configurations and other spinning tops.}
By direct comparison of the symmetry equations, it is clear that none of the other highly-symmetric configurations for $\BB>4$ described in Section \ref{sec:database} will appear as points in the moduli space of spinning tops \eqref{general-top}. However, the spinning tops we have just described are only some of the simplest in a more general family. In general, it follows from \eqref{torus-symmetry} (and \eqref{hedgehog-symmetry} when $\BB_i=1$) that we could construct rank $\BB=2\BB_1+\BB_2$ data with the $\mathrm{C}_{\BB_1+\BB_2}$-symmetry
\begin{align}
    \begin{aligned}
        L&=p(\vec{e}_3,\sfrac{2\BB_1\pi}{\BB_1+\BB_2})Lq(\vec{e}_3,\sfrac{2\pi}{\BB_1+\BB_2})^{-1}O_{\BB_1,\BB_2,\BB_1}^{-1},\\
    M&=O_{\BB_1,\BB_2,\BB_1}q(\vec{e}_3,\sfrac{2\pi}{\BB_1+\BB_2})Mq(\vec{e}_3,\sfrac{2\pi}{\BB_1+\BB_2})^{-1}O_{\BB_1,\BB_2,\BB_1}^{-1},
    \end{aligned}\label{B_1-B_2-B_1-top}
\end{align}
with $O_{\BB_1,\BB_2,\BB_1}=\diag\{O_{\BB_1},-O_{\BB_2},O_{\BB_1}\}$ the alternating direct sum of c.g.ts for the $\BB_i$-tori, evaluated at $\theta=\sfrac{2\pi}{\BB_1+\BB_2}$. Such configurations may be called \textit{spinning tops of type} $(\BB_1,\BB_2,\BB_1)$, and we have just discussed the $(1,\BB-2,1)$ case in detail.\footnote{There is also the $(0,\BB,0)$ case which arises by restricting \eqref{torus-symmetry} to the $\mathrm{C}_\BB$-subgroup, and this will describe $\BB$ constituents on a regular $\BB$-gon in an attractive channel.} It should be clear from the construction of the $\BB=6$ and $\BB=8$ solutions in Section \ref{sec:database} that these represent special dihedrally-symmetric solutions inside the sequence of spinning tops of type $(2,\BB-4,2)$. By comparing the $\mathrm{C}_5$-symmetry for the dodecahedron in \eqref{dodecahedron-symmetry} with \eqref{B_1-B_2-B_1-top}, and the corresponding c.g.t \eqref{dodecahdedron-gts}, the same is also true for the $\BB=7$ dodecahedron. All of these examples, including those plotted in Figure \ref{fig:spinningtopcs}, lead us to conjecture that there should always be ${\rm D}_{(\BB_1+\BB_2)d}$ and ${\rm D}_{(\BB_1+\BB_2)h}$-symmetric subfamilies inside the spinning tops of type $(\BB_1,\BB_2,\BB_1)$.
\subsection{Further decompositions of the tetrahedron and cube}
In Section \ref{sec:Central-data-in-tops} we showed that the $\BB=3$ tetrahedron \eqref{tetrahedron} and $\BB=4$ cube \eqref{cube} appear as central points in the larger moduli space of spinning tops. The spinning tops are not the only way to decompose these solutions, and in this section we discuss some further examples at great depth.
\subsubsection{\texorpdfstring{$\BB=3$}{B=3} as \texorpdfstring{$1+2$}{1+2}, and \texorpdfstring{$\BB=4$}{B=4} as \texorpdfstring{$2+2$}{2+2} and \texorpdfstring{$3+1$}{3+1}}
First we shall consider the $\BB=3$ solution describing a $\BB=2$ torus and $\BB=1$ hedgehog aligned along an axis of symmetry within the attractive channel. As with previous examples and the general discussion in Section \ref{sec:sym-decom}, inverting the orientation of one constituent breaks the symmetry to $\mathrm{C}_3$. By fixing the axis of symmetry as $\vec{e}_3$, and by comparing \eqref{hedgehog-symmetry} and \eqref{torus-symmetry-B=2}, we should consider data invariant under
\begin{align}
\begin{aligned}
    L&=p(\vec{e}_3,\sfrac{2\pi}{3})Lq(\vec{e}_3,\sfrac{2\pi}{3})^{-1}O_{1,2}^{-1},&M&=O_{1,2}q(\vec{e}_3,\sfrac{2\pi}{3})Mq(\vec{e}_3,\sfrac{2\pi}{3})^{-1}O_{1,2}^{-1},
\end{aligned}\label{2+1-sym}
\end{align}
where the compensating gauge transformation takes the block-diagonal form
\begin{align}
    O_{1,2}=\diag\{1,-Q_1(\sfrac{2\pi}{3})\}=\begin{pmatrix}
    1&0&0\\
    0&-\sfrac{1}{2}&\sfrac{\sqrt{3}}{2}\\
    0&-\sfrac{\sqrt{3}}{2}&-\sfrac{1}{2}
    \end{pmatrix}.
\end{align}
Up to gauge-equivalence, orientation, and translations, we obtain the ansatz
\begin{align}
    \begin{aligned}
    L&=\begin{pmatrix}
    \lambda_1\omega(\phi)&\lambda_2\i&-\lambda_2\j
    \end{pmatrix},\\
    M&=\begin{pmatrix}
    R\k+\eta\1&c_1\i+c_2\j&c_2\i-c_1\j\\
    c_1\i+c_2\j&\mu\i-R\k-\eta\1&\mu\j\\
    c_2\i-c_1\j&\mu\j&-\mu\i-R\k-\eta\1
    \end{pmatrix},
    \end{aligned}\label{B=3_2+1}
\end{align}
where $\omega(\phi)=\1\cos\phi+\k\sin\phi$. The reality condition \eqref{reality-cond} reduces to the equations
\begin{align}
    \begin{aligned}
    2\eta c_1+2Rc_2+\lambda_1\lambda_2\cos\phi&=0,\\
    2Rc_1-2\eta c_2+\lambda_1\lambda_2\sin\phi&=0,\\
    c_1^2+c_2^2+\lambda_2^2-2\mu^2&=0.
    \end{aligned}\label{B=3-reality-1,1,1}
\end{align}
Assuming non-zero separation $r=R\k+\eta\1$, these are resolved by setting
\begin{align}
\begin{aligned}
    \begin{pmatrix}
    c_1\\c_2
    \end{pmatrix}&=-\dfrac{\lambda_1\lambda_2}{2|r|^2}\begin{pmatrix}
    \cos\phi&\sin\phi\\-\sin\phi&\cos\phi
    \end{pmatrix}\begin{pmatrix}
    \eta\\R
    \end{pmatrix},&\mu^2&=\dfrac{\lambda_2^2}{2}\left(1+\dfrac{\lambda_1^2}{4|r|^2}\right).
\end{aligned}\label{B=3_2+1-soln}
\end{align}
We thus have a five-parameter family: $\lambda_1,\lambda_2$ control the sizes of the clusters, the parameters $R,\eta$ form the `separation vector' $r=R\k+\eta\1\in\H$, and the angle $\phi$ represents a relative orientation around the axis of symmetry.

An analogous solution exists in the case $\BB=4$, with the $\BB=1$ constituent above replaced by another $\BB=2$ torus. The unbroken symmetry is $\mathrm{C}_4$-symmetry, namely
\begin{align}
    L=p(\vec{e}_3,\pi)Lq(\vec{e}_3,\sfrac{\pi}{2})O_{2,2}^{-1},\quad M=O_{2,2}q(\vec{e}_3,\sfrac{\pi}{2})Mq(\vec{e}_3,\sfrac{\pi}{2})O_{2,2}^{-1},\label{two-tori-symmetry}
\end{align}
with compensating transformation
\begin{align}
    O_{2,2}=\diag\{-Q_1(\sfrac{\pi}{2}),Q_1(\sfrac{\pi}{2})\}=\frac{\sqrt{2}}{2}\begin{pmatrix}
    -1&1&0&0\\
    -1&-1&0&0\\
    0&0&1&-1\\
    0&0&1&1
    \end{pmatrix}.\label{C4-compensating}
\end{align}
Assuming non-zero separation, similarly to the $\BB=3$ case, up to gauge, orientation, and position, we obtain a five-parameter family of $\mathrm{C}_4$-symmetric ADHM data
\begin{align}
    \begin{aligned}
    L&=\begin{pmatrix}
    \lambda_1\i&-\lambda_1\j&\lambda_2\1&\lambda_2\k
    \end{pmatrix},\\
    M&=\begin{pmatrix}
    \mu_1\i+r&\mu_1\j&\sfrac{\lambda_1\lambda_2}{2|r|^2}r\i&-\sfrac{\lambda_1\lambda_2}{2|r|^2}r\j\\
    \mu_1\j&-\mu_1\i+r&-\sfrac{\lambda_1\lambda_2}{2|r|^2}r\j&-\sfrac{\lambda_1\lambda_2}{2|r|^2}r\i\\
    \sfrac{\lambda_1\lambda_2}{2|r|^2}r\i&-\sfrac{\lambda_1\lambda_2}{2|r|^2}r\j&\mu_2\omega(\phi)\i-r&\mu_2\omega(\phi)\j\\
    -\sfrac{\lambda_1\lambda_2}{2|r|^2}r\j&-\sfrac{\lambda_1\lambda_2}{2|r|^2}r\i&\mu_2\omega(\phi)\j&-\mu_2\omega(\phi)\i-r
    \end{pmatrix},
    \end{aligned}\label{two-tori}
\end{align}
with $r=R\k+\eta\1$, $\omega(\phi)=\1\cos\phi+\k\sin\phi$, and $\mu_1$ and $\mu_2$ determined via
\begin{align}
    \mu_1^2=\dfrac{\lambda_1^2}{2}\left(1+\frac{\lambda_2^2}{2|r|^2}\right),\quad
    \mu_2^2=\dfrac{\lambda_2^2}{2}\left(1+\frac{\lambda_1^2}{2|r|^2}\right),\label{mu-conditions}
\end{align}
with all different roots gauge-equivalent.

As with the spinning tops, in both of the above families there will be points which are gauge-equivalent to the highly-symmetric configurations: the tetrahedron and the cube. Extracting the tetrahedron from \eqref{B=3_2+1} is more subtle than previous examples. Not only do the axes of rotation and isorotation need to be changed, but we also must translate the solution; for the data \eqref{B=3_2+1} we have $\tr(M)=-r$ whereas $\tr(M_{{\rm T}_d})=0$. To this end, we require an isorotation $\omega$, rotation $\rho$, gauge transformation $\Omega$, and translation $a$, such that
\begin{align}
    L_{{\rm T}_d}=\omega L\rho^{-1}\Omega^{-1},\quad M_{{\rm T}_d}=\Omega\rho (M+a\ind_3)\rho^{-1}\Omega^{-1}.\label{gt-C3-tetra}
\end{align}
We find that this occurs for all points of the form
\begin{align}
    (\lambda_1,\lambda_2,R,\eta,\phi)=(\lambda,\lambda,\sfrac{\sqrt{3}}{2}\lambda,0,\phi)
\end{align}
with translation $a=\sfrac{\sqrt{3}}{6}\lambda\k$, and $(\Omega,\omega,\rho)$ dependent on $\phi$, which may be determined straightforwardly.
Similarly we may determine all points in the $\BB=4$ solution \eqref{two-tori} where (after rotating and isorotating) the cube \eqref{cube} appears up to gauge. These are
\begin{align}
    (\lambda_1,\lambda_2,R,\eta,\phi)=(\lambda,\lambda,\lambda,0,\phi),
\end{align}
and again the corresponding transformations required for \eqref{central-config-gt} depend on the relative orientation $\phi$.

The above analysis for the $\BB=3$ data \eqref{B=3_2+1} immediately allows us to extract a $3+1$ decomposition of the $\BB=4$ spinning top \eqref{general-top}. It is straightforward to see that when $\lambda_1=\kappa=\lambda$,\footnote{Note that by fixing $\kappa$, this means $\chi$ is determined via \eqref{kappa-fix-B-general}.} and $R_1=\pm\sqrt{3}\lambda$, for sufficiently large $|r_2|$ compared to $\lambda$ and $\lambda_2$ (and suitably chosen $\eta_1$) the solution will look like a cluster decomposition consisting of a tetrahedron and a hedgehog, aligned on the $\mathrm{C}_3$-symmetry axis of the tetrahedron, with $\pm$ giving the two dual tetrahedra. Different choices of $\phi$ and $\psi$ will alter the relative orientations.
\subsubsection{Large and small \texorpdfstring{$|r|$}{r} limits}
It is easy to see from the formulae \eqref{B=3_2+1}, \eqref{B=3_2+1-soln}, \eqref{two-tori}, and \eqref{mu-conditions} that when the `separation parameter' $|r|$ is large compared to $\lambda_1$ and $\lambda_2$, the above solutions are approximately in the block-diagonal form \eqref{diagonal-ADHM-data}, with the leading order terms corresponding precisely to the expected clusters, in support of the framework outlined in Section \ref{sec:CSW}. We have also seen above that when $|r|$ is comparable to $\lambda_1$ and $\lambda_2$ (at least in the case $\lambda_1=\lambda_2$), this interpretation in terms of a separation parameter breaks down, and the identities of the constituents are lost within a more highly-symmetric configuration.

It is possible to move beyond the highly-symmetric configurations, and consider a limit where $|r|$ is small compared to the scales $\lambda_1,\lambda_2$. For the sake of this illustration, we shall look at the $\BB=4$ data \eqref{two-tori}, and for simplicity set $\eta=0$ since this only affects the holographic direction. Assuming a gauge and orientation where $\mu_i,\lambda_i>0$, using \eqref{mu-conditions} we note that for $R\sim 0$,
\begin{align}
    \mu_i&=\frac{\lambda_i\lambda_j}{2R}+\frac{\lambda_i}{\lambda_j}\frac{R}{2}+O(R^3),
\end{align}
so that we have
\begin{align}
    M=\frac{\lambda_1\lambda_2}{2R}\begin{pmatrix}
    \i&\j&\j&\i\\
    \j&-\i&\i&-\j\\
    \j&\i&\i\cos\phi+\j\sin\phi&\j\cos\phi-\i\sin\phi\\
    \i&-\j&\j\cos\phi-\i\sin\phi&-(\i\cos\phi+\j\sin\phi)
    \end{pmatrix}+O(R).
\end{align}
The matrices corresponding to the components of $\i$ and $\j$ above both have the eigenvalues $\pm\sqrt{2}\sqrt{1+\sin\sfrac{\phi}{2}}$ and $\pm\sqrt{2}\sqrt{1-\sin\sfrac{\phi}{2}}$, and so may be simultaneously diagonalised with an orthogonal gauge transformation. This is an unwieldy expression in general, but in the case $\phi=0$ it is easy to see that the gauge transformation
\begin{align}
    \OO=\dfrac{\sqrt{2+\sqrt{2}}}{4}\begin{pmatrix}
        -\sqrt{2}& 2 - \sqrt{2}&     \sqrt{2}& \sqrt{2} - 2\\
 \sqrt{2} - 2&    -\sqrt{2}& 2 - \sqrt{2}&     \sqrt{2}\\
    -\sqrt{2}& \sqrt{2} - 2&    -\sqrt{2}& \sqrt{2} - 2\\
 2 - \sqrt{2}&    -\sqrt{2}& 2 - \sqrt{2}&    -\sqrt{2}
    \end{pmatrix},
\end{align}
transforms the data to the form
\begin{align}
\begin{aligned}
    L&=\sqrt{\frac{\lambda_1^2+\lambda_2^2}{2}}\begin{pmatrix}
    \omega_1&\omega_2&\omega_3&\omega_4
    \end{pmatrix},\\
    M&=\frac{\lambda_1\lambda_2}{2R}\diag\{\i-\j,\i+\j,-\i-\j,-\i+\j\}+O(R),
\end{aligned}\label{small-R-parameters}
\end{align}
where
\begin{align}
    \begin{pmatrix}
    \omega_1\\\omega_2\\\omega_3\\\omega_4
    \end{pmatrix}=\frac{1}{2}\sqrt{\frac{2+\sqrt{2}}{\lambda_1^2+\lambda_2^2}}\begin{pmatrix}
    \lambda_2\1-\lambda_1\i+(2-\sqrt{2})\lambda_1\j+(2-\sqrt{2})\lambda_2\k\\
    \lambda_2\1+\lambda_1\i+(2-\sqrt{2})\lambda_1\j-(2-\sqrt{2})\lambda_2\k\\
    -(2-\sqrt{2})\lambda_2\1-(2-\sqrt{2})\lambda_1\i+\lambda_1\j-\lambda_2\k\\
    -(2-\sqrt{2})\lambda_2\1+(2-\sqrt{2})\lambda_1\i+\lambda_1\j+\lambda_2\k
    \end{pmatrix}.
\end{align}
This represents four $\BB=1$ constituents, each with scale $\lambda=\sqrt{\sfrac{\lambda_1^2+\lambda_2^2}{2}}$, positioned on the vertices of a square, with orientations $\omega_i$. We therefore see that the data \eqref{two-tori} interpolates between a $2+2$ and $1+1+1+1$ decomposition for large and small `separation' $|r|$ respectively. Importantly, note that for the case considered, a relative size difference between the two tori at $R\gg\lambda_i$ is transferred into a relative orientation for the four hedgehogs at $R\ll\lambda_i$. It is only when $\lambda_1=\lambda_2=\lambda$ that the orientations are in the attractive channel, with the orientations given by $\omega(\1,\i,\j,\k)$, where $\omega=\sfrac{1}{2}\sqrt{\sfrac{2+\sqrt{2}}{2}}(\1-\i+(2-\sqrt{2})(\j+\k))$. This phenomenon is known to exist in the $1+1$ sector \cite{HalcrowWinyard2021consistent}, but this is the first time it has been observed for higher charges.

We remark that a similar limit may be considered for the $\BB=3$ data \eqref{B=3_2+1}-\eqref{B=3_2+1-soln}, with the outgoing constituents lying on the vertices of an equilateral triangle.
\subsubsection{Summary of decompositions of the cube}
\begin{figure}[htbp]
	\centering
	\includegraphics[scale=0.6]{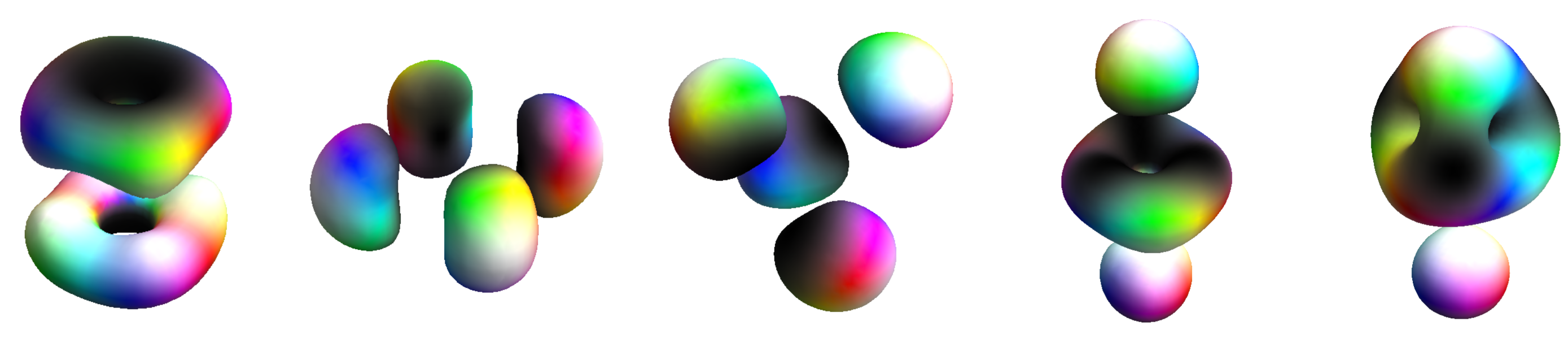}
	\caption{Energy density plots of the considered decompositions of the cube \eqref{cube}. From left to right: $2+2$, $1+1+1+1$ ($\mathrm{C}_4$-symmetric), $1+1+1+1$ (${\rm T}_d$-symmetric), $1+2+1$ and $3+1$ ($\mathrm{C}_3$-symmetric).}
	\label{fig:the4s}
\end{figure}

By using ADHM data, we have seen how to split the cubic 4-skyrmion into different decompositions of smaller skyrmions. We plot energy densities of these decompositions in Figure \ref{fig:the4s}. The data used for each configuration is given by\footnote{For the $1+2+1$ and $3+1$ case, since we have fixed $\kappa=1$, we have used \eqref{kappa-fix-B-general} to find the missing parameter by solving a polynomial equation; specifically for $\eta=\eta_1=\eta_2$ in the $1+2+1$, and for $\chi$ in the $3+1$. In each case there is a unique real solution.}
\begin{align*}
2+2: \quad &\text{\eqref{two-tori} with }(\lambda_i,R,\eta,\phi) = \left(1,\sfrac{5}{2},0,0\right); \\
1+1+1+1:\quad&\text{\eqref{two-tori} with }(\lambda_i,R,\eta,\phi)=\left(1,\sfrac{7}{10},0,0\right);\\
1+1+1+1: \quad & \text{Tetrahedral data in \cite{LeeseManton1994stable} with } (a,b) = \left(\sfrac{1}{2},\sqrt{2}\right).\\
1+2+1: \quad &\text{\eqref{general-top} with }\BB=4\text{ and }
(\lambda_i,R_i,\eta_i,\phi,\psi,\chi,\kappa) = \left(1,\sfrac{5}{2},\eta,\sfrac{\pi}{2},0,\sfrac{\sqrt{2}}{2},1\right).\\
3+1:\quad &\text{\eqref{general-top} with }\BB=4\text{ and }
(\lambda_i,R_1,R_2,\eta_1,\eta_2,\phi,\psi,\kappa) = \left(1,\sqrt{3},3,\sfrac{1}{2},1,\sfrac{\pi}{2},0,1\right).
\end{align*}
Figure \ref{fig:the4s} demonstrates the flexibility and power of the instanton approximation. In the same simple framework, we can describe how the 4-skyrmion can fission in many different ways. We note that it is common for authors to informally describe the cube as being made from two tori, or four skyrmions on a tetrahedron. The framework developed here takes this informal notion and makes it firm, and helps us to see other equally valid decompositions. Each of these descriptions is correct, and as we have shown, they all share a point in common, namely the cube \eqref{cube}, related to each other by gauge transformations and isometries of instantons.
\subsubsection{A potential on the \texorpdfstring{$\BB=4$}{B=4} i-skyrmion moduli space} \label{sec:energypot}

In the previous sections we have found several families of i-skyrmions. We now round off our discussions with a view of how to apply our work to nuclear physics in the Skyrme model. To make contact with nuclear data the first task is to understand how the energy varies across these families. This has not been done before, likely because it is numerically challenging. With our new numerical technique, we can tackle this problem. 

Consider the previously discussed family \eqref{two-tori}. A restricted version of the family, when the tori have equal sizes, and a specific relative position and orientation, is given by
\begin{align} \label{eq:4as22and1111}
\begin{aligned}
	L &= \begin{pmatrix}  \lambda \i & -\lambda \j & \lambda\1 & \lambda \k \end{pmatrix} \\
	M &= \begin{pmatrix} 
		\mu \i +R\k & \mu \j & \frac{\lambda^2}{2R}\j & \frac{\lambda^2}{2R}\i \\
		\mu \j & -\mu\i + R \k & \frac{\lambda^2}{2R}\i &-\frac{\lambda^2}{2R} \j \\
		\frac{\lambda^2}{2R}\j & \frac{\lambda^2}{2R}\i & \mu \i -R\k & \mu \j \\
		\frac{\lambda^2}{2R}\i & -\frac{\lambda^2}{2R}\j & \mu\j & -\mu\i -R\k \end{pmatrix} ,
\end{aligned}
\end{align}
with the reality condition satisfied when 
\begin{equation}
	\mu = \frac{\lambda}{\sqrt{2}}\sqrt{1 + \frac{\lambda^2}{2R^2}} .
\end{equation}
As was previously discussed, the cube forms at $R=\lambda$. When $R>\lambda$ we can interpret the configuration as two stacked tori, and when $R<\lambda$ as four 1-skyrmions on the vertices of a square. As such, we expect that the energy should asymptote to the energy of these configurations when $R$ is large and small. 

\begin{figure}[htbp]
	\begin{center}
		\includegraphics[scale=0.7]{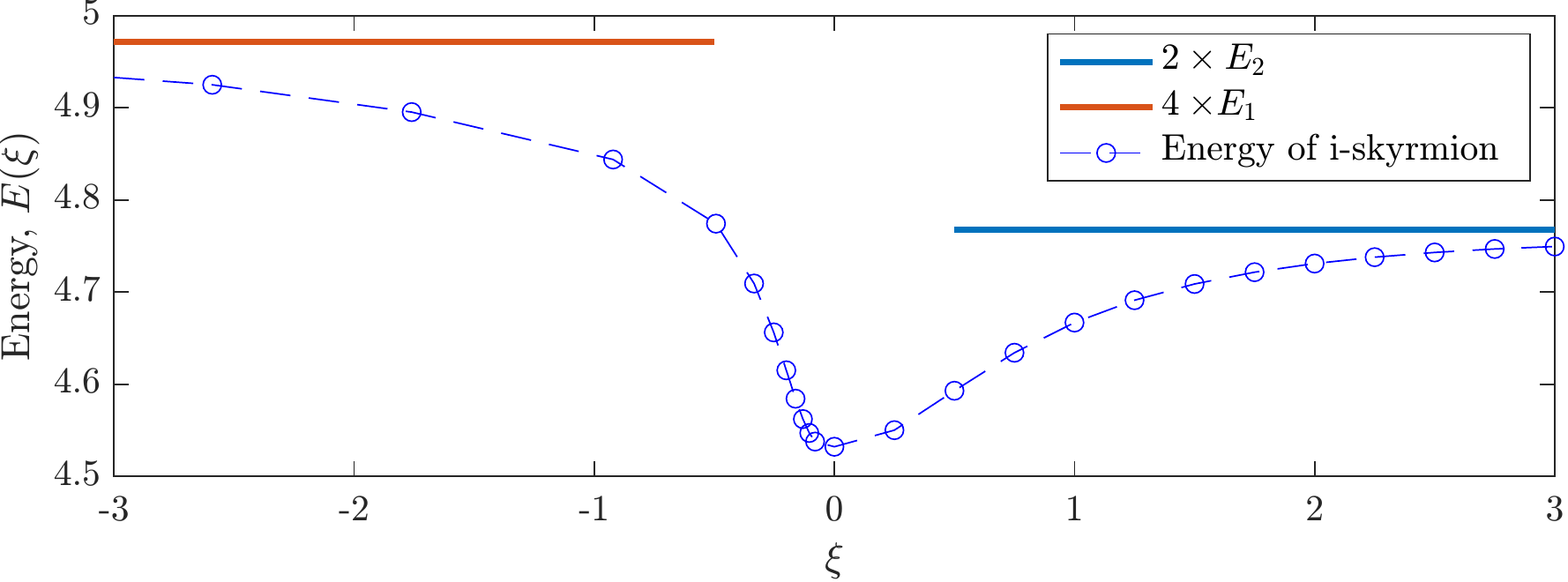}
		\caption{The energy of the optimal i-skyrmions for the family \eqref{eq:4as22and1111} in terms of the coordinate \eqref{xi-coordinate}. We also plot the energies of four separated $1$-i-skyrmions ($4E_1$) and two separated $2$-i-skyrmions ($2E_2$),
		as may be found in Table \ref{tab:energy-comparison}. }
		\label{fig:VRr}
	\end{center}
\end{figure}

To see if the energy acts as expected, we find the energy-minimising i-skyrmion for each fixed $R$. This requires varying $\lambda$ at each value of $R$ to minimise energy, which we do using a Newton--Raphson method. We use a $50\times 50\times 200$ box for when $R>\lambda$ and a $200\times200\times 50$ box when $R<\lambda$. We know from \eqref{small-R-parameters} that when $R\ll\lambda$ the shortest distance between any two skyrmions is of order $\sfrac{\lambda^2}{R}$. To reflect this behaviour we introduce the coordinate
\begin{equation}
	\xi(R) = \begin{cases}
		R-\lambda, &\text{for } R\geq\lambda, \\
		R-\sfrac{\lambda^2}{R}, &\text{for } R\leq \lambda.
	\end{cases}\label{xi-coordinate}
\end{equation}
The coordinate $\xi(R)$ is continuous in $R$ and its magnitude is proportional to the skyrmion separations for large and small $R$. The energy results are plotted in Figure \ref{fig:VRr} as a function of $\xi$. We also plot
twice the energy of the 2-i-skyrmion ($2\times2.384 = 4.768$) and four times the value of the 1-i-skyrmion ($4\times1.243=4.972$). We see that the energy does, as expected, approach these values at large and small $R$.

Potentials which depend on parameters such as separation are difficult to calculate in the Skyrme model; this is usually done by stitching together results from gradient flow \cite{Halcrow:2015rvz}. Within the instanton approximation, this is a much simpler calculation. We note that if one uses the i-skyrmions, one can also calculate the metric on this family. If we have the metric and potential, we can form a Schr\"odinger equation which describes part of the $4$-nucleon system within the Skyrme model. The potential found here is the first step in such a calculation.
\pagebreak
\section{Conclusions}

It is clear that instantons provide a useful tool for approximating skyrmions. We have clarified and added to the dictionary between the two systems. Using a mixture of representation theory and intuition from skyrmion phenomenology we have been able to find the ADHM data corresponding to the well-known skyrmion solutions for $1\leq\BB\leq8$, alongside some families of solutions for infinitely many $\BB$, such as the tori, and spinning tops. Until now, analytic expressions for solutions to the reality condition \eqref{reality-cond} have only been achieved for very low charges, or isolated cases with very high symmetry. In contrast, here we have given multi-parameter closed-form solutions to the reality condition for arbitrary charges, and with relatively low symmetry. Further, we have used these to describe how to model the break-up of large skyrmions into constituent clusters; it is often said that skyrmions ``look like" clusters of lower-charge skyrmions, and our work provides a way to make these types of statements concrete. To illustrate this explicitly, we showed how to split up the $3$- and $4$-skyrmions, for example showing that the $4$ decomposes into $2+2$, $1+1+1+1$, $1+2+1$, and $3+1$ clusters.

This work provides a solid foundation for future progress. To describe bound states of nuclei in the Skyrme model, one must consider a static skyrmion and its vibrations. Small vibrations can be studied using a linear analysis and are classified by irreducible representations of the skyrmions' symmetry group. The classical picture was recently studied systematically \cite{GudnasonHalcrow2018vibrational} for skyrmions with $1\leq\BB\leq8$. These harmonic vibrational modes have been included in quantisation of the 4-skyrmion \cite{Walhout:1992gr, Rawlinson:2019xsn}. However, it costs little energy to break skyrmions apart and different directions in configuration space can behave disparately. An example of this is seen in Figure \ref{fig:VRr}. Hence a realistic description of nuclei must go beyond the harmonic analysis. To do so properly, we must construct the full nonlinear vibrational manifold, building upon the linear space near the minimiser. We have started this process in Section \ref{sec:energypot}, extending a one-dimensional submanifold of the 2-dimensional $E_{g}$ normal mode of the $\BB=4$ skyrmion \cite{GudnasonHalcrow2018vibrational}. To complete the analysis, we must calculate ADHM data which describe this entire 2-dimensional space.

Another important problem in nuclear physics is nucleus-nucleus scattering. To describe the quantum scattering of nuclei in the Skyrme model, one must be able to describe the skyrmions when well-separated and in arbitrary (iso)orientations. In this paper, we were able to describe separated clusters in special symmetric channels. To go beyond this we may need to forgo the hope of analytic solutions of the reality condition and instead develop a numerical technique to find approximate ADHM data. We are currently working on this project and the numerical method relies heavily on the work presented here. If this works, it will provide a general framework to describe all nuclear processes from instantons arising from ADHM data.

Recently, hundreds of new local skyrmion solutions have been found numerically \cite{gudnasonHalcrow2022smorgaasbord}. Many of the solutions have similar energies, revealing that the landscape of low energy skyrmions is highly complex. Knowledge of the energy barriers and distances between solutions is vital if one is to consider quantum mechanics on this landscape. We propose that the instanton approximation is the only technique with enough flexibility to be able to model a large number of the new solutions and probe the structure of the landscape analytically; one may even be able to understand paths between solutions.

Using instantons to describe nuclear theory is compelling as so much is known about instanton moduli spaces, which is yet to be applied to the Skyrme model. For example, in \cite{boyerhurtubisemannmilgram1993topology}, various theorems regarding the topology of the instanton moduli spaces $\II_\BB$ are proven, including the Atiyah--Jones conjecture, which says that for all $k\leq\lfloor{\sfrac{\BB}{2}}\rfloor-2$, there is an isomorphism $\pi_k(\II_\BB)\cong \pi_{k+3}(\SU(2))$; in particular $\pi_1(\II_\BB)=\Z_2$ for $\BB>5$. Probing this topology is important for questions of quantisation, such as the writing down of Finkelstein--Rubenstein constraints. Another example is that the metric is (reasonably) simple to calculate from ADHM data, and is known analytically for 2-instantons \cite{AllenSmith2013}. However, it is not obvious how or if this metric can be used to approximate skyrmion dynamics. It is possible that the most realistic Skyrme model is very close to BPS. If so, we may be able to ignore the potential energy on the space of skyrmion configurations. Then there is a possibility of describing nuclear scattering as a semi-classical scattering on $\II_\BB$. This proposition is tantalising and demands further investigation.

\section*{Acknowledgements}
The authors have benefited greatly from the lively online communities which arose over the last couple of years,  including the Scheme 3 London Mathematical Society lectures on solitons, and the ongoing \href{http://solitonsatwork.net/}{solitons at work} network. Special thanks should also go to Derek Harland for many helpful conversations regarding the ideas and results of this paper. While this work was completed, CH was supported by The Leverhulme Trust as an Early Career Fellow and by the University of Leeds as an Academic Development Fellow.
\appendix
\section{Properties of ADHM data}\label{appendix:properties-ADHM}
In this appendix we provide an overview of important properties of ADHM data. In particular, we give a proof of Lemmas \ref{lem:free-action}-\ref{lem:reps}, stated earlier, and used throughout this paper.

It is convenient here to extract the various real components of the ADHM data. Specifically, let $l_\nu:\R^\BB\lto\R$ and $m_\nu:\R^\BB\lto\R^\BB$ for $\nu=1,2,3,4$, be defined by
\begin{align}
\begin{aligned}
    L&=l_1\otimes\i+l_2\otimes\j+l_3\otimes\k+l_4\otimes \1,\\
    M&=m_1\otimes\i+m_2\otimes\j+m_3\otimes\k+m_4\otimes \1,
\end{aligned}\label{real-maps}
\end{align}
with $m_j$ all symmetric matrices. In this form, the action of gauge transformations \eqref{orthog-gt} is only on these components of $L$ and $M$, i.e.
\begin{align}
    (l_\nu,m_\nu)\mapsto(l_\nu O^{-1},Om_\nu O^{-1}),\quad\nu=1,2,3,4.\label{orthog-gt-real}
\end{align}
In terms of these real matrices, the reality condition \eqref{reality-cond} takes the form
\begin{align}
    l_4^tl_j-l_j^tl_4+[m_4,m_j]-\sum_{l,k=1}^3(l_k^tl_l+\sfrac{1}{2}[m_k,m_l])\epsilon_{jkl}=0,\label{reality-real}
\end{align}
for $j=1,2,3$, and the irreducibility condition \eqref{non-singularity-cond} is equivalent to
\begin{align}
\begin{aligned}
    \bigcap_{\nu=1}^4\ker l_\nu\cap\ker(m_\nu-x_\nu\ind_\BB)=0,\quad\text{and}\quad
    \bigoplus_{\nu=1}^4\im(l_\nu^t)\oplus\im(m_\nu-x_\nu\ind_\BB)=\R^\BB,\label{non-sing}
\end{aligned}
\end{align}
for all $(x_1,x_2,x_3,x_4)\in\R^4$. All of this can be rephrased more concisely in terms of complex maps. Define $N_1,N_2:\C^\BB\to\C^\BB$, and $\lambda_1,\lambda_2:\C^\BB\to\C$ by
\begin{equation}
    N_1=m_2+{\rm i} m_1,\quad N_2=m_4+{\rm i} m_3,\quad
    \lambda_1=l_2+{\rm i} l_1,\quad \lambda_2=l_4+{\rm i} l_3.\label{complex-data}
\end{equation}
Then \eqref{reality-real} is equivalent to the equations
\begin{align}
\begin{aligned}
    {}[N_1,N_2]+\lambda_1^t\lambda_2-\lambda_2^t\lambda_1&=0,\\
    [N_1^\dagger,N_1]+[N_2^\dagger,N_2]+\lambda_1^\dagger\lambda_1-\lambda_1^t\ol{\lambda}_1+\lambda_2^\dagger\lambda_2-\lambda_2^t\ol{\lambda}_2&=0.
\end{aligned}\label{reality-complex}
\end{align}
Furthermore, setting $z_1=x_2+{\rm i} x_1$ and $z_2=x_4+{\rm i} x_3$, we see that \eqref{non-sing} are equivalent to
\begin{align}
\begin{aligned}
    \ker\lambda_1\cap\ker\lambda_2\cap\ker(N_1-z_1\ind_\BB)\cap\ker(N_2-z_2\ind_\BB)&=0,\\
    \im(\lambda_1^\dagger)\oplus\im(\lambda_2^\dagger)\oplus\im(N_1-z_1\ind_\BB)\oplus\im(N_2-z_2\ind_\BB)&=\C^\BB.\end{aligned}\label{non-sing-complex}
\end{align}
\begin{remark}
From this, we see how to identify quaternionic ADHM data with the \textit{complex ADHM data} as in \cite{donaldson1984instantons,atiyah1978geometry}. Indeed, by setting $I:\C^2\to\C^\BB$ and $J:\C^\BB\to\C^2$ as
\begin{align}
    I=\begin{pmatrix}
    \lambda_2^t&\lambda_1^t\end{pmatrix},\quad J=\begin{pmatrix}-\lambda_1\\\lambda_2\end{pmatrix},\label{complex-vectors}
\end{align}
the equations \eqref{reality-complex} are the complex ADHM equations
\begin{align}
\begin{aligned}
    {}[N_1,N_2]+IJ&=0,\\
    [N_1^\dagger,N_1]+[N_2^\dagger,N_2]+J^\dagger J-II^\dagger&=0.
\end{aligned}
\end{align}
\end{remark} 
This equivalent form of the data is convenient for proving some key results.\footnote{Some of these results appear in a similar form in \cite{FurutaHashimoto1990}, in the context of the complex ADHM data, and we have adapted them to the quaternionic data.} For these, let $\mathcal{W}$ denote the set of all non-commuting words in $2$ variables.
\begin{lemma}\label{lem:words}
Let $(L,M)\in\A_\BB$ be a representative of the moduli space of rank $\BB$ ADHM data, decomposed as \eqref{real-maps}, and define $N_1,N_2,\lambda_1,\lambda_2$ via \eqref{complex-data}. Then for all $v\in\C^\BB\setminus\{0\}$, there exist $a_1,a_2\in\C$, and $\Pi\in\mathcal{W}$ such that
\begin{align}
    (a_1\lambda_1+a_2\lambda_2)\Pi(N_1,N_2)v\neq0.\label{words-id}
\end{align}
\end{lemma}
\noindent\textit{Proof}. Let $V=\ker(\lambda_1)\cap\ker(\lambda_2)$. Define the subspace $V_\infty\subset V$ as
\begin{align}
    V_\infty=\{u\in V\::\:\Pi(N_1,N_2)u\in V,\;\forall\;\Pi\in\mathcal{W}\}.
\end{align}
By \eqref{reality-complex}, we must have $[N_1,N_2](V_\infty)=0$. By definition, $N_1(V_\infty)\subset V_\infty$ and $N_2(V_\infty)\subset V_\infty$. So, if $V_\infty\neq0$, it follows that there exists a common eigenvector $u$ of $N_1$ and $N_2$ such that $u\in V_\infty$. But this violates \eqref{non-sing-complex} (take $z_1$ and $z_2$ as the eigenvalues of $N_1$ and $N_2$ respectively for $u$). So it holds that $V_\infty=0$. Now let $v\in\C^\BB\setminus\{0\}$. If $v\notin\ker\lambda_1$, then let $a_1=1$ and $a_2=0$ and $\Pi=\ind$, and then \eqref{words-id} holds. Similarly if $v\notin\ker\lambda_2$, do the same but with $a_1=0$ and $a_2=1$. Finally, if $v\in V$, then since $V_\infty=0$, there exists $\Pi\in\mathcal{W}$ such that $\Pi(N_1,N_2)v\notin V$. So either $\Pi(N_1,N_2)v\notin\ker\lambda_1$, and we can choose $a_1=1$ and $a_2=0$, or $\Pi(N_1,N_2)v\notin\ker\lambda_2$, and we can choose $a_1=0$ and $a_1=1$, and in both cases we obtain \eqref{words-id}.\hfill$\square$
\begin{corollary}\label{cor:L,M=0}
Let $(L,M)\in\A_\BB$. Then $L\neq0$.
\end{corollary}
\begin{lemma}\label{lem:sim-diag}
Let $(N_1,N_2,I,J)\in\A_\BB$ be defined as in \eqref{complex-data} and \eqref{complex-vectors}. Then the set of matrices $\{J\Pi(N_1,N_2)I\::\:\Pi\in\mathcal{W}\}$ cannot be simultaneously diagonalised.
\end{lemma}
\textit{Proof}. The tuple $(N_1,N_2,I,J)$ give rise to a complex vector bundle $E$ over $\CP^2$ with $c_1(E)=0$ and $c_2(E)=\BB$ (see \cite{atiyah1978geometry}). If $\{J\Pi(N_1,N_2)I\::\:\Pi\in\mathcal{W}\}$ could be simultaneously diagonalised, then $E$ would be the direct sum of line bundles, and hence $c_2(E)=0$, a contradiction.\hfill$\square$\\

Now we are in a position to prove Lemmas \ref{lem:free-action}-\ref{lem:reps}.\\

\noindent \textit{Proof of Lemma \ref{lem:free-action}.} The ``if" part of the statement is clear. Conversely, suppose $\omega\in\SU(2)$, and $\Omega\in\O(\BB)$ are such that
\begin{align}
(L,M)=(\omega L\Omega^{-1},\Omega M \Omega^{-1}).\label{omega-inv}
\end{align}
From the definition of the complex data $(N_1,N_2,I,J)$ via \eqref{real-maps}, \eqref{complex-data}, and \eqref{complex-vectors}, the action $(L,M)\mapsto(\omega L,M)$ extends to the action $(N_1,N_2,I,J)\mapsto(N_1,N_2,I\omega^{-1},\omega J)$, where here we are viewing $\omega\in\SU(2)$ in its standard $2$-dimensional complex representation. For all $\Pi\in\mathcal{W}$, the quantity $J\Pi(N_1,N_2)I$ is invariant under the action \eqref{orthog-gt-real} of gauge transformations. Thus \eqref{omega-inv} is equivalent to the condition
\begin{align}
    J\Pi(N_1,N_2)I=\omega J\Pi(N_1,N_2)I\omega^{-1},\quad\forall\;\Pi\in\mathcal{W}.
\end{align}
If $\omega\neq\pm\1$, this implies that the set $\{J\Pi(N_1,N_2)I\::\:\Pi\in\mathcal{W}\}$ is simultaneously diagonalisable, in contradiction to Lemma \ref{lem:sim-diag}. Thus $\omega=\pm\1$, and we have
\begin{align}
(L,M)=(\pm LO^{-1},OMO^{-1}).\label{gt-pm-action}
\end{align}
\eqref{gt-pm-action} implies that for all $a_1,a_2\in\C$ and $\Pi\in\mathcal{W}$, $a_1\lambda_1+a_2\lambda_2=\pm(a_1\lambda_1+a_2\lambda_2)O^{-1}$ and $\Pi(N_1,N_2)=O\Pi(N_1,N_2)O^{-1}$. Now let $v\in\C^\BB\setminus\{0\}$ and $u=(\ind_\BB\mp O)v$. Then for all $a_1,a_2\in\C$ and $\Pi\in\mathcal{W}$ we have
\begin{multline}
    (a_1\lambda_1+a_2\lambda_2)\Pi(N_1,N_2)u=(a_1\lambda_1+a_2\lambda_2)\Pi(N_1,N_2)v\mp(a_1\lambda_1+a_2\lambda_2)\Pi(N_1,N_2)Ov\\
    =(a_1\lambda_1+a_2\lambda_2)\Pi(N_1,N_2)v-(a_1\lambda_1+a_2\lambda_2)O^{-1}O\Pi(N_1,N_2)O^{-1}Ov=0.
\end{multline}
Thus, by Lemma \ref{lem:words} it follows that $u=0$, i.e. $\im(\ind_\BB\mp O)=0$. Therefore $O=\pm\ind_\BB$.\hfill$\square$\\

\noindent \textit{Proof of Lemma \ref{lem:reps}.} Since $G$ is the binary group of a subgroup of $\SO(3)$, it has even order, and so every such group may be presented by at most two generators $\SS,\TT\in G$ which satisfy relations of the form
\begin{align}
    \SS^a=\TT^b=(\SS\TT)^c=-\id,
\end{align}
for $a,b,c\in\Z^+$. Therefore, the representation $p$ satisfies
\begin{align}
    p(\SS)^a=p(\TT)^b=p(\SS\TT)^c=\varepsilon,
\end{align}
with $\varepsilon=\pm1$ denoting the sign of the representation. It is clear that ADHM data $(L,M)\in\A_\BB$ will be $\RR=\{(q,p(q))\::\:q\in G\}$-invariant if and only if it satisfies \eqref{ADHM-symm} for each generator $\SS$ and $\TT$. Letting $O_\SS$ and $O_\TT$ be the corresponding c.g.ts, applying \eqref{ADHM-symm} enough times we see that
\begin{align}
\begin{aligned}
L&=-\varepsilon LO_\SS^{-a},&M&=O_\SS^aMO_\SS^{-a},\\
L&=-\varepsilon LO_\TT^{-b},&M&=O_\TT^bMO_\TT^{-b},\\
L&=-\varepsilon L(O_\SS O_\TT)^{-c},&M&=(O_\SS O_\TT)^cM(O_\SS O_\TT)^{-c},
\end{aligned}
\end{align}
and thus by Lemma \ref{lem:free-action}, we have
\begin{align}
    O_\SS^a=O_\TT^b=(O_\SS O_\TT)^c=-\varepsilon\ind_\BB.
\end{align}
\hfill$\square$
\newpage
\bibliographystyle{unsrt}
\bibliography{refs}

\begin{thebibliography}{10}

\bibitem{ADHM1978construction}
M~F Atiyah, V~G Drinfeld, N~J Hitchin, and Y~I Manin.
\newblock Construction of instantons.
\newblock {\em Phys. Lett. A}, 65(3):185--187, 1978.

\bibitem{AtiyahManton1989}
M~F Atiyah and N~S Manton.
\newblock Skyrmions from instantons.
\newblock {\em Phys. Lett. B}, 222(3):438--442, 1989.

\bibitem{Skyrme1962nucl}
T~H~R Skyrme.
\newblock A unified field theory of mesons and baryons.
\newblock {\em Nucl. Phys.}, 31:556, 1962.

\bibitem{witten1979baryons}
E~Witten.
\newblock Baryons in the {1N} expansion.
\newblock {\em Nucl. Phys. B}, 160(1):57--115, 1979.

\bibitem{AdkinsNappiWitten1983static}
G~S Adkins, C~R Nappi, and E~Witten.
\newblock Static properties of nucleons in the skyrme model.
\newblock {\em Nucl. Phys. B}, 228(3):552--566, 1983.

\bibitem{Braaten:1985np}
E~Braaten and L~Carson.
\newblock {The Deuteron as a Soliton in the Skyrme Model}.
\newblock {\em Phys. Rev. Lett.}, 56:1897, 1986.

\bibitem{Irwin:1998bs}
P~Irwin.
\newblock {Zero mode quantization of multi - Skyrmions}.
\newblock {\em Phys. Rev. D}, 61:114024, 2000.

\bibitem{Lin:2008uf}
W~T Lin and B~Piette.
\newblock {Skyrmion Vibration Modes within the Rational Map Ansatz}.
\newblock {\em Phys. Rev. D}, 77:125028, 2008.

\bibitem{Halcrow:2016spb}
C~J Halcrow, C~King, and N~S Manton.
\newblock {A dynamical $\alpha$-cluster model of $^{16}$O}.
\newblock {\em Phys. Rev. C}, 95(3):031303, 2017.

\bibitem{Halcrow:2015rvz}
C~J Halcrow.
\newblock {Vibrational quantisation of the B = 7 Skyrmion}.
\newblock {\em Nucl. Phys. B}, 904:106--123, 2016.

\bibitem{Rawlinson:2017rcq}
J~I Rawlinson.
\newblock {An Alpha Particle Model for Carbon-12}.
\newblock {\em Nucl. Phys. A}, 975:122--135, 2018.

\bibitem{AtiyahManton1993}
M~F Atiyah and N~S Manton.
\newblock Geometry and kinematics of two skyrmions.
\newblock {\em Commun. Math. Phys.}, 153(2):391--422, 1993.

\bibitem{Leese:1994hb}
R~A Leese, N~S Manton, and B~J Schroers.
\newblock {Attractive channel skyrmions and the deuteron}.
\newblock {\em Nucl. Phys. B}, 442:228--267, 1995.

\bibitem{LeeseManton1994stable}
R~A Leese and N~S Manton.
\newblock Stable instanton-generated {Skyrme} fields with baryon numbers three
  and four.
\newblock {\em Nucl. Phys. A}, 572(3-4):575--599, 1994.

\bibitem{houghton1999-3Skyrme}
C~J Houghton.
\newblock Instanton vibrations of the {3-Skyrmion}.
\newblock {\em Phys. Rev. D}, 60(10):105003, 1999.

\bibitem{SingerSutcliffe1999}
M~A Singer and P~M Sutcliffe.
\newblock Symmetric instantons and {Skyrme} fields.
\newblock {\em Nonlinearity}, 12(4):987, 1999.

\bibitem{sutcliffe2004Buckyball}
P~M Sutcliffe.
\newblock Instantons and the buckyball.
\newblock In {\em Proc. R. Soc. Lon. A: Mathematical, Physical and Engineering
  Sciences}, volume 460, pages 2903--2912. The Royal Society, 2004.

\bibitem{sutcliffe2010skyrmions}
P~Sutcliffe.
\newblock Skyrmions, instantons and holography.
\newblock {\em J. High Energ. Phys.}, 2010(8):19, 2010.

\bibitem{SakaiSugimoto2005}
T~Sakai and S~Sugimoto.
\newblock Low energy hadron physics in holographic {QCD}.
\newblock {\em Prog. Th. Phys.}, 113(4):843--882, 2005.

\bibitem{HalcrowWinyard2021consistent}
C~Halcrow and T~Winyard.
\newblock A consistent two-skyrmion configuration space from instantons.
\newblock {\em J. High Energ. Phys.}, 2021(12):1--24, 2021.

\bibitem{CorkHarlandWinyard2021gaugedSkyrmelowbinding}
J~Cork, D~Harland, and T~Winyard.
\newblock A model for gauged skyrmions with low binding energies.
\newblock {\em J. Phys. A: Math. Th.}, 55(1):015204, 2021.

\bibitem{AtiyahSutcliffe2005skyrmions}
M~Atiyah and P~Sutcliffe.
\newblock Skyrmions, instantons, mass and curvature.
\newblock {\em Phys. Lett. B}, 605(1-2):106--114, 2005.

\bibitem{cork2018skyrmions}
J~Cork.
\newblock Skyrmions from calorons.
\newblock {\em J. High Energ. Phys.}, 2018(11):137, 2018.

\bibitem{Uhlenbeck1982}
K~K Uhlenbeck.
\newblock Removable singularities in {Yang-Mills} fields.
\newblock {\em Commun. Math. Phys.}, 83(1):11--29, 1982.

\bibitem{donaldson1992boundary}
S~K Donaldson.
\newblock Boundary value problems for {Yang--Mills} fields.
\newblock {\em J. Geom. Phys.}, 8(1-4):89--122, 1992.

\bibitem{donaldson1984instantons}
S~K Donaldson.
\newblock Instantons and geometric invariant theory.
\newblock {\em Commun. Math. Phys.}, 93(4):453--460, 1984.

\bibitem{atiyah1978geometry}
M~F Atiyah.
\newblock Geometry of {Yang-Mills} fields.
\newblock In {\em Mathematical problems in theoretical physics}, pages
  216--221. Springer, 1978.

\bibitem{christWeinbergStanton1978general}
N~H Christ, E~J Weinberg, and N~K Stanton.
\newblock General self-dual {Yang-Mills} solutions.
\newblock {\em Phys. Rev. D}, 18(6):2013, 1978.

\bibitem{CorriganFairlieTempletonGoddard1978green}
E~F Corrigan, D~B Fairlie, S~Templeton, and P~Goddard.
\newblock A green function for the general self-dual gauge field.
\newblock {\em Nucl. Phys. B}, 140(1):31--44, 1978.

\bibitem{CorriganGoddard1984}
E~Corrigan and P~Goddard.
\newblock Construction of instanton and monopole solutions and reciprocity.
\newblock {\em Annal. Phys.}, 154(1):253--279, 1984.

\bibitem{FurutaHashimoto1990}
M~Furuta and Y~Hashimoto.
\newblock Invariant instantons on {$S^4$}.
\newblock {\em J. Fac. Sci. Univ. Tokyo}, 37:585--600, 1990.

\bibitem{AllenSutcliffe2013}
J~P Allen and P~M Sutcliffe.
\newblock {ADHM} polytopes.
\newblock {\em J. High Energ. Phys.}, 2013(5):1--36, 2013.

\bibitem{FeistLauManton2013skyrmions}
D~T~J Feist, P~H~C Lau, and N~S Manton.
\newblock Skyrmions up to baryon number {108}.
\newblock {\em Phys. Rev. D}, 87(8):085034, 2013.

\bibitem{mankomantonwood2007}
O~V Manko, N~S Manton, and S~W Wood.
\newblock Light nuclei as quantized skyrmions.
\newblock {\em Phys. Rev. C}, 76(5):055203, 2007.

\bibitem{battye2001solitonic}
R~A Battye and P~M Sutcliffe.
\newblock Solitonic fullerene structures in light atomic nuclei.
\newblock {\em Phys. Rev. Lett.}, 86(18):3989, 2001.

\bibitem{GudnasonNitta2015baryonic}
S~B Gudnason and M~Nitta.
\newblock Baryonic torii: toroidal baryons in a generalized {Skyrme} model.
\newblock {\em Phys. Rev. D}, 91(4):045027, 2015.

\bibitem{Jackiw1977}
R~Jackiw, C~Nohl, and C~Rebbi.
\newblock Conformal properties of pseudoparticle configurations.
\newblock {\em Phys. Rev. D}, 15(6):1642, 1977.

\bibitem{BraamAustin1990boundary}
P~J Braam and D~M Austin.
\newblock Boundary values of hyperbolic monopoles.
\newblock {\em Nonlinearity}, 3(3):809, 1990.

\bibitem{cockburn2014symmetric}
A~Cockburn.
\newblock Symmetric hyperbolic monopoles.
\newblock {\em J. Phys. A: Math. Th.}, 47(39):395401, 2014.

\bibitem{HoughtonSutcliffe1996monopole}
C~J Houghton and P~M Sutcliffe.
\newblock Monopole scattering with a twist.
\newblock {\em Nucl. Phys. B}, 464(1-2):59--84, 1996.

\bibitem{walet1996quantising}
N~R Walet.
\newblock Quantising the {$B= 2$} and {$B= 3$ Skyrmion} systems.
\newblock {\em Nucl. Phys. A}, 606(3-4):429--458, 1996.

\bibitem{mantonPiette2001SkyRat}
N~S Manton and B~M A~G Piette.
\newblock Understanding skyrmions using rational maps.
\newblock In {\em European Congress of Mathematics}, pages 469--479. Springer,
  2001.

\bibitem{kato2021symmetric}
T~Kato, A~Nakamula, and K~Takesue.
\newblock Symmetric calorons of higher charges and their large period limits.
\newblock {\em J. Geom. Phys.}, 162:104071, 2021.

\bibitem{Ward2004}
R~S Ward.
\newblock Symmetric calorons.
\newblock {\em Phys. Lett. B}, 582(3):203--210, 2004.

\bibitem{GudnasonHalcrow2018vibrational}
S~B Gudnason and C~Halcrow.
\newblock Vibrational modes of skyrmions.
\newblock {\em Phys. Rev. D}, 98(12):125010, 2018.

\bibitem{Walhout:1992gr}
T~S Walhout.
\newblock {Quantizing the four baryon skyrmion}.
\newblock {\em Nucl. Phys. A}, 547:423--446, 1992.

\bibitem{Rawlinson:2019xsn}
J~I Rawlinson.
\newblock {Coriolis terms in Skyrmion Quantization}.
\newblock {\em Nucl. Phys. B}, 949:114800, 2019.

\bibitem{gudnasonHalcrow2022smorgaasbord}
S~B Gudnason and C~Halcrow.
\newblock A {sm\"org\aa sbord} of skyrmions.
\newblock {\em arXiv preprint arXiv:2202.01792}, 2022.

\bibitem{boyerhurtubisemannmilgram1993topology}
C~P Boyer, J~C Hurtubise, B~M Mann, and R~J Milgram.
\newblock The topology of instanton moduli spaces, {I: The Atiyah-Jones}
  conjecture.
\newblock {\em Annal. Math.}, 137(3):561--609, 1993.

\bibitem{AllenSmith2013}
J~P Allen and D~J Smith.
\newblock The low energy dynamics of charge two dyonic instantons.
\newblock {\em J. High Energ. Phys.}, 2013(2):113, 2013.

\end{thebibliography}
\end{document}